\documentclass[a4paper]{scrartcl}
\usepackage[utf8]{inputenc}
\usepackage[T1]{fontenc}
\usepackage{amsmath}
\usepackage{amssymb}
\usepackage{amsfonts}
\usepackage{biblatex}
\usepackage{mathtools}
\usepackage{tikz}
\usetikzlibrary{matrix}
\numberwithin{equation}{section}
\usepackage{multirow}
\usepackage{hyperref}

\usepackage{varioref}
\usepackage[noabbrev]{cleveref}
\usepackage{float}
\usepackage{lmodern}
\usepackage{palatino}
\usepackage[english]{babel}
\usepackage{mathptmx}
\usepackage[scaled=.90]{helvet}
\usepackage{titlesec}
\titleformat{\section}{\normalfont\rmfamily\Large\bfseries}
{\thesection}{1em}{}
\titleformat{\subsection}{\normalfont\rmfamily\Large\bfseries}
{\thesubsection}{1em}{}
\titleformat{\subsubsection}{\normalfont\rmfamily\Large\bfseries}
{\thesubsubsection}{1em}{}
\usepackage[affil-it]{authblk}
\usepackage{etoolbox}
\usepackage{indentfirst}
\usepackage{easybmat}
\usepackage{xcolor}

\begin{document}
\title{\textrm{ Superpotentials of D-branes in  Calabi-Yau Manifolds with Several Moduli by Mirror Symmetry and Blown-up}}
\date{}
\author[]{ Xuan Li, Yuan-Chun Jing and Fu-Zhong Yang  \thanks{corresponding author \hspace{1cm} E-mail:\href{fzyang@ucas.ac.cn}{fzyang@ucas.ac.cn}}}
\affil[]{School of Physical Sciences,University of Chinese Academy of Sciences,\\No.19(A) Yuquan Road, Shijingshan District, Beijing, P.R.China 100049}
\maketitle

\begin{abstract}
\textbf{Abstract}: 
We study  B-brane superpotentials depending on several closed- and open- moduli on Calabi-Yau hypersurfaces and complete intersections. By blowing up the ambient space along a curve wrapped by B-branes in a Calabi-Yau manifold, we obtain a blow-up new manifold and the period integral satisfying the GKZ-system.  Via mirror symmetry to A-model,  we calculate the  superpotentials and extract Ooguri-Vafa invariants for concrete examples of several  open-closed  moduli in Calabi-Yau manifolds .  
\end{abstract}

\newpage
\tableofcontents
\newpage
\section{Introduction}  
The type IIB compactification with branes  can be described by an effective $N=1$ supergravity theory with a non-trivial superpotential on the open-closed moduli space,  because the D-branes wrapping  supersymmetric cycles reduce the N=2 supersymmetry to N=1 supersymmetry. In topological string theory, there are two type D-branes. A-branes wrap special lagrangian cycles and B-branes wrap holomorphic cycles that can be even dimensional in a Calabi-Yau threefold.   

On the A-model side, the superpotential is related to  the topological string amplitudes, which counts disk instantons\parencites{Katz2001}{Li2001}. On the B-model side, the topological string is related to holomorphic Chern-Simons theory\cite{Witten1992}. The B-brane superpotential is given by a integral over 3-chain with boundary consisting of  2-cycles $\gamma$ around the B-branes, which is the sections of a line bundle over the open-closed moduli space  described by the holomorphic N=1 special geometry \parencites{Lerche2002}{Lerche2002a}{Mayr2001}.
The B-brane superpotential can be  expressed as a  linear combination of the integral of the basis of relative period. When considering B-brane  wrapping two curves within the same homology class, the superpotential changes on the two sides of the domain wall whose tension is in terms of the Able-Jacobi map. 

 In physics, the D-brane superpotential is topological sector of the d=4,N=1 superpymmetric spacetime effective Lagrangian, as well as the generating function of open string topological field theory correlators. It encode the instanton correction and its derivatives  determine structure constants of 2d chiral ring and Gauss-Manin connection of the vacuum bundle on the moduli space. The flatness of this connection determines the Picard-Fuchs equations satisfied by period vector.  The  expansion of superpotentials at large volume phase underlies the Ooguri-Vafa invariants counting the holomorphic disks ending on a lagrangian submanifold on the A-model side.  These invariants  closely relate to space of the states, non-pertubative effects, and geometric properties of the moduli space.

From the perspective of the deformation theory, the deformations of a curve in Calabi-Yau threefold are given by the  sections of the normal bundle. The holomorphic sections lead to massless or light fields in the effective theory and the non-holomorphic sections lead to  massive fields whose masses are given by the volume change under infinitesimal deformations.  The superpotentials are related to the deformations with masses vanishing at some point in the closed moduli space. In order words, determining the B-brane superpotentials is equivalent to solve the deformation theory of a pair $(X,S)$  with  curve $S$ and  Calabi-Yau manifold $X$. When a D5  brane wrapping a rational curve, the non-trivial superpotential is defined on a family of  curves $S$, whose members are in general non-holomorphic except at some critical points where $S$ is  holomorphic curve.  The critical locus corresponds to the supersymmetric vacus,  other points in moduli space correspond to the obstructed deformation of the rational curve and excitation about the supersymmetric minimum.

 The computation of off-shell superpotential  for a toric brane, has been presented by local case in \parencites{Aganagic2000}{Lerche2002}{Lerche2002a}, and extended to compact Calabi-Yaus in \parencites{Alim2009}{Jockers2008}{Jockers2009}. For this brane, the onshell superpotentials and flat coordinates are the solutions to a system of open-closed Picard-Fuchs equations, which arise as a consequence of the  N = 1 special geometry. These equation can be obtained by Griffish-Dwork reduction method or GKZ system. When a  B-brane wrapping a curve $S$ in $X$, the blowing up $X$ along $S$ lead to a new manifold $X^\prime$  with an exceptional divisor $E$ \parencites{Grimm2010}{Grimm2008}. Meanwhile, the deformation theory of $(X,S)$ is equivalent to that of $(X^\prime,E)$\footnote{ Section 4.4 in \cite{Grimm2008} and footnote 17 in \cite{Grimm2010}}.  The B-brane superpotential on the Calabi-Yau threefold $X$ can be calculated in terms of period vector of  manifold $X^\prime$. In this note, we first calculate B-brane superpotentials  in Calabi-Yau manifolds with several moduli for via blowing up method, then  extract Ooguri-Vafa invariants from A-model side at large volume phase by mirror symmetry.  

  The organization of this paper is as follows. In section 2, we introduce the background and formalism. To begin with, we  review D-brane superpotential in the Type II string theory and relative cohomology description, recall the basic toric geometry about constructing Calabi-Yau manifold, generalized GKZ system and its local solutions, and outline the procedure to blow up a curve on a Calabi-Yau manifold.  In section 3, for degree -9,-8,-12 Calabi-Yau hypersurface and degree-$(3,3)$ complete intersection Calabi-Yau manifold, we apply the blow up method to the mainfold with a curve on it and obtain a new Kahler manifold an exceptional divisor. The Picard-Fuchs equations and their solutions are derived by GKZ hypergeometric system from toric data of the enhanced polyhedron. The superpotential are identified as double-logarithmic solutions of the Picard-Fuchs equations and Ooguri-Vafa invariants are extracted at large volume phase. The last section is a brief summary and further discussions.In Appendix A, we summarize the GKZ-system for two complete intersections Calabi-Yau manifolds $X^{(112|112)}_{[4,4]}$ and $X^{(123|123)}_{[6,6]}$. In Appendix B, we present the compact instanton invariants of above models for first several orders.

\section{Toric geometry and Blowing up}
\subsection{Toric geometry and GKZ system}\label{sec:2.2}
A Calabi-Yau manifold can be defined as a hypersurface or complete intersection of several hyperserfaces in ambient toric variety. We refer to \cite{Cox2011} for the background of toric geometry,construction of Calabi-Yau manifolds \parencites {Batyrev1993}{Batyrev1994}, and GKZ system \parencites{Hosono1995}{Hosono1993}{Hosono1994}{Stienstra2005}{Cox1999}.
  
  For a mirror pair of compact hypersurfaces $(X^*,X)$, one may associate a pair of integral polyhedra $(\Delta^*,\Delta)$ in a four-dimensional integral lattice  and its dual.  The n integral points of the polyhedron correspond to homogeneous coordinates $x_i$ on the toric ambient space and satisfy  linear relations
\begin{equation*}
 \sum_i l^j_i v_i=0,\quad a=1,...,h^{2,1}
\end{equation*}
where $l^j_i$ is the ith component of the charge vector $l^j$.
The integral entries of the vectors $l^j$  define the weights $l^j_i$
of the coordinates $x_i$ under the $\mathbb{C}^*$ action
\begin{equation*}
x_i\rightarrow (\lambda_j)^{l^j_i}x_i,\quad \lambda_j\in \mathbb{C}^*
\end{equation*}
 and  $l^j_i$ 's are the $U(1)$  charges of the fields in the gauged linear sigma model (GLSM)\cite{Witten1993}.
 In above description, the mirror Calabi-Yau threefold is determined as a hypersurface in the dual toric ambient space with constraints
 \begin{equation}
 P=\sum_i a_i y_i,\quad \prod_i y_i^{l^j_i}=z_j
 \end{equation}
 Here $z_j$ denotes the complex structure moduli of $X$. In terms of vertices $v_j^*\in\Delta^*,v_i\in\Delta$
 \begin{equation}\label{eq:2.16}
 P(x;a)=\sum_{v_j^* \in \Delta^*} a_j \prod_i x_i^{\langle v_j^*,v_i\rangle+1}
 \end{equation}
The complete intersection Calabi-Yau threefolds can be constructed similarly and we omit the details.
Equivalently,  Calabi-Yau hypersurfaces  is also given by the zero loci of certain sections of the anticanonical bundle. The toric variety contains a canonical Zariski open torus $\mathbb{C}^4$ with coordinates $X=(X_1,X_2,X_3,X_4)$.                                                                                          The sections are
\begin{equation}
P_{\Delta^*}(X,a)=\sum_{v^*_i\in \Delta^* \cap \mathbb{Z}^{5}}a_iX^{v_i^*}
\end{equation}
After homogenization, above equation is the same as \ref{eq:2.16}.
On these manifolds, a mirror pairs of branes, defined in \cite{Aganagic2000}  by another N charge vectors $\hat{l}^j$. The special Lagrangian submanifold
wrapped by the A-brane is described in terms of the vectors
$\hat{l}^j$ satisfying
\begin{equation*}
\sum_i \hat{l}^j_i|x_i|^2=c_j,\quad\sum_i \hat{l}^j_i=0
\end{equation*}
where $c_j$ parametrize the brane position. And the holomorphic submanifold wrapped by mirror B-brane is defined by the following equation
\begin{equation*}
\prod_i y_i-\hat{z}_a=0,\quad \hat{z}_a=\epsilon_a e^{-c_a}
\end{equation*}
The $N=2$ case, a toric curve, is more interesting to us and it is the geometry setting we are studying in this notes. 
  To handle the toric curve case, we consider the enhanced polyhedron method  proposed in \cite{Alim2009}.  It is possible in a simple manner to construct the enhanced polyhedron from the original polyhedra and the toric curve specified by two charge vectors.  We denote the vertices of $\Delta$ by $v_i$, $i=1,...,n $, with $v_0$ the origin, its charge vectors by $l^{i}$, and two brane vectors by $\hat{l}^1$ and $\hat{l}^2$. We add 4 points to $\Delta^*$ to define a new polyhedron $\Delta^e$ with vertices
\begin{equation}\label{eq:2.19}
\begin{aligned}
&X: v_i =(1,0,v_i,0),\quad i=0,...,n\\
&\hat{l}^1: v_{n+1}=(0,1,v_1^-,-1),\quad v_{n+2}=(0,1,v_1^+,-1)\\
&\hat{l}^2: v_{n+3}=(0,1,v_2^-,-1),\quad v_{n+4}=(0,1,v_2^+,-1)
\end{aligned}
\end{equation}
where we use the abbreviation 
\begin{equation}
\begin{gathered}
v_1^+=\sum_{\hat{l}_i^1>0}\hat{l}_i^1 v_i,\quad v_1^-=-\sum_{\hat{l}_i^1<0}\hat{l}_i^1 v_i\\
v_2^+=\sum_{\hat{l}_i^2>0}\hat{l}_i^2\ v_i,\quad v_2^-=-\sum_{\hat{l}_i^2<0}\hat{l}_i^2 v_i
\end{gathered}
\end{equation}
The first line of \ref{eq:2.19} simply embeds the original toric data associated to $\Delta$ into $\Delta^e$, whereas the second and third line translate the brane data into geometric data of $\Delta^e$. 

Given the toric data , the GKZ-system on the complex structure moduli space of $X$ is given by the standard formula
\begin{equation}\label{eq:2.21}
\begin{gathered}
\mathcal{L}_i=\prod_{l^i_j>0}(\frac{\partial}{\partial a_j})^{l^i_j}-\prod_{l^i_j<0}(\frac{\partial}{\partial a_j})^{-l^i_j},\quad i=1,...,h^{2,1}+2\\
\mathcal{Z}_i=\sum_j(\bar{v}_j)^i \vartheta_j-\beta_i,\quad j=0,...,6
\end{gathered}
\end{equation}
Here $\beta=(-1,0,0,0,0)$ is the so-called exponent of GKZ system, $\vartheta_j=a_j \frac{\partial}{\partial a_j}$  are the logarithmic derivative and $\bar{v}_j=(1,v_j)$.  The operators $\mathcal{L}_i$'s express the trivial algebraic relations  among monomials entering  hypersurface constraints, $\mathcal{Z}_0$ expresses the infinitesimal generators of overall rescaling,$\mathcal{Z}_i,i\neq 0$'s eare the infinitesimal generators of rescalings of coordinates $x_j$.  All GKZ operators can annihilate the period matrix, thus determine the mirror maps and superpotentials.

This immediately yields a natural choice of complex coordinates given by
\begin{equation}\label{eq:2.22}
z^j=(-)^{l^j_0} \prod_{i} a_i^{l^j_i}
\end{equation}
 And from the operators $\mathcal{L}_i$ ,it is easy to obtain a complete set of Picard-Fuchs operators $\mathcal{D}_i$. Using monodromy information and knowledge of the classical terms, their solution can be associated to integrals over an integral basis of cycles
in $H^3(X,\mathbb{Z})$ and given the flux quanta explicit superpotentials can be written down. 

For appropriate choice of basis vector $l^j$, solutions to the GKZ system can be written interm of the generating functions in these variables
\begin{equation*}
\varpi(z;\rho)=\sum \frac{\Gamma(1-\sum_j l^j_0(n_j+\rho_j))}{\prod_{i>0}\Gamma(1+\sum_j l^j_i(n_j+\rho_j))}\prod_k z_k^{n_j+\rho_j}
\end{equation*}
 then we have a natural basis for the period vector
 \begin{equation*}
 \begin{aligned}
 \omega_0(z)&=\varpi(z;\rho)|_{\rho \rightarrow 0},\\
 \omega_{1,i}(z)&=\partial_{\rho_i}\varpi(z;\rho)|_{\rho \rightarrow 0},\\\omega_{2,i}(z)&=\sum_{j,k}K_{ijk}\partial_{\rho_j}\partial_{\rho_k}\varpi(z;\rho)|_{\rho \rightarrow 0}\\
 ...
 \end{aligned}
 \end{equation*}
 For a maximal triangulation corresponding to a large complex structure point centered at $z = 0$,$ \forall a$, $\omega_0(z)= 1 + \mathcal{O}(z)$ and $\omega_{1,i}(z)\sim \log(z_i)$ that define the open-closed mirror maps 
\begin{equation}
t_i(z)=\frac{\omega_{1,i}(z)}{\omega_0(z)}=\frac{1}{2 \pi i}\log(z_i)+S(z),\quad q_i=e^{2 \pi i t_i}
\end{equation}
where $S(z)$ is a series in the coordinates $z$.In addition, the special solution $\Pi=\mathcal{W}_{open}(z)$ has further property that its instanton expansion near a large volume/large complex structure point encodes the Ooguri-Vafa invariants of the brane geometry.
\begin{equation}
\mathcal{W}_{inst}(q)=\sum_\beta G_\beta q^\beta=\sum_\beta \sum_{k=1}^\infty N_\beta \frac{q^{k \cdot\beta}}{k^2}
\end{equation}

\subsection{Blowing Up and Hodge Structure}
Blowing up in algebraic geometry is an important tool in this work. Now, we review the  construction and properties  of blowing up a manifold along its submanifold. Given $S$ be a curve in a Calabi-Yau threefold $X\in Z$ with $Z$ an ambient toric variety,  we can blow up along $S$ to obtain a new manifold.

According to Section 2.2 (d) in \cite{Shafarevich2016}, for this case that  $X\subset Z$ is a closed irreducible non-singular subvariety of $Z$ and $X$ is transversal to $S$ at every point $S \cap X$, $\pi: Z^\prime \rightarrow Z$ be the blowup of $S$. 
 Then the subvariety $\pi^{-1}(X)$ consist of two irreducible components,
\begin{equation*}
\pi^{-1}(X)=\pi^{-1}(S \cap X) \cup X^\prime
\end{equation*}
and $\pi:X^\prime \rightarrow X$ defines the blow-up  of $X$ with center in $S \cap X = S$,i.e.$X^\prime$ is the manifold obtained from blowing up $X$ along $S$. The subvariety $X^\prime \subset Z^\prime$ is called the \textit{
birational transform} of $X \subset Z$ under the blowup. 

 First, by the local construction, we consider an three dimensional multidisk in $X$ $\Delta$ with holomorphic coordinates $x_i,i=1,2,3$, and $V$ is specified by $x_1=x_2=0$ on each $\Delta$. Then we define the smooth variety
\begin{equation*}
\tilde{\Delta}\subset \Delta \times \mathbb{P}^1
\end{equation*}
as follows 
\begin{equation*}
\tilde{\Delta}=\{(x_1,x_2,x_3,(y_1:y_2))\subset \Delta\times \mathbb{P}^1:x_2y_1-x_1y_2=0\}
\end{equation*}
Here $y_1,y_2$ are the homogeneous coordinates on $\mathbb{P}^1$. The projection map $\pi:\tilde{\Delta}\rightarrow \Delta$ on the first factor is clearly  an isomorphism away from $V$, while the inverse image of a point $z \in V$ is a projective space  $\mathbb{P}^1$. The manifold $\tilde{\Delta}$, together with the projection map $\pi$ is the blow-up of $\delta$ along $V$; The inverse image $E=\pi^{-1}(V)$ is an exceptional divisor of the blow-up. For two coordinates patches $U_i=(y_i \neq 0),i=1,2$, they have holomorphic coordinates respectively
\begin{equation*}
\begin{gathered}
z^{(1)}_1=x_1,\quad z_2^{(1)}=\frac{y_2}{y_1}=\frac{x_2}{x_1},\quad y_3=x_3\\
z^{(2)}_1=\frac{y_1}{y_2}=\frac{x_1}{x_2},\quad z_2^{(1)}=x_2,\quad y_3=x_3\\
\end{gathered}
\end{equation*} 
with transition function on $U_1 \cap U_2$given by $g_{ij}=z^{(j)}_i=\frac{y_i}{y_j}=\frac{x_i}{x_j}$. Next we consider the global construction of the blow-up manifold.  Let $X$ be a complex manifold of dimension three and $S \subset X$ be a curve. Let $\{U_\alpha\}$ be a collection of disks in $X$ covering $S$ such that in each disk $\Delta_\alpha $ the subvariety $S\cap \Delta_\alpha$ may be given as the locus $(x_1=x_2=0)$, and let $\pi_\alpha:\tilde{\Delta}_\alpha \rightarrow \Delta_\alpha$ be the blow-up of $\Delta_\alpha$ along $S \cap \Delta_\alpha$. We then have 
\begin{equation*}
\pi_{\alpha \beta}:\pi_{\alpha \beta}^{-1}(U_\alpha \cap U_\beta)\rightarrow \pi_\beta^{-1}(U_\alpha \cap U_\beta)
\end{equation*}
and using them, we can patch together the local blow-ups $\tilde{\Delta}_\alpha $ to form a manifold
\begin{equation*}
\tilde{\Delta}=\cup_{\pi_{\alpha \beta}}\tilde{\Delta}_\alpha
\end{equation*}
Finally, sinve $\pi$ is an isomorphism away from $X \cap (\cup \Delta_\alpha)$, we can take
\begin{equation*}
X^\prime=\tilde{\Delta}\cup_\pi X-S
\end{equation*}
$X^\prime$, together with the projection map $\pi:X^\prime \rightarrow X$ extending $\pi$ on $\tilde{\Delta}$ and the identity on $X-S$, is called the blow-up of $X$ along $S$, and the inverse image $\pi^{-1}(S)$ is an exceptional divisor.

From the excision theorem of cohomology in algebraic topology\cite{Hatcher2001},\begin{equation}
H^3(X,S)\cong H^3(X-S)\cong H^3(X^\prime-E) \cong H^3(X^\prime,E)
\end{equation}
which means that the variation of the mixed Hodge structures of $H^3(X,S)$ and $H^3(X^\prime,E)$ over the corresponding  moduli space are equivalent.
The mixed Hodge structure as follow,
\begin{equation}
\phi:H^3(X^\prime-E)\tilde{\longrightarrow}\oplus_{p+q=3}H^q(X^\prime,\Omega^\prime)
\end{equation}
where $\Omega^\prime$ denotes the holomorphic p-forms on $X^\prime$. The filtrations have the form
\begin{equation*}
F^m H^3=\oplus_{p\leq m}H^{3-p}(X^\prime,\Omega^\prime)
\end{equation*}
and 
\begin{equation*}
W_{-1} H^3=0,\quad W_0H^3=H^3(X^\prime),\quad W_1H^3=H^3(X^\prime-E)
\end{equation*}
Additionally, the mixed Hodge structure has graded weights 
\begin{equation*}
Gr^W_k H^3= Gr^W_{-k+3} H^3 / Gr^W_{-(k+1)+3} H^3
\end{equation*}
that take the following form for the divisor E
\begin{equation*}
\begin{gathered}
Gr^W_3 H^3= Gr^W_0 H^3 / Gr^W_{-1} H^3 \cong H^3(X^\prime),\\Gr^W_2 H^3= Gr^W_1 H^3 / Gr^W_{0} H^3 \cong H^2(E)
\end{gathered}
\end{equation*}
The reason to consider these (graded) weights is the following: The mixed Hodge structure is defined such that the Hodge filtration $F^m H^3$ induces a pure Hodge structure on each graded weight, i.e.
on $Gr^W_2 H^3$ and $Gr^W_3 H^3$. Thus, the following two induced filtrations on $Gr^W_3 H^3$
\begin{equation}
H^3(X^\prime)\cap F^3H^3 \subset H^3(X^\prime)\cap F^2H^3 \subset H^3(X^\prime)\cap F^1H^3 \subset H^3(X^\prime)\cap F^0H^3=H^3(X^\prime)
\end{equation}
and on $Gr^W_2 H^3$
\begin{equation}
H^2(E)\cap F^2H^3 \subset H^2(E)\cap F^1 H^3 \subset H^2(E)\cap F^0H^3 =H^2(E)
\end{equation}
lead to pure Hodge structures on $H^3(X^\prime)$ and $H^2(E)$. $H^3(X^\prime-E)$ forms a bundle $\mathcal{H}^3$ over the open-closed moduli space $\mathcal{M}$ with the Gauss-Manin connection $\nabla$ satisfying the Griffish transversality condition 
\begin{equation}
\nabla \mathcal{F}^p\in \mathcal{F}^{p-1}\otimes \Omega^1_{\mathcal{M}}
\end{equation}
The flatness of the Gauss-Manin connection leads to N=1 special geometry and a Picard-Fuchs system of differential equations that govern the mirror maps and superpotentials.

 The geometric setting we are interested in is a hypersurface $X:P=0$ with a curve $S$ on it, $S:P=0,h_1=h_2=0$. After blowing up along $S$, the blow-up manifold $X^\prime$ is given globally as the complete intersection in the total space of the projective bundle $\mathcal{W}=\mathbb{P}(\mathcal{O}(D_1)\oplus\mathcal{O}(D_2))$,
\begin{equation}\label{eq:2.42}
P=0, \quad Q \equiv y_1 h_2-  y_2 h_1=0
\end{equation}
where $(y_1,y_2)\sim\lambda(y_1,y_2)$ is the projective coordinates on the $\mathbb{P}^1$ -fiber of the blow-up $X^b$. We have to emphasize that  $X^\prime$ is not Calabi-Yau since the first Chern class is nonzero. In addition, the blow-up procedure do not introduce  new degrees of freedom associated to deformations of $ E$. 
 Under blowing up map, the the open-closed moduli space of $(X,S)$ is mapped into the complex structure deformation of $X^\prime$.
This  enable us to calculate the superpotential $W_ {\text{brane}}$ for B- branes wrapping  rational curves via the periods on the complex structure moduli space of  $X^\prime$ determined by Picard-Fuchs equations.

\section{Two Closed and Two Open Moduli Case}
\subsection{Open-Closed GKZ-system: Branes on $X_9^{(1,1,1,3,3)}$}
\subsubsection{Five Branes Wrapping Lines}
The Calabi-Yau threefold $X^{(1,1,1,3,3)}_9$  is defined as the mirror of the Calabi-Yau hypersurface $X^*$ in $\mathbb{P}_{(1,1,1,3,3)}^4$ with $h^{2,1} = 2$ complex structure moduli and the charge vectors of the GLSM for the A model manifold are given by:
\begin{center}
\begin{tabular}{cc|cccccc}
$\quad$&0&1&2&3&4&5&6\\
\hline
$l^1$&-3&0&0&0&1&1&1 \\
$l^2$&0&1&1&1&0&0&-3
\end{tabular}
\end{center}
The hypersurface constraint for the mirror manifold, written in homogeneous coordinates of $\mathbb{P}_{(1,1,1,3,3)}$, is
\begin{equation*}
P=x_1^9+x_2^9+x_3^9+x_4^3+x_5^3+\psi(x_1x_2x_3x_4x_5)+\phi(x_1x_2x_3)^3
\end{equation*}
where $\psi = z_1^{-\frac{1}{3}}z_2^{-\frac{1}{9}}$ and $\phi= z_2^{-\frac{1}{3}}$. 
The Greene-Plesser orbifold group G acts as $x_i \rightarrow \lambda^{g_{k,i}}_k x_i$ with $ \lambda^9_1=\lambda^9_2=1$,$\lambda^3_3=1$ and weights

\begin{equation*}
\mathbb{Z}_9:g_1=(1,-1,0,0,0),\quad\mathbb{Z}_9:g_2=(1,0,-1,0,0),\quad\mathbb{Z}_3:g_3=(0,0,0,1,-1)
\end{equation*} 

Next,we add a five-brane wrapping a rational curve on a toric curve $S$ 
\begin{equation}
\begin{gathered}
S :P=0,\quad h_1 \equiv \alpha^3 \gamma^3 x_2^9-\beta^6 (x_1x_2x_2)^3=0,\quad h_2\equiv \beta^3 \gamma^3 x_3^9-\alpha^6 (x_1x_2x_3)^3=0\\
\hat{l}^1=(0,0,1,0,0,0,-1)$, $\hat{l}^2=(0,0,0,1,0,0,-1)
\end{gathered}
\end{equation}

An equivalent and convenient form is obtained after simple algebraic manipulations
\begin{equation}\label{eq:3.4}
S:P=0,\quad\alpha^9x_2^9-\beta^9x_3^9=0,\quad \alpha^9x_1^9-\gamma^9x_3^9=0
\end{equation}

For generic values of the moduli in \ref{eq:3.4},  $S$ is an irreducible high genus Riemann surface.
But we can  make a linearization by following steps:
To begin with, we inserted $h_1$ and $h_2$ into $P$,
\begin{equation}\label{eq:3.5}
\tilde{\mathbb{P}}^1: \quad \eta_1 x_4+\sqrt[3]{x_5^3+ m(x_3,x_4,x_5)}=0,\quad \eta_2 \alpha x_2-\beta x_3=0,\quad \eta_3 \alpha x_1-\gamma x_3=0
\end{equation}
Here $\eta_1^3=\eta_2^9=\eta_3^9=1$ and 
\begin{equation*}
m(x_3,x_4,x_5)=(\frac{\alpha^9+\beta^9+\gamma^9}{\alpha^9}+\frac{\alpha^3 \beta^3
\gamma^3}{\alpha^9}\phi)x_3^9+\frac{\alpha \beta
\gamma}{\alpha^3}\psi x_3^3x_4x_5
\end{equation*}
Due to the non-trivial branching of the roots of unity, \ref{eq:3.5} is non-holomorphic, i.e. it is a non-holomorphic family of rational curves on $X$.

For special loci of $\mathcal{M}(S)$ where $m(x_1,x_2,x_5)$ vanishes identically, 
\begin{equation}
\mathcal{M}_{\mathbb{P}^1}(S):\quad \alpha^9+\beta^9+\gamma^9+\phi \alpha^3 \beta^3\gamma^3=0,\quad \psi\alpha \beta\gamma=0
\end{equation}
the Riemann surface $S$ in \ref{eq:3.5} degenerates to
\begin{equation}\label{eq:3.8}
S:\quad h_0 \equiv x_4^3+x_5^3,\quad h_1=\alpha^9x_2^9-\beta^9x_3^9,\quad h_2= \alpha^9x_1^9-\gamma^9x_3^9
\end{equation}
Under the action of  $G =\mathbb{Z}_9^2 \times \mathbb{Z}_3$, \ref{eq:3.8} describes a single line,
\begin{equation}
\mathbb{P}^1:\quad \eta_1 x_4+x_5=0,\quad  \alpha x_2-\beta x_3=0,\quad \alpha x_1-\gamma x_3=0
\end{equation}
In other words, these lines in $\mathbb{P}^4$ have a parametrization by homogeneous coordinates $U, V$ on $\mathbb{P}^1$ as the Veronese mapping
\begin{equation*}
\begin{aligned}
 \mathcal{M}_{\mathbb{P}^1}(S)& \hookrightarrow \mathcal{M}(S)\\
(U,V)& \mapsto (\gamma U,\beta U,\alpha U, -\eta_1 V,V),\quad\eta_1^3=1
\end{aligned}
\end{equation*}

Thus all obstructed deformations locate at $\mathcal{M}(S)-\mathcal{M}_{\mathbb{P}^1}(S)$, inducing a non-trivial superpotential, which plays an important role in research on obstruction deformation, especially for a manifold with a submanifold on it. As we know, blowing up is very effective method to handle such case. According to  \ref{eq:2.42}, we construct  the blow-up manifold $X^b$ given by the complete intersection in projective bundle 
\begin{equation*}
X^\prime: P=0,\quad Q=y_1 h_2-y_2 h_1
\end{equation*}
It is obvious from above defining equations  that the moduli of $S$ described by the coefficients of the monomials in $h_i,i=1,2$ turn into complex structure moduli of $X^\prime$.  We obtain the embedding of  the obstructed deformation space of $(X,\tilde{\mathbb{P}}^1)$ into the complex structure moduli space of $X^\prime$, which is crucial for the following superpotential calculations.

\subsubsection{Toric Branes and Blowing up Geometry}
Now, we study the A-model manifold, whose toric polyhedron is denoted by $\Delta^*$ and charge vectors are denoted by $l^1$ and $l^2$. The integral vertices of polyhedron $\Delta^*$ and the charge vectors $l^1$, $l^2$ for A-model manifold, $\hat{l}^1$, $\hat{l}^2$ for A-branes is as the following table.

 \begin{table}[H]
\centering
 \begin{tabular}{c|cccc|cc|c|cc}
 $\quad $& \multicolumn{4}{|c|}{$\Delta^*$}&$l^1$&$l^2$&$\quad$&$\hat{l}^1$&$\hat{l}^2$\\
 \hline
 $v^*_0$&$0$&$0$&$0$&$0$&$-3$&$0$&$x_1x_2x_3x_4x_5$&$0$&$0$\\
 $v^*_1$&$-1$&$-1$&$-3$&$-3$&$0$&$1$&$x_1^9$&$0$&$0$\\
 $v^*_2$&$1$&$0$&$0$&$0$&$0$&$1$&$x_2^9$&$1$&$0$\\
 $v^*_3$&$0$&$1$&$0$&$0$&$0$&$1$&$x_3^9$&$0$&$1$\\
 $v^*_4$&$0$&$0$&$1$&$0$&$1$&$0$&$x_4^3$&$0$&$0$\\
 $v^*_5$&$0$&$0$&$0$&$1$&$1$&$0$&$x_5^3$&$0$&$0$\\
 $v^*_6$&$0$&$0$&$-1$&$-1$&$1$&$-3$&$(x_1x_2x_3)^3$&$-1$&$-1$
 \end{tabular}
 \caption{ Toric Data of A-model side}
\end{table}

From above toric data of $\Delta^*$ and its dual polyhedron $\Delta$,
\begin{equation*}
\begin{gathered}
v_1=(-1,-1,-1,-1),\quad v_2=(8,-1,-1,-1), \quad v_3=(-1,8,-1,-1),\\
v_4=(-1,-1,2,-1), \quad v_5=(-1,-1,-1,2)
\end{gathered}
\end{equation*}

 In B-model, the defining equations of the mirror manifold $X$ and the curve $S$ as follow in torus coordinates
 \begin{equation}\label{eq:3.7}
\begin{aligned}
X:& \quad P=a_0+a_1(X_2X_3X_4^3X_5^3)^{-1}+a_2X_2+a_3X_3+a_4 X_4+a_5 X_5+a_6(X_4X_5)^{-1}\\
S:&\quad h_1=a_7X_2+a_8(X_4X_5)^{-1},\quad h_2=a_9 X_3+a_{10}(X_4X_5)^{-1}
\end{aligned}
\end{equation}
where $a_i$'s are free complex-valued coefficients. With the abbreviation of logarithmic derivatives $\vartheta_i=a_i \frac{\partial}{\partial a_i}$,  the  GKZ-system of $X$  by \ref{eq:2.22} is,
\begin{equation}
\begin{gathered}
\mathcal{Z}_0=\sum_{i=0}^6\vartheta_i+1,\quad\mathcal{Z}_i=-\vartheta_1+\vartheta_{i+1},\quad i=1,2,\\
\mathcal{Z}_i=-3\vartheta_1+\vartheta_{i+1}-\vartheta_6,\quad i=3,4,\\
\mathcal{L}_1=\prod^6_{i=4}\frac{\partial}{\partial a_i}-(\frac{\partial}{\partial a_0})^3,\quad \mathcal{L}_2=\prod
^3_{i=1}\frac{\partial}{\partial a_i}-(\frac{\partial}{\partial a_6})^3
\end{gathered}
\end{equation}
Here $\mathcal{Z}_0$ represent the invariance of $P$ under overall rescaling and other $\mathcal{Z}_i$'s relate to the invariance of $P$ under the rescaling of torus coordinates $X_i$'s combined with the rescaling of coefficients $a_i$'s.
 \begin{equation*}
 \begin{aligned}
\mathcal{Z}_i:&\quad X_{i+1} \mapsto \lambda X_{i+1},\quad (a_1,a_{i+1})\mapsto (\lambda a_1,\lambda^{-1}a_{i+1}),\quad i=1,2\\ 
\mathcal{Z}_i:&\quad X_{i+1} \mapsto \lambda X_{i+1},\quad (a_1,a_{i+1},a_6)\mapsto (\lambda^3 a_1,\lambda^{-1}a_{i+1},\lambda a_6),\quad i=3,4
 \end{aligned}
 \end{equation*}
Operators $\mathcal{L}_i$'s relate to the symmetries among the Laurent monomials in $P$ \ref{eq:3.7}, 
\begin{equation*}
\begin{aligned}
\mathcal{L}_1 :&\quad X_4 X_5 (X_4X_5)^{-1}=1\\
\mathcal{L}_2 :&\quad (X_2^{-1}X_3^{-1}X_4^{-3}X_5^{-3})X_2X_3=((X_4X_5)^{-1})^3
\end{aligned}
\end{equation*}

By blowing up $X$ along $S$, the blow-up manifold $X^\prime$ is obtained.
\begin{equation}
X^\prime:P=0,\quad Q=y_1(a_9 X_3+a_{10}(X_4X_5)^{-1})-y_2(a_7X_2+a_8(X_4X_5)^{-1})
\end{equation}

After careful observation on the torus symmetry of $X^\prime$, we can obtain the infinitesimal generators which are belong to GKZ system associated to $X^\prime$,
\begin{equation}\label{eq:3.18}
\begin{gathered}
\mathcal{Z}_0^\prime=\sum_{i=0}^6 \vartheta_i+1,\quad\mathcal{Z}_1^\prime=\sum_{i=7}^{10} \vartheta_i,\quad\mathcal{Z}_2^\prime=-\vartheta_1+\vartheta_2+\vartheta_{7},\quad\mathcal{Z}_3^\prime=-\vartheta_1+\vartheta_3+\vartheta_9,\\
\mathcal{Z}_i^\prime=-3\vartheta_1+\vartheta_i-\vartheta_6-\vartheta_8-\vartheta_{10}\quad i=4,5,\quad \mathcal{Z}_6^\prime=-\vartheta_7-\vartheta_8+\vartheta_9+\vartheta_{10}\\
\mathcal{L}_1^\prime=\prod^6_{i=4}\frac{\partial}{\partial a_i}-(\frac{\partial}{\partial a_0})^3,\quad \mathcal{L}_2^\prime=\prod
^3_{i=1}\frac{\partial}{\partial a_i}-(\frac{\partial}{\partial a_6})^3 \\
\mathcal{L}_3^\prime=\frac{\partial}{\partial a_6} \frac{\partial}{\partial a_7}-\frac{\partial}{\partial a_2} \frac{\partial}{\partial a_8},\quad \mathcal{L}_4^\prime=\frac{\partial}{\partial a_3} \frac{\partial}{\partial a_{10}}-\frac{\partial}{\partial a_6} \frac{\partial}{\partial a_{9}}
\end{gathered}
\end{equation}
Here $\mathcal{Z}_0,\mathcal{Z}_1$ are associated with the overall rescaling with respect to $P=0,Q=0$ respectively. $\mathcal{Z}_i,i=2,...5$ are related to the torus symmetry as before. 
\begin{equation*}
\begin{aligned}
\mathcal{Z}_2^\prime:&\quad X_2\mapsto \lambda X_2,\quad(a_1,a_2,a_7)\mapsto(\lambda a_1,\lambda^{-1}a_2,\lambda^{-1}a_7)\\
 \mathcal{Z}_3^\prime:&\quad X_3\mapsto \lambda X_3,\quad(a_1,a_3,a_9)\mapsto (\lambda a_1,\lambda^{-1}a_3,\lambda^{-1}a_9) \\
 \mathcal{Z}_4^\prime:&\quad X_4\mapsto \lambda X_4,\quad (a_1,a_4,a_6,a_8,a_{10})\mapsto (\lambda ^3 a_1,\lambda^{-1}a_4,\lambda a_6,\lambda a_8 ,\lambda a_{10})\\
 \mathcal{Z}_5^\prime:&\quad X_5\mapsto \lambda X_5,\quad(a_1,a_5,a_6,a_8,a_{10})\mapsto (\lambda ^3 a_1,\lambda^{-1}a_5,\lambda a_6,\lambda a_8 ,\lambda a_{10})\\ 
 \end{aligned}
\end{equation*} 
 In addition,$\mathcal{Z}_6^\prime$ is related to the torus symmetry $(y_1,y_2)\mapsto(\lambda y_1,\lambda^{-1} y_2) $. 
The new $\mathcal{L}_3,\mathcal{L}_4$ incorporate the parameter $a_7,...,a_{10}$ that are associated with the open-closed moduli of the curve $S$. All these GKZ operators annihilate the holomorphic three form $\Omega^\prime$ on $X^\prime$ that is the pull back of the homomorphic three form $\Omega$ on $X$, i.e. $\Omega^\prime=\pi^* \Omega$.

 Now, we formulate the GKZ-system\ref{eq:3.18} on an enhanced polyhedron $\Delta^\prime$, by adding additional vertices on the original polyhedron $\Delta^*$.

\begin{table}[H]\label{table:2}
\centering
 \begin{tabular}{c|ccccccc|cccc|l}
 $\quad$& \multicolumn{7}{|c|}{$\Delta^\prime$} & $l^\prime_1$&$l^\prime_2$&$l^\prime_3$&$l^\prime_4$\\
 \hline
$v^\prime_0$&$1$&$0$&$0$&$0$&$0$&$0$&$0$&$-3$&$0$&$0$&$0$&$w^\prime_0=x_1x_2x_3x_4x_5$\\
$v^\prime_1$&$1$&$0$&$-1$&$-1$&$-3$&$-3$&$0$&$0$&$1$&$0$&$0$&$w^\prime_1=x_1^9$\\
$v^\prime_2$&$1$&$0$&$1$&$0$&$0$&$0$&$0$&$1$&$1$&$-1$&$0$&$w^\prime_2=x_2^9$\\
$v^\prime_3$&$1$&$0$&$0$&$1$&$0$&$0$&$0$&$0$&$0$&$0$&$1$&$w^\prime_3=x_3^9$\\
$v^\prime_4$&$1$&$0$&$0$&$0$&$1$&$0$&$0$&$1$&$0$&$0$&$0$&$w^\prime_4=x_4^3$\\
$v^\prime_5$&$1$&$0$&$0$&$0$&$0$&$1$&$0$&$1$&$0$&$0$&$0$&$w^\prime_5=x_5^3$\\
$v^\prime_6$&$1$&$0$&$0$&$0$&$-1$&$-1$&$0$&$0$&$-2$&$1$&$-1$&$w^\prime_6=(x_1x_2x_3)^3$\\
$v^\prime_7$&$0$&$1$&$1$&$0$&$0$&$0$&$-1$&$-1$&$0$&$1$&$0$&$w^\prime_7=y_1 w^\prime_2$\\
$v^e_\prime$&$0$&$1$&$0$&$0$&$-1$&$-1$&$-1$&$1$&$0$&$-1$&$0$&$w^\prime_8=y_1 w^\prime_6$\\
$v^\prime_\prime$&$0$&$1$&$0$&$1$&$0$&$0$&$1$&$0$&$1$&$0$&$-1$&$w^\prime_9=y_2 w^\prime_3$\\
$v^\prime_{10}$&$0$&$1$&$0$&$0$&$-1$&$-1$&$1$&$0$&$-1$&$0$&$1$&$w^\prime_{10}=y_2 w^\prime_6$\\
 \end{tabular}
\end{table}
where $v^\prime_i$'s are the integral vertices of $\Delta^\prime$ and their  corresponding monomials in homogeneous coordinates of $\mathbb{P}^4$ are $w^\prime_i$. The A-model closed string charge vectors and A-branes charge vectors relate to the maximal triangulation of $\Delta^\prime$ and satisfy the relations, 
\begin{equation*}
l^1=l^\prime_1+l^\prime_3, l^2=l^\prime_2+l^\prime_4, \hat{l}^1=l^\prime_3, \hat{l}^2=l^\prime_4
\end{equation*}

The coordinates $z_j$ by \ref{eq:2.22} on the complex structure moduli space of $X^\prime$.
\begin{equation}
z_1=\frac{a_2 a_4 a_5 a_8}{a_0^3 a_7},\quad z_2=\frac{a_1 a_2 a_{9}}{a_6^2 a_{10}},\quad z_3=\frac{a_6 a_7}{a_2 a_8},\quad z_4=\frac{a_3 a_{10}}{a_6 a_{9}}
\end{equation}

Next, we convert the $\mathcal{L}_i$ operators to Picard-Fuchs operators $\mathcal{D}_i$, from differential equations about  $a_j(j=0,\dots,10)$ to those about $z_j(j=1,\dots,4)$ of $X^b$.
From Table \ref{table:2}, we obtain the identity
\begin{equation*}
\begin{gathered}
\vartheta_0=-3\theta_1,\vartheta_1=\theta_2,
\quad\vartheta_2=\theta_1+\theta_2-\theta_3,
\quad\vartheta_3=\theta_4,\\
\vartheta_4=\theta_1,
\quad\vartheta_5=\theta_1,
\quad\vartheta_6=-2\theta_2+\theta_3-\theta_4,\\
\vartheta_7=-\theta_1+\theta_3,
\quad\vartheta_8=\theta_1-\theta_3,
\quad\vartheta_9=\theta_2-\theta_4,
\quad\vartheta_{10}=\theta_2+\theta_{10}
\end{gathered}
\end{equation*}
 Inserting above r elations between the logarithmic derivatives $\vartheta_j$ w.r.t $a_j$ and  the logarithmic derivatives $\theta_j$ w.r.t  $z_j$ into $\mathcal{L}$ operators in \ref{eq:3.18}, the full set of Picard-Fuchs operators are obtained 
\begin{equation}\label{eq:3.22}
\begin{aligned}
\mathcal{D}_1&=(\theta_1+\theta_2-\theta_3)\theta_1^2 (\theta_1 -\theta_3)-z_1
 (-\theta_1+\theta_3) \prod^3_{i=1}(-3\theta_1-i),\\
 \mathcal{D}_2&=\theta_2 (\theta_1+\theta_2-\theta_3) (\theta_2-\theta_4)-z_2
 (-2\theta_1+\theta_3-\theta_4)  (-\theta_2+\theta_4)\\
 \mathcal{D}_3&=(-2\theta_1+\theta_3-\theta_4)(-\theta_1+\theta_3)-z_3 (\theta_1+\theta_2-\theta_3) (\theta_1 -\theta_3),\\
 \mathcal{D}_4&=\theta_4 (-\theta_2+\theta_4)-z_4(-2\theta_1+\theta_3-\theta_4)(\theta_2-\theta_4)\\
 \cdots
 \end{aligned}
 \end{equation}
where $\theta_i=z_i \frac{\partial}{\partial z_i}$'s are the logarithmic derivatives and each operator $\mathcal{D}_a$ corresponds to a linear combination of the charge vectors among $l^\prime_1,l^\prime_2,l^\prime_3,l^\prime_4$.

\subsubsection{Brane Superpotential and Disk Instantons }

Now, we  solve the Picard-Fuchs equations \ref{eq:3.22} at $z_i\rightarrow 0$  and identified the mirror maps and superpotentials. By the method  introduced in Section \ref{sec:2.2} , the fundamental period of $X$ as power series solution as follow
\begin{equation*}
\omega_0=1 + 6 z_1 z_3 + 90 z_1^2 z_3^2 + 1680 z_1^3 z_3^3 + 34650 z_1^4 z_3^4  + 10080 z_1^3 z_2 z_3^3 z_4 +\mathcal{O}(z^8)
\end{equation*}.
There are four logarithmic solutions  with leading term $\omega_0\log(Z_i)$
\begin{equation*}
\begin{aligned}
\omega_{1,1}=&\omega_0\log(z_1)-6 z_1 + 45 z_1^2 - 560 z_1^3 + \frac{17325}{2} z_1^4 - \frac{756756}{5} z_1^5 + 
 21 z_1 z_3 - 180 z_1^2 z_3 + 2520 z_1^3 z_3\\
 & - 46200 z_1^4 z_3 + \frac{783}{2} z_1^2 z_3^2 - 5040 z_1^3 z_3^2+\mathcal{O}(z^5)\\
\omega_{1,2}=&\omega_0 \log(z_2)+12 z_1 z_3 + 270 z_1^2 z_3^2 - 180 z_1^2 z_2 z_3^2 - 4 z_2 z_4 + 6 z_1 z_3 z_4 + 
 12 z_1 z_2 z_3 z_4 + 180 z_1^2 z_3^2 z_4 \\
 &+ 30 z_2^2 z_4^2+\mathcal{O}(z^5)\\
\omega_{1,3}=&\omega_0\log(z_3)+6 z_1 - 45 z_1^2 + 560 z_1^3 - \frac{17325}{2} z_1^4 + \frac{756756}{5} z_1^5 - 6 z_1 z_3 + 180 z_1^2 z_3 - 2520 z_1^3 z_3\\
& + 46200 z_1^4 z_3 - 135 z_1^2 z_3^2 + 5040 z_1^3 z_3^2 + 2 z_2 z_4 - 6 z_1 z_2 z_3 z_4 -  15 z_2^2 z_4^2+\mathcal{O}(z^5)\\
\omega_{1,4}=&\omega_0 \log(z_4)6 z_1 z_3 + 135 z_1^2 z_3^2 + 180 z_1^2 z_2 z_3^2 - 2 z_2 z_4 - 6 z_1 z_3 z_4 + 
 6 z_1 z_2 z_3 z_4 - 180 z_1^2 z_3^2 z_4 \\
 &+ 15 z_2^2 z_4^2+\mathcal{O}(z^5)
\end{aligned}
\end{equation*}.

By the definition of the flat coordinates and mirror maps from Kahler moduli space to complex structure moduli space 
\begin{equation}
t_j=\frac{\omega_{1,j}}{\omega_0}
 \end{equation}
 \begin{equation}
q _j = e^{2\pi it_j}
 \end{equation}
we obtain the $z_j$  as a series of $q_j$ upon inversion of the mirror maps
\begin{equation*}
\begin{aligned}
z_1=&q_1 + 6 q_1^2 + 9 q_1^3 + 56 q_1^4 - 300 q_1^5 - 21 q_1^2 q_3 - 
 108 q_1^3 q_3 - 225 q_1^4 q_3 + 270 q_1^3 q_3^2 \\
 &+ 42 q_1^2 q_2 q_3 q_4+\mathcal{O}(q^5)\\
z_2=&q_2 - 12 q_1 q_2 q_3 + 54 q_1^2 q_2 q_3^2 + 4 q_2^2 q_4 - 6 q_1 q_2 q_3 q_4 - 
 132 q_1 q_2^2 q_3 q_4 + 2 q_2^3 q_4^2+\mathcal{O}(q^5)\\
z_3=&q_3 - 6 q_1 q_3 + 27 q_1^2 q_3 - 164 q_1^3 q_3 + 1377 q_1^4 q_3 + 6 q_1 q_3^2 - 54 q_1^2 q_3^2 + 414 q_1^3 q_3^2 + 27 q_1^2 q_3^3\\
& - 2 q_2 q_3 q_4 + 12 q_1 q_2 q_3 q_4 - 54 q_1^2 q_2 q_3 q_4 + 30 q_1 q_2 q_3^2 q_4+\mathcal{O}(q^5)\\
z_4=&q_4 - 6 q_1 q_3 q_4 + 9 q_1^2 q_3^2 q_4 + 2 q_2 q_4^2 + 6 q_1 q_3 q_4^2 - 
 54 q_1 q_2 q_3 q_4^2 - q_2^2 q_4^3+\mathcal{O}(q^5)
\end{aligned}
\end{equation*}
In addition, we abbreviate the double logarithmic solutions by their leading terms
\begin{equation*}
\begin{gathered}
\frac{1}{2} \ell_1^2+2\ell_3\ell_4,\quad\frac{1}{2}\ell_2^2,\quad\frac{1}{2}\ell_3^2+\ell_3\ell_4,\quad\frac{1}{2}\ell_4^2,\\
\ell_1\ell_2+\ell_2\ell_4,\quad\ell_1\ell_3-3\ell_3\ell_4,
\quad\ell_1\ell_4+\ell_3\ell_4,\quad\ell_2\ell_3
\end{gathered}
\end{equation*}
where $\log(z_i)$'s are abbreviated as $\ell_i$'s. According to above, a specific linear combination of double logarithmic solutions is constructed, 
\begin{equation}
\mathcal{W}_{\mathrm{brane}}=3t_1t_2-3t_3t_4+\sum_{N}N_{d_1,d_2,d_3,d_4}\mathrm{Li}_2(q_1^{d_1}q_2^{d_2}q_3^{d_3}q_4^{d_4})
\end{equation}
where $\text{Li}_2(q_1^{d_1}q_2^{d_2}q_3^{d_3}q_4^{d_4})=\sum_{n=1}^\infty \frac{q_1^{n d_1}q_2^{n d_2}q_3^{n d_3}q_4^{n d_4}}{n^2}$ is the dilogarithm function. We extract the disk instantons $N_{d_1,d_2,d_3,d_4}$ from $\mathcal{W}_{\text{brane}}$ and present first a few invariants of the form $N_{k,m,k,m+n}$ in Table \ref{table:13}. When focusing only on the invariants of the form $N_{k,m,k,m+n}$,  
\begin{equation}\label{eq:3.16}
\mathcal{W}_{\mathrm{brane}}=\dots+\sum_{N}N_{k,m+n,k,n}\mathrm{Li}_2((\tilde{q_1})^{k}(\tilde{q_2})^{m}\tilde{q_3}^{n})
\end{equation}
where $\dots$ are terms independent of invariants, and $\tilde{q_1}=q_1 q_3, \tilde{q_2}=q_2q_4,\tilde{q_3}=q_4$. 
 The superpotential \ref{eq:3.16} can be written as 
\begin{equation}
\mathcal{W}_{\mathrm{brane}}=\dots+\sum_{N}N_{a,b,c}\mathrm{Li}_2((\tilde{q_1})^{a}(\tilde{q_2})^{b}\tilde{q_c}^{c})
\end{equation} which are essentially the superpotential of the model in \cite{Alim2009}. 
Therefore, our invariants at first several order exactly match with the data of Table 5 in \cite{Alim2009} and these invariants are marked by blue color in Table \ref{table:13} . In addition, we also calculate the invariants at higher order and put them into Table \ref{table:13}.

\subsubsection{Another Curve on X}
Following the similar logic, we consider another toric curve $S$ on X with charge vectors
\begin{equation*}
\begin{aligned}
P&=0,\\
 h_1& \equiv a_7(x_1x_2x_3x_4x_5)+a_8 x_4^3=0,\quad\hat{l}_1 =(-1,0,0,0,1,0,0)\\
 h_2&\equiv a_9 x_3^9+a_{10} x_4^3=0
,\quad \hat{l}_2=(0,0,0,1,-1,0,0)
\end{aligned}
\end{equation*}
After blowing up, the toric data of  mirror manifold of $X^\prime$ is determined by charge vectors
\begin{equation}
\begin{aligned}
l^e_1&=(-2,0,0,0,0,1,1,-1,1,0,0),\quad l^e_2=(0,1,1,0,1,0,-3,0,0,1,-1),\\
l^e_3&=(-1,0,0,0,1,0,0,1,-1,0,0),\quad l^e_4=(0,0,0,1,-1,0,0,0,0,-1,1)
\end{aligned}
\end{equation}
satisfying $l^1=l^e_1+l^e_3$, $l^2=l^e_2+l^e_4$, from which we obtain the local coordinates on complex structure moduli of $X^\prime$
\begin{equation}
z_1=\frac{a_5a_6a_8}{a_0^2 a_7},\quad z_2=\frac{a_1 a_2 a_4 a_{9}}{a_6^3 a_{10}},\quad z_3=\frac{a_4 a_7}{a_0 a_8},\quad z_4=\frac{a_6 a_{10}}{a_4 a_{9}}
\end{equation}
and Picard-Fuchs operators
\begin{equation}
\begin{aligned}
\mathcal{D}_1&=\theta_1(\theta_1-3\theta_2)(\theta_1-\theta_3)-z_1(-\theta_1+\theta_3)\prod_{i=1}^2(-2\theta_1-\theta_3-i)\\
\mathcal{D}_2&=(\theta_2+\theta_3-\theta_4)(\theta_2)^2(\theta_2-\theta_4)-z_2(\theta_1-3\theta_2)^3(-\theta_2-\theta_4)\\\mathcal{D}_3&=(\theta_2+\theta_3-\theta_4)(-\theta_1+\theta_3)-z_3(-2\theta_1-\theta_3-1)(\theta_1-\theta_3)\\
\mathcal{D}_4&=\theta_4(-\theta_2+\theta_4)-z_4(\theta_2+\theta_3-\theta_4)(\theta_2-\theta_4)\\
\cdots
\end{aligned}
\end{equation}
As before, we find single logarithmic solutions 
\begin{equation*}
\begin{aligned}
\omega_{1,1}=&\omega_0\log(z_1)-2 z_1 + 3 z_1^2 - \frac{20}{3} z_1^3 + \frac{35}{2} z_1^4 - \frac{252}{5} z_1^5 + z_3 + 10 z_1 z_3 - 30 z_1^2 z_3 + 70 z_1^3 z_3 \\
 &- 210 z_1^4 z_3 - \frac{1}{2}z3^2 + 12 z_1 z_3^2 + 171 z_1^2 z_3^2 - 560 z_1^3 z_3^2 + \frac{1}{3}z_3^3 - 10 z_1 z_3^3 + 210 z_1^2 z_3^3 - \frac{1}{4}z_3^4 \\
 &+ 10 z_1 z_3^4 + \frac{1}{5}z_3^5 + 2 z_2 z_4 - 40 z_1^3 z_2 z_4 - 6 z_1 z_2 z_3 z_4+\mathcal{O}(z^5)\\
\omega_{1,2}=&\omega_0\log(z_2)+12 z_1 z_3 + 270 z_1^2 z_3^2 - 6 z_2 z_4 + 6 z_1 z_3 z_4 + 18 z_1 z_2 z_3 z_4 + 
 180 z_1^2 z_3^2 z_4\\
 & + 45 z_2^2 z_4^2+\mathcal{O}(z^5)\\
\omega_{1,3}=&\omega_0\log(z_3)+2 z_1 - 3 z_1^2 + \frac{20}{3} z_1^3 -\frac{35}{2} z_1^4 + \frac{252}{5} z_1^5 - z_3 + 
 5 z_1 z_3 + 30 z_1^2 z_3 - 70 z_1^3 z_3\\
 & + 210 z_1^4 z_3 + \frac{1}{2}z_3^2 - 
 12 z_1 z_3^2 + \frac{171}{2} z_1^2 z_3^2 + 560 z_1^3 z_3^2 - \frac{1}{3}z_3^3 + 
 10 z_1 z_3^3 - 210 z_1^2 z_3^3 +\frac{1}{4} z_3^4\\
 & - 10 z_1 z_3^4 - \frac{1}{5}z_3^5+ 40 z_1^3 z_2 z_4+\mathcal{O}(z^5)\\
\omega_{1,4}=&\omega_0\log(z_4)+6 z_1 z_3 + 135 z_1^2 z_3^2 - 6 z_1 z_3 z_4 - 180 z_1^2 z_3^2 z_4+\mathcal{O}(z^5)
\end{aligned}
\end{equation*}
and mirror maps 
\begin{equation*}
\begin{aligned}
z_1=&q_1 + 2 q_1^2 + 3 q_1^3 + 4 q_1^4 + 5 q_1^5 - q_1 q_3 - 12 q_1^2 q_3 - 
 24 q_1^3 q_3 - 36 q_1^4 q_3 + 9 q_1^2 q_3^2 + 108 q_1^3 q_3^2 + q_1^2 q_3^3 \\
 &- 2 q_1 q_2 q_4 - 8 q_1^2 q_2 q_4 - 18 q_1^3 q_2 q_4 + 2 q_1 q_2 q_3 q_4 + 
 102 q_1^2 q_2 q_3 q_4+\mathcal{O}(q^5)\\
z_2=&q_2 - 12 q_1 q_2 q_3 + 54 q_1^2 q_2 q_3^2 + 6 q_2^2 q_4 - 6 q_1 q_2 q_3 q_4 - 
 210 q_1 q_2^2 q_3 q_4 + 9 q_2^3 q_4^2+\mathcal{O}(q^5)\\
z_3=&q_3 - 2 q_1 q_3 + q_1^2 q_3 + q_3^2 - 7 q_1 q_3^2 + 12 q_1^2 q_3^2 - 
 7 q_1^3 q_3^2 + q_1^4 q_3^2 + q_3^3 - 6 q_1 q_3^3 + 42 q_1^2 q_3^3 + q_3^4\\
 & - 6 q_1q_3^4+ q_3^5 + 4 q_1 q_2 q_3 q_4 - 4 q_1^2 q_2 q_3 q_4 + 14 q_1 q_2 z_3^2 q_4+\mathcal{O}(q^5)\\
z_4=&q_4 - 6 q_1 q_3 q_4 + 9 q_1^2 q_3^2 q_4 + 6 q_1 q_3 q_4^2 + 12 q_1 q_2 q_3 q_4^2+\mathcal{O}(q^5)
\end{aligned}
\end{equation*} as single logarithmic solutions.

The superpotentials is constructed as follow 
\begin{equation}
\mathcal{W}_{\mathrm{brane}}=9t_1^2+6t_1t_2+6t_1t_3+\sum_{N}N_{d_1,d_2,d_3,d_4}\mathrm{Li}_2(q_1^{d_1}q_2^{d_2}q_3^{d_3}q_4^{d_4})
\end{equation} as linear combination of double logarithmic solutions.
We extract and summarize the Ooguri-Vafa invariants of the form $N_{k,m,k,m+n}$  and $N_{k,m+n,k,n}$  in Table \ref{table:15} and Table\ref{table:16}. At first several order, our result marked by blue color exactly agree with Table 6 in\cite{Alim2009} and we also present higher order result in the table.

\subsection{Open-Closed GKZ-system: Branes on $X_8^{(1,1,2,2,2)}$}
\subsubsection{Branes Wrapping Rational Curves and Blowing up Geometry}
Now, we study the A-model manifold, whose toric polyhedron is denoted by $\Delta^*$ and charge vectors are denoted by $l^1$ and $l^2$. A-model manifold and A-brane are specified by the following toric data.
\begin{table}[H]\label{table:3}
\centering
 \begin{tabular}{c|cccc|cc|c|cc}
 $\quad$&\multicolumn{4}{|c|}{$\Delta^*$}&$l^1$&$l^2$&$\quad$&$\hat{l}^1$&$\hat{l}^2$\\
 \hline
 $v^*_0$&$0$&$0$&$0$&$0$&$-4$&$0$&$x_1x_2x_3x_4x_5$&$-1$&$0$\\
 $v^*_1$&$-1$&$-2$&$-2$&$-2$&$0$&$1$&$x_1^8$&$0$&$-1$\\
 $v^*_2$&$1$&$0$&$0$&$0$&$0$&$1$&$x_2^8$&$0$&$1$\\
 $v^*_3$&$0$&$1$&$0$&$0$&$1$&$0$&$x_3^4$&$1$&$0$\\
 $v^*_4$&$0$&$0$&$1$&$0$&$1$&$0$&$x_4^4$&$0$&$0$\\
 $v^*_5$&$0$&$0$&$0$&$1$&$1$&$0$&$x_5^4$&$0$&$0$\\
 $v^*_6$&$0$&$-1$&$-1$&$-1$&$1$&$-2$&$(x_1x_2)^4$&$0$&$0$
 \end{tabular}
 \caption{ Toric Data of A-model side}
\end{table}
The hypersurface equation for $X$, written in homogeneous coordinates of $\mathbb{P}_{(1,1,2,2,2)}$, is
\begin{equation*}
P=x_1^8+x_2^8+x_3^4+x_4^4+x_5^4+\psi(x_1x_2x_3x_4x_5)+\phi(x_1x_2)^4
\end{equation*}
where $\psi = z_1^{-\frac{1}{4}}z_2^{-\frac{1}{8}}$ and $\phi= z_2^{-\frac{1}{4}}$. The toric curve $S$ on $X$ is defined as  the complete intersection 
\begin{equation*}
S : \quad P=0,\quad h_1 \equiv \gamma^8 (x_1x_2x_3x_4x_5)-\alpha^8  x_3^4=0,\quad h_2\equiv \alpha^8 x_1^8-\beta^8 x_2^8=0
\end{equation*}
The Greene-Plesser orbifold group G acts as $x_i \rightarrow \lambda^{g_{k,i}}_k x_i$ with $ \lambda^8_1=\lambda^4_2=1$,$\lambda^4_3=1$ and weights

\begin{equation*}
\mathbb{Z}_8:g_1=(1,-1,0,0,0),\quad\mathbb{Z}_4:g_2=(1,0,-1,0,0),\quad\mathbb{Z}_4:g_3=(1,0,0,-1,0)
\end{equation*} 
 Insert $h_1$ and $h_2$ into $P=0$
\begin{equation}\label{eq:4.4}
\begin{gathered}
\tilde{\mathbb{P}}_1:\eta_1 x_4+\sqrt[4]{x_5^4+m(x_1,x_3)}=0,\quad \eta_1^4=1\\
m(x_1,x_3)=(\frac{\alpha^8+\beta^8}{\alpha^8}+\phi \frac{\beta^8}{\alpha^8})x_1^8+(1+\frac{\gamma^8}{\alpha^8}\psi)x_3^4
\end{gathered}
\end{equation}
$\tilde{\mathbb{P}}^1$  is a non-holomorphic family due to fourth roots of unity.  At critical loci of the parameter space $\alpha,\beta,\gamma$ 
\begin{equation}
\mathcal{M}_{\mathbb{P}^1}(S):\quad \alpha^8+\beta^8+\phi  \beta^8=0,\quad \alpha^8+ \gamma^8 \psi=0
\end{equation}
 $m(x_1,x_4)$ vanishes identically and $S$ degenerates to 
\begin{equation}\label{eq:4.6}
S:\quad h_0 \equiv x_4^4+x_5^4,\quad h_1=\gamma^8(x_1x_2x_3x_4x_5)-\alpha^8 x_3^4=0,\quad h_2 =\alpha^8x_1^8-\beta^8x_2^8=0
\end{equation}
Modulo the action of $G$, \ref{eq:4.4} can be solved holomorphically. Thus the anholomorphic deformation of $\mathbb{P}^1$ \ref{eq:4.6} can be used to described holomorphic deformation of $S$.  

 From the toric data in Table \ref{table:3}, we  obtain the  GKZ-system of $X$ by \ref{eq:2.21},
\begin{equation}
\begin{gathered}
\mathcal{Z}_0=\sum_{i=0}^6\vartheta_i+1,\quad\mathcal{Z}_1=-\vartheta_1+\vartheta_2,\quad\mathcal{Z}_i=-2\vartheta_i+\vartheta_{i+1}-\vartheta_{6}(i=2,3,4),\\
\mathcal{L}_1=\prod^6_{i=3}\frac{\partial}{\partial a_i}-(\frac{\partial}{\partial a_0})^4,\quad \mathcal{L}_2=
\frac{\partial}{\partial a_1}\frac{\partial}{\partial a_2}-(\frac{\partial}{\partial a_6})^2
\end{gathered}
\end{equation}
where $\mathcal{Z}_0$ represents the invariant of $P$ under overall rescaling, $\mathcal{Z}_{i}$'s relate to the torus symmetry, and  $\mathcal{L}_i$'s relate to the symmetries among monomials consisting of $P$. As before, all GKZ operators  annihilate the period integrals and determine the mirror maps and superpotentials.

After blowing up $X$ along $S$, we obtain the blow-up manifold defined by,
\begin{equation}
\begin{aligned}
X^\prime:P&=a_0x_1x_2x_3x_4x_5+a_1x_1^{8}+a_2x_8+a_3x_4+a_4x_4^4+a_5x_5^4+a_6(x_1x_2)^{4},\\ 
Q&=y_1(a_9 x_1^9+a_{10} x_2^9)-y_2(a_7x_1x_2x_3x_4x_5+a_8 x_3^4)
\end{aligned}
\end{equation}

As before, the corresponding infinitesimal generators are obtained , which are belong to the GKZ system of $X^\prime$ the GKZ system of $X^\prime$ by observation on the torus symmetry,
\begin{equation} \label{eq:4.9}
\begin{gathered}
\mathcal{Z}_0^\prime=\sum_{i=0}^6 \vartheta_i+1,\quad\mathcal{Z}_1^\prime=\sum_{i=7}^{10} \vartheta_i,\quad\mathcal{Z}_2^\prime=-\vartheta_1+\vartheta_2-\vartheta_9+\vartheta_{10},\quad\mathcal{Z}_3^\prime=-2\vartheta_1+\vartheta_3-\vartheta_6+\vartheta_8-2\vartheta_9,
\\
\mathcal{Z}_4^\prime=-2\vartheta_1+\vartheta_4-\vartheta_6-2\vartheta_9,\quad\mathcal{Z}_5^\prime=-2\vartheta_1+\vartheta_5-\vartheta_6-2\vartheta_{9},\quad\mathcal{Z}_6^\prime=-\vartheta_7-\vartheta_8+\vartheta_9+\vartheta_{10
}\\
\mathcal{L}_1^\prime=\prod^6_{i=3}\frac{\partial}{\partial a_i}-(\frac{\partial}{\partial a_0})^4,\quad \mathcal{L}_2^\prime=\prod
^2_{i=1}\frac{\partial}{\partial a_i}-(\frac{\partial}{\partial a_6})^2 \\
\mathcal{L}_3^\prime=\frac{\partial}{\partial a_3} \frac{\partial}{\partial a_7}-\frac{\partial}{\partial a_0} \frac{\partial}{\partial a_8},\quad \mathcal{L}_4^\prime=\frac{\partial}{\partial a_2} \frac{\partial}{\partial a_9}-\frac{\partial}{\partial a_1} \frac{\partial}{\partial a_{10}}
\end{gathered}
\end{equation}
where $\mathcal{Z}_0^\prime,\mathcal{Z}_1^\prime$ are associated with the overall rescaling with respect to $P=0,Q=0$ respectively. $\mathcal{Z}^\prime_i,i=2,...,6$ are related to the torus symmetry and The $\mathcal{L}_3^\prime,\mathcal{L}_4^\prime$ incorporate the parameter $a_7,...,a_{10}$ that are associated with the moduli of the  curve $S$.

Then we formulate above GKZ-system in an enhanced polyhedron $\Delta
^\prime$,
 \begin{table}[H]
\centering
 \begin{tabular}{c|ccccccc|cccc|l}
 $\quad$& \multicolumn{7}{|c|}{$\Delta^\prime$} & $l^\prime_1$&$l^\prime_2$&$l^\prime_3$&$l^\prime_4$\\
 \hline
$v^\prime_0$&$1$&$0$&$0$&$0$&$0$&$0$&$0$&$-3$&$0$&$-1$&$0$&$w^\prime_0=x_1x_2x_3x_4x_5$\\
$v^\prime_1$&$1$&$0$&$-1$&$-2$&$-2$&$-2$&$0$&$0$&$2$&$0$&$-1$&$w^\prime_1=x_1^8$\\
$v^\prime_2$&$1$&$0$&$1$&$0$&$0$&$0$&$0$&$0$&$0$&$0$&$1$&$w^\prime_2=x_2^8$\\
$v^\prime_3$&$1$&$0$&$0$&$1$&$0$&$0$&$0$&$0$&$0$&$1$&$0$&$w^\prime_3=x_3^4$\\
$v^\prime_4$&$1$&$0$&$0$&$0$&$1$&$0$&$0$&$1$&$0$&$0$&$0$&$w^\prime_4=x_4^4$\\
$v^\prime_5$&$1$&$0$&$0$&$0$&$0$&$1$&$0$&$1$&$0$&$0$&$0$&$w^\prime_5=x_5^4$\\
$v^\prime_6$&$1$&$0$&$0$&$-1$&$-1$&$-1$&$0$&$1$&$-2$&$0$&$0$&$w^\prime_6=(x_1x_2)^4$\\
$v^\prime_7$&$0$&$1$&$0$&$0$&$0$&$0$&$-1$&$-1$&$0$&$1$&$0$&$w^\prime_7=y_1 w^\prime_1$\\
$v^\prime_8$&$0$&$1$&$0$&$1$&$0$&$0$&$-1$&$1$&$0$&$-1$&$0$&$w^\prime_8=y_1 w^\prime_2$\\
$v^\prime_9$&$0$&$1$&$-1$&$-2$&$-2$&$-2$&$1$&$0$&$-1$&$0$&$1$&$w^\prime_9=y_2 w^e_0$\\
$v^\prime_{10}$&$0$&$1$&$1$&$0$&$0$&$0$&$1$&$0$&$1$&$0$&$-1$&$w^\prime_{10}=y_2 w^\prime_3$\\
 \end{tabular}
 \caption{Toric Data of $X^\prime$'s Mirror Manifold }
\end{table}
Here we  present the integral points $v^\prime_ i$ of the enhanced polyhedron $\Delta^\prime$ and their the corresponding monomials $w^\prime_i$'s. $l^\prime_i$'s are the generators of Mori cone $l^\prime_i$ satisfying  $l^1=l^\prime_1+l^\prime_3$, $l^2=l^\prime_2+l^\prime_4$ and they are the maximal triangulation of $\Delta^\prime$. 

By definition \ref{eq:2.22}, the local coordinates of complex structure moduli space of $X^\prime$ are
\begin{equation}
z_1=\frac{a_4 a_5 a_6 a_8}{a_0^3 a_7},\quad z_2=\frac{a_1^2 a_{10}}{a_6^2 a_9},\quad z_3=\frac{a_3 a_7}{a_0 a_8},\quad z_4=\frac{a_2 a_9}{a_1 a_{10}}
\end{equation}

Next, we convert the $\mathcal{L}_i$ operators in \ref{eq:4.9} to Picard-Fuchs operators $\mathcal{D}_i$, from differential equations about $a_j(j=0,\dots,10)$ to those about  $z_j(j=1,\dots,4)$ ,
\begin{equation}\label{eq:4.11}
\begin{aligned}
\mathcal{D}_1&=\theta_1^2 (\theta_1-2\theta_2)(\theta_1 -\theta_3)-z_1
 (-\theta_1+\theta_3) \prod^3_{i=1}(-3\theta_1-\theta_3-i),\\
 \mathcal{D}_2&= (2\theta_2-\theta_4)^2 (\theta_2-\theta_4)-z_2
 (\theta_1-2\theta_2)^2 (-\theta_2+\theta_4)\\
 \mathcal{D}_3&=\theta_3(-\theta_1+\theta_3)-z_3 (-3\theta_1-\theta_3-1) (\theta_1 -\theta_3),\\
 \mathcal{D}_4&=\theta_4 (-\theta_2+\theta_4)-z_4(2\theta_2-\theta_4)(\theta_2-\theta_4)\\
 \cdots
 \end{aligned}
 \end{equation}
where $\theta_i$'s are the logarithmic derivatives with respect to
$z_i$'s and each $D_i$ corresponds to a specific linear combination among $l^\prime_1,l^\prime_2,l^\prime_3,l^\prime_4$.
\subsubsection{Brane Superpotential and Disk Instantons}
Now, we  solve the Picard-Fuchs equations \ref{eq:4.11} at $z_i\rightarrow 0$  and identified the mirror maps and superpotentials. By the techniques we introduced in Section \ref{sec:2.2}

The unique power series solution, as well as the fundamental period of $X$,is
\begin{equation*}
\begin{aligned}
\omega_0=&1 + 24 z_1 z_3 + 2520 z_1^2 z_3^2 + 369600 z_1^3 z_3^3 + 
 63063000 z_1^4 z_3^4 + 5040 z_1^2 z_2 z_3^2 z_4 \\
& + 2217600 z_1^3 z_2 z_3^3 z_4+\mathcal{O}(z^5)
 \end{aligned}
\end{equation*} 
The single logarithmic solutions are
\begin{equation*}
\begin{aligned}
\omega_{1,1}=&\omega_0\log(z_1)-6 z_1 + 45 z_1^2 - 560 z_1^3 + \frac{17325}{2} z_1^4 - \frac{756756}{5} z_1^5 + z_3 + 
 78 z_1 z_3 - 630 z_1^2 z_3 \\
& + 8400 z_1^3 z_3 - 150150 z_1^4 z_3 - \frac{1}{2}z_3^2 + 
 60 z_1 z_3^2 + 9207 z_1^2 z_3^2 - 92400 z_1^3 z_3^2 + \frac{1}{3}z_3^3 - 
 60 z_1 z_3^3\\
 & + 7560 z_1^2 z_3^3 -\frac{1}{4} z_3^4 + 70 z_1 z_3^4 + \frac{1}{5}z_3^5 - z_2 z_4 +  90 z_1^2 z_2 z_4 - 3360 z_1^3 z_2 z_4 + 24 z_1 z_2 z_3 z_4 \\
 &-  1260 z_1^2 z_2 z_3 z_4 - \frac{3}{2} z_2^2 z_4^2+\mathcal{O}(z^5)\\
\omega_{1,2}=&\omega_0\log(z_2)+48 z_1 z_3 + 7560 z_1^2 z_3^2 - 2520 z_1^2 z_2 z_3^2 + 2 z_2 z_4 - 
 48 z_1 z_2 z_3 z_4 + 3 z_2^2 z_4^2+\mathcal{O}(z^5)\\
\omega_{1,3}=&\omega_0\log(z_3)+6 z_1 - 45 z_1^2 + 560 z_1^3 - \frac{17325}{2} z_1^4 + \frac{756756}{5} z_1^5 - z_3 + 
 26 z_1 z_3 + 630 z_1^2 z_3 \\
 &- 8400 z_1^3 z_3 + 150150 z_1^4 z_3 + \frac{1}{2}z_3^2 - 60 z_1 z_3^2 + 3069 z_1^2 z_3^2 + 92400 z_1^3 z_3^2 -\frac{1}{3} z_3^3 + 
 60 z_1 z_3^3\\
 & - 7560 z_1^2 z_3^3 +\frac{1}{4} z_3^4 - 70 z_1 z_3^4 +\frac{1}{5} z_3^5 -  90 z_1^2 z_2 z_4 + 3360 z_1^3 z_2 z_4 + 1260 z_1^2 z_2 z_3 z_4+\mathcal{O}(z^5)\\
\omega_{1,4}=&\omega_0\log(z_4)+2520 z_1^2 z_2 z_3^2 + 1108800 z_1^3 z_2 z_3^3 + 378378000 z_1^4 z_2 z_3^4 + 
 252252000 z_1^4 z_2^2 z_3^4 z_4 \\
 &- 2520 z_1^2 z_2 z_3^2 z_4^2 - 
 1108800 z_1^3 z_2 z_3^3 z_4^2+\mathcal{O}(z^{10})
\end{aligned}
\end{equation*}
such that single logarithmic period of $X$ can be reproduced by $\Pi^1_1=\omega_{1,1}+\omega_{1,3}$, $\Pi^1_2\omega_{1,2}+\omega_{1,4}$,and open-closed mirror maps are inverse series of flat coordinates,
\begin{equation*}
\begin{aligned}
z_1=&q_1 + 6 q_1^2 + 9 q_1^3 + 56 q_1^4 - 300 q_1^5 - q_1 q_3 - 84 q_1^2 q_3 - 
 432 q_1^3 q_3 - 900 q_1^4 q_3 + 68 q_1^2 q_3^2 \\
 &+ 4182 q_1^3 q_3^2 + 
 12 q_1^2 q_3^3 + q_1 q_2 q_4 + 12 q_1^2 q_2 q_4 - 63 q_1^3 q_2 q_4 - 
 q_1 q_2 q_3 q_4 - 264 q_1^2 q_2 q_3 q_4+\mathcal{O}(q^5)\\
z_2=&q_2 - 48 q_1 q_2 q_3 - 264 q_1^2 q_2 q_3^2 - 2 q_2^2 q_4 + 240 q_1 q_2^2 q_3 q_4 +3 q_2^3 q_4^2+\mathcal{O}(q^5)\\
z_3=&q_3 - 6 q_1 q_3 + 27 q_1^2 q_3 - 164 q_1^3 q_3 + 1377 q_1^4 q_3 + q_3^2 - 
 32 q_1 q_3^2 + 138 q_1^2 q_3^2 - 184 q_1^3 q_3^2 + q_3^3\\
 & - 22 q_1 q_3^3 + 675 q_1^2 q_3^3 + q_3^4 - 24 q_1 q_3^4 + q_3^5 - 6 q_1 q_2 q_3 q_4 + 
 144 q_1^2 q_2 q_3 q_4 - 32 q_1 q_2 q_3^2 q_4+\mathcal{O}(q^5)\\
z_4=&q_4 - 2520 q_1^2 q_2 q_3^2 q_4 - 403200 q_1^3 q_2 q_3^3 q_4 - 
 53701200 q_1^4 q_2 q_3^4 q_4 - 403200 q_1^3 q_2^2 q_3^3 q_4^2\\
 & + 2520 q_1^2 q_2 q_3^2 q_4^3 + 403200 q_1^3 q_2 q_3^3 q_4^3+\mathcal{O}(q^{10})
\end{aligned}
\end{equation*}
The double logarithmic solutions are denoted by their leading term 
\begin{equation*}
\begin{gathered}
\frac{1}{2}\ell_1^2,\quad\frac{1}{2}\ell_2^2,\quad\frac{1}{2}\ell_3^2+\ell_1\ell_3,\quad\frac{1}{2}\ell_4^2+\ell_2\ell_4,\\
\ell_1\ell_2,\quad\ell_1\ell_4,\quad\ell_2\ell_3,\quad\ell_3\ell_4
\end{gathered}
\end{equation*}
with abbreviations $\ell_i=\log(z_i)$. 
The brane superpotential is constructed as linear combination of double logarithmic solutions, 
\begin{equation}
\mathcal{W}_{\mathrm{brane}}=t_1^2+\sum_{n_i}N_{d_1,d_2,d_3,d_4}\mathrm{Li}_2(q_1^{d_1}q_2^{d_2}q_3^{d_3}q_3^{d_4})
\end{equation}
This has the expected integrality properties of the Ooguri-Vafa $\text{Li}_2$ multicover formula. The Ooguri-Vafa invariants  of the form $N_\mathrm{(m,k,m+n,k)}$ are exactly match the data in Table 5 in \cite{Jiang2017}. In addition, we also extract the invariants of the form $N_\mathrm{(m+n,k,n,k)}$ and summarize them into Table \ref{table:19}, where the rows and columns are labelled by m and n, respectively.

\subsection{Open-Closed GKZ-system: Branes on $X_{12}^{(1,1,2,2,6)}$}
\subsubsection{Five Branes Wrapping Rational Curves and Blowing Up Geometry}

 In this section, we study the A-model manifold, whose toric polyhedron is denoted by $\Delta^*$ and charge vectors are denoted by $l^1$ and $l^2$. A-model manifold and A-brane are specified by the following toric data.
  \begin{table}[H]\label{table:5}
\centering
 \begin{tabular}{c|cccc|cc|c|cc}
 $\quad$&\multicolumn{4}{|c|}{$\Delta^*$}&$l^1$&$l^2$&$\quad$&$\hat{l}^1$&$\hat{l}^2$\\
 \hline
 $v^*_0$&$0$&$0$&$0$&$0$&$-6$&$0$&$x_1x_2x_3x_4x_5$&$-1$&$0$\\
 $v^*_1$&$-1$&$-2$&$-2$&$-6$&$0$&$1$&$x_1^{12}$&$0$&$-1$\\
 $v^*_2$&$1$&$0$&$0$&$0$&$0$&$1$&$x_2^{12}$&$0$&$1$\\
 $v^*_3$&$0$&$1$&$0$&$0$&$1$&$0$&$x_3^6$&$0$&$0$\\
 $v^*_4$&$0$&$0$&$1$&$0$&$1$&$0$&$x_4^6$&$0$&$0$\\
 $v^*_5$&$0$&$0$&$0$&$1$&$3$&$0$&$x_5^2$&$1$&$0$\\
 $v^*_6$&$0$&$-1$&$-1$&$-3$&$1$&$-2$&$(x_1x_2)^6$&$0$&$0$
 \end{tabular}
 \caption{Toric Data of A-model side}
\end{table}

The mirror hypersurface $X$ is determined by the constraint, 
\begin{equation*}
P=x_1^{12}+x_2^{12}+x_3^6+x_4^6+x_5^2+\psi(x_1x_2x_3x_4x_5)+\phi(x_1x_2)^6
\end{equation*}
where $x_i$'s are homogeneous coordinates in $\mathbb{P}_{(1,1,2,2,6)}$ and $\psi = z_1^{-\frac{1}{6}}z_2^{-\frac{1}{12}}$, $\phi= z_2^{-\frac{1}{2}}$. 
The Greene-Plesser orbifold group G acts as $x_i \rightarrow \lambda^{g_{k,i}}_k x_i$ with $ \lambda^6_1=\lambda^6_2=\lambda^2_3=1$ and weights
\begin{equation*}
\mathbb{Z}_6:g_1=(1,0,-1,0,0),\quad\mathbb{Z}_6:g_2=(1,0,0,-1,0),\quad\mathbb{Z}_2:g_3=(1,0,0,0,-1)
\end{equation*} 
The toric curve $S$ is described by the complete intersection
\begin{equation*}
S : \quad P=0,\quad h_1 \equiv \gamma^{12} (x_1x_2x_3x_4x_5)-\alpha^{12} x_5^2=0,\quad h_2\equiv \alpha^{12}x_1^{12}-\beta^{12} x_2^{12}=0
\end{equation*}
Insert $h_1$ and $h_2$ into $P=0$ 
\begin{equation}\label{eq:5.4}
\begin{gathered}
\tilde{\mathbb{P}}^1:\quad \eta_1 x_3+\sqrt[6]{x_4^6+ m(x_1,x_5)}=0,\eta_1^6=1\\
m(x_1,x_5)= (\frac{\alpha^{12}+\beta^{12}}{\alpha^{12}}+\frac{\beta^{12}}{\alpha^{12}}\phi)x_1^{12}+(1+\frac{\gamma^{12}}{\alpha^{12}}\phi)x_5^{2}
\end{gathered}
\end{equation}
Here $\tilde{\mathbb{P}^1}$is evidently non-holomorphic because of the sixth roots of unity and hence, a non-holomorphic family of rational curves on. However, at special loci 
\begin{equation}
\alpha^{12}+\beta^{12}+\phi \beta^{12}=0,\quad \alpha^{12}+\gamma^{12}\psi=0
\end{equation}
we see that $S$ degenerates as follows,
\begin{equation}\label{eq:5.6}
\Sigma:\quad h_0 \equiv x_3^6+x_4^6,\quad h_1 \equiv \gamma^{12} (x_1x_2x_3x_4x_5)-\alpha^{12} x_5^2=0,\quad h_2\equiv \alpha^{12}x_1^{12}-\beta^{12} x_2^{12}=0
\end{equation}
Modulo the action of G, \ref{eq:5.4} can be solve holomorphically. Thus the anholomorphic deformation of $\mathbb{P}^1$ \ref{eq:5.6} can be used to describe holomorphic deformation of $S$.

From the toric data in Table \ref{table:5}, the full set of GKZ operators are derived by \ref{eq:2.21}
\begin{equation}
\begin{gathered}
\mathcal{Z}_0=\sum_{i=0}^6 \vartheta_i+1,\quad\mathcal{Z}_1=\vartheta_1-\vartheta_2,\quad\mathcal{Z}_2=2\vartheta_1-\vartheta_3+\vartheta_6,\\
\mathcal{Z}_3=2\vartheta_1-\vartheta_4+\vartheta_6,\quad\mathcal{Z}_4=6\vartheta_1-\vartheta_5+3\vartheta_6,\\
\mathcal{L}_1=\frac{\partial}{\partial a_1} \frac{\partial}{\partial a_2}-(\frac{\partial}{\partial a_6})^6,\quad \mathcal{L}_2=\frac{\partial}{\partial a_3}\frac{\partial}{\partial a_4}(\frac{\partial}{\partial a_5})^3\frac{\partial}{\partial a_6}-(\frac{\partial}{\partial a_0})^6
\end{gathered}
\end{equation}
where $\mathcal{Z}_0$ represents the invariance of $P$ under overall rescaling, $\mathcal{Z}_{i}$'s relate to the torus symmetry, and $\mathcal{L}_i$'s relate to the symmetries among monomials consisting of $P$.  And all  GKZ operators above annihilate the period matrix and determine the mirror maps and superpotential.

After blowing up $X$ along $S$, the blow-up manifold $X^\prime$ is obtained as the complete intersection in $\mathcal{W}=\mathbb{P}(\mathcal{O}(12)\oplus \mathbb{P}\mathcal{O}(12))$
\begin{equation}
\begin{aligned}
X^\prime:P &= a_1x_1^{12}+1_2x_2^{12}+a_3x_3^6+a_4x_4^6+a_5x_5^2+a_0(x_1x_2x_3x_4x_5)+a_6(x_1x_2)^6=0\\
Q & =y_1(a_9x_1^{12}+a_{10}x_2^{12})-y_2(a_7(x_1x_2x_3x_4x_5)+a_8x_3^6)
\end{aligned}
\end{equation}
where $a_i$'s  are free complex-valued coefficients. By simple observation, we can obtain the GKZ system of $X^\prime$ as complement to GKZ system of $X$.
\begin{equation}\label{eq:5.10}
\begin{gathered}
\mathcal{Z}_0^\prime=\sum_{i=0}^6 \vartheta_i+1,\quad\mathcal{Z}_1^\prime=\sum_{i=7}^{10} \vartheta_i,\quad\mathcal{Z}_2^\prime=-\vartheta_1+\vartheta_2-\vartheta_9+\vartheta_{10},\quad\mathcal{Z}_3^\prime=-2\vartheta_1+\vartheta_3-\vartheta_6-2\vartheta_{9},\\
\mathcal{Z}_4^\prime=-2\vartheta_1+\vartheta_4-\vartheta_6-2\vartheta_9,\quad\mathcal{Z}_5^\prime=-6\vartheta_{1}+\vartheta_5-3\vartheta_6+\vartheta_8-6\vartheta_9,\quad\mathcal{Z}_6^\prime=-\vartheta_7-\vartheta_8+\vartheta_9+\vartheta_{10}\\
\mathcal{L}_1^\prime=\prod^3_{i=1}\frac{\partial}{\partial a_i}-(\frac{\partial}{\partial a_0})^3,\quad \mathcal{L}_2^\prime=\prod
^6_{i=4}\frac{\partial}{\partial a_i}-(\frac{\partial}{\partial a_3})^3 \\
\mathcal{L}_3^\prime=\frac{\partial}{\partial a_3} \frac{\partial}{\partial a_8}-\frac{\partial}{\partial a_5} \frac{\partial}{\partial a_7},\quad \mathcal{L}_4^\prime=\frac{\partial}{\partial a_6} \frac{\partial}{\partial a_9}-\frac{\partial}{\partial a_3} \frac{\partial}{\partial a_{10}}
\end{gathered}
\end{equation}
where $\mathcal{Z}_0^\prime,\mathcal{Z}_1^\prime$ are associated with the overall rescaling with respect to $P=0,Q=0$, $\mathcal{Z}_i^\prime,i=2,...,6$ are related to the torus symmetry, and  $\mathcal{L}_3^\prime,\mathcal{L}_4^\prime$ incorporate the parameter $a_7,...,a_{10}$ that are associated with the moduli of the curve $S$.

Then we formulate above GKZ-system on an enhanced polyhedron $\Delta^\prime$
 \begin{table}[H]
\centering
 \begin{tabular}{c|ccccccc|cccc|l}
 $\quad$& \multicolumn{7}{|c|}{$\Delta^\prime$} & $l^\prime_1$&$l^\prime_2$&$l^\prime_3$&$l^\prime_4$\\
 \hline
$v^\prime_0$&$1$&$0$&$0$&$0$&$0$&$0$&$0$&$-1$&$-3$&$0$&$0$&$w^\prime_0=x_1x_2x_3x_4x_5$\\
$v^\prime_1$&$1$&$0$&$-1$&$-2$&$-2$&$-6$&$0$&$0$&$0$&$2$&$-1$&$w^\prime_1=x_1^{12}$\\
$v^\prime_2$&$1$&$0$&$1$&$0$&$0$&$0$&$0$&$0$&$0$&$0$&$1$&$w^\prime_2=x_2^{12}$\\
$v^\prime_3$&$1$&$0$&$0$&$1$&$0$&$0$&$0$&$0$&$1$&$0$&$0$&$w^\prime_3=x_3^6$\\
$v^\prime_4$&$1$&$0$&$0$&$0$&$1$&$0$&$0$&$0$&$1$&$0$&$0$&$w^\prime_4=x_4^6$\\
$v^\prime_5$&$1$&$0$&$0$&$0$&$0$&$1$&$0$&$1$&$0$&$0$&$0$&$w^\prime_5=x_5^2$\\
$v^\prime_6$&$1$&$0$&$0$&$-1$&$-1$&$-3$&$0$&$0$&$1$&$-2$&$0$&$w^\prime_6=(x_1x_2)^6$\\
$v^\prime_7$&$0$&$1$&$0$&$0$&$0$&$0$&$-1$&$1$&$-3$&$0$&$0$&$w^\prime_7=y_1 w^\prime_1$\\
$v^\prime_8$&$0$&$1$&$0$&$0$&$0$&$1$&$-1$&$-1$&$3$&$0$&$0$&$w^\prime_8=y_1 w^\prime_2$\\
$v^\prime_9$&$0$&$1$&$-1$&$-2$&$-2$&$-6$&$1$&$0$&$0$&$-1$&$1$&$w^\prime_9=y_2 w^\prime_0$\\
$v^\prime_{10}$&$0$&$1$&$1$&$0$&$0$&$0$&$1$&$0$&$0$&$1$&$-1$&$w^\prime_{10}=y_2 w^e_5$\\
 \end{tabular}
\end{table}
Here we present the integral points $v^\prime_i$'s and their corresponding monomials $w^\prime_i$, as well as $l^\prime_i$'s as basis of  Mori cone satisfying relations $l^1=3l^\prime_1+l^\prime_2$ and $l^2=l^\prime_3+l^\prime_4$.

  The local coordinates $z^i$'s on the complex structure moduli space of $X^\prime$ by \ref{eq:2.22},
\begin{equation}\label{eq:5.12}
z_1=-\frac{ a_5 a_7}{a_0^3 a_8},\quad z_2=\frac{a_3a_4 a_6a_8^3}{a_0^3 a_7^3},\quad z_3=\frac{a_1^2 a_{10}}{a_6^2 a_9},\quad z_4=\frac{a_2 a_9}{a_1 a_{10}}
\end{equation}
Next, we convert the $\mathcal{L}_i$ in \ref{eq:5.10}operators to Picard-Fuchs operators $\mathcal{D}_i$, from differential equations about $a_j(j=0,\dots,10)$ to those about  $z_j(j=1,\dots,4)$ ,
\begin{equation}
\begin{aligned}
\mathcal{D}_1&=\theta_1 (\theta_1-3\theta_2-1)-z_1
 (-\theta_1-3\theta_2)(-\theta_1+3\theta_2),\\
 \mathcal{D}_2&=\theta_2^2 (\theta_2-2\theta_3) (-\theta_1+3\theta_2)^3-z_2
 (\theta_1-3\theta_2)^3  \prod_{i=1}^3(-\theta_1-3\theta_2-i)\\
 \mathcal{D}_3&=(2\theta_3-\theta_4)^2(\theta_3-\theta_4)-z_3 (\theta_2-2\theta_3)^2 (-\theta_3+\theta_4),\\
 \mathcal{D}_4&=\theta_4 (-\theta_3+\theta_4)-z_4(2\theta_3-\theta_4)(\theta_3-\theta_4)\\
 \cdots
 \end{aligned}
 \end{equation}
where $\theta_i=z_i \frac{\partial}{\partial z_i}$'s are the logarithmic derivatives and each $\mathcal{D}_i$ corresponds to a specific linear combination among $l^\prime_i$'s.

\subsubsection{Brane Superpotential and Disk Instantons}
Along the line of \ref{sec:2.2}, we solve the differential equations \ref{eq:5.12}  at $z_i\rightarrow 0$ and identify the mirror maps and superpotential. The fundamental period of $X$ as power series solution is
\begin{equation*}
\omega_0=1 - 120 z_1^3 z_2 + 83160 z_1^6 z_2^2 + 166320 z_1^6 z_2^2 z_3 z_4+\mathcal{O}(z^{10})
\end{equation*}.
 The single logarithmic solutions are
\begin{equation*}
\begin{aligned}
\omega_{1,1}=& \omega_0 \log(z_1)+z_1 + \frac{1}{2} z_1^2 + \frac{1}{3} z_1^3 + \frac{1}{4}z_1^4 + \frac{1}{5}z_1^5 + 2 z_2 + 12 z_1 z_2 + 60 z_1^2 z_2 - 74 z_1^3 z_2 - 210 z_1^4 z_2\\
& - 15 z_2^2 - 126 z_1 z_2^2 - 
 630 z_1^2 z_2^2 - 2520 z_1^3 z_2^2 + \frac{560}{3} z_2^3 + 2100 z_1 z_2^3 + 
 13200 z_1^2 z_2^3 - \frac{5775}{2} z_2^4 \\
 &- 40950 z_1 z_2^4 + \frac{252252}{5} z_2^5 - 
 30 z_2^2 z_3 z_4 - 252 z_1 z_2^2 z_3 z_4 + 1120 z_2^3 z_3 z_4+\mathcal{O}(z^6)
 \\
\omega_{1,2}=&\omega_0\log(z_2)-3 z_1 - \frac{3}{2} z_1^2 - z_1^3 - \frac{3}{4} z_1^4 - \frac{3}{5} z_1^5 - 6 z_2 - 
 36 z_1 z_2 - 180 z_1^2 z_2 - 522 z_1^3 z_2 + 630 z_1^4 z_2 \\
 &+ 45 z_2^2 + 
 378 z_1 z_2^2 + 1890 z_1^2 z_2^2 + 7560 z_1^3 z_2^2 - 560 z_2^3 - 
 6300 z_1 z_2^3 - 39600 z_1^2 z_2^3 + \frac{17325}{2} z_2^4 \\
 &+ 122850 z_1 z_2^4 -\frac{756756}{5} z_2^5 - z_3 z_4 + 90 z_2^2 z_3 z_4 + 756 z_1 z_2^2 z_3 z_4 - 
 3360 z_2^3 z_3 z_4 - \frac{3}{2} z_3^2 z_4^2+\mathcal{O}(z^6)\\
\omega_{1,3}=&\omega_0\log(z_3)-240 z1^3 z2 + 2 z3 z4 + 240 z1^3 z2 z3 z4 + 3 z3^2 z4^2 + 120 z1^3 z2 z3^2 z4^2 + \frac{20}{3} z3^3 z4^3 \\
&+ 160 z1^3 z2 z3^3 z4^3 + \frac{35}{2} z3^4 z4^4 + \frac{252}{5} z3^5 z4^5++\mathcal{O}(z^{10})\\
\omega_{1,4}=&\omega_0\log(z_4)
\end{aligned}
\end{equation*}
such that the single logarithmic periods of $X$ are reproduced by $\Pi^1_1=\omega_{1,1}+\omega_{1,2}$, $\Pi^1_2=\omega_{1,3}+\omega_{1,4}$. With single logarithmic solutions, open-closed mirror maps are inverse series of flat coordinates     
\begin{equation*}
\begin{aligned}
z_1=&q_1 - q_1^2 + q_1^3 - q_1^4 + q_1^5 - 2 q_1 q_2 - 14 q_1^2 q_2 - 60 q_1^3 q_2 + 
 134 q_1^4 q_2 + 5 q_1 q_2^2 + 58 q_1^2 q_2^2 \\
 &+ 270 q_1^3 q_2^2 - 32 q_1 q_2^3 - 546 q_1^2 q_2^3 + 286 q_1 q_2^4 - 2 q_1 q_2 q_3 q_4 - 14 q_1^2 q_2 q_3 q_4 + 40 q_1 q_2^2 q_3 q_4+\mathcal{O}(q^5)\\
z_2=&q_2 + 3 q_1 q_2 + 3 q_1^2 q_2 + q_1^3 q_2 + 6 q_2^2 + 66 q_1 q_2^2 + 
 384 q_1^2 q_2^2 + 1338 q_1^3 q_2^2 + 9 q_2^3 + 222 q_1 q_2^3\\
 & + 2940 q_1^2 q_2^3 + 56 q_2^4 + 1350 q_1 q_2^4 - 300 q_2^5 + q_2 q_3 q_4 + 
 3 q_1 q_2 q_3 q_4 + 3 q_1^2 q_2 q_3 q_4 \\
 &+ 12 q_2^2 q_3 q_4 +  132 q_1 q_2^2 q_3 q_4 - 63 q_2^3 q_3 q_4+\mathcal{O}(q^5)\\
z_3=&q_3 + 240 q_1^3 q_2 q_3 - 2 q_3^2 q_4 - 1200 q_1^3 q_2 q_3^2 q_4 + 
 3 q_3^3 q_4^2 + 3120 q_1^3 q2 q_3^3 q_4^2 - 4 q_3^4 q_4^3 + 5 q_3^5 q_4^4+\mathcal{O}(q^{10})\\
z_4=&q_4
\end{aligned}
\end{equation*}
The double logarithmic solutions are denoted by their leading term 
\begin{equation}
\begin{gathered}
\frac{3}{2}\ell_1^2+\ell_1\ell_2,\quad\frac{1}{2}\ell_2^2,\quad\frac{1}{2}\ell_3^2,\quad\frac{1}{2}\ell_4^2+\ell_3\ell_4,\\
\ell_1\ell_3,\quad\ell_1\ell_4,\quad\ell_2\ell_3,\quad\ell_2\ell_4
\end{gathered}
\end{equation}
with abbreviations $\ell_i=\log(z_i)$. 
Then we construct two linear combination of double logarithmic  solutions and  insert the inverse mirror maps to match the disk instantons in  \cite{Jockers2009}.

\begin{equation*}
\begin{aligned}
\mathcal{W}_{\mathrm{brane}}^I&=\frac{3}{2}t_1^2+t_1t_2+\sum_{N^I}N^I_{d_1,d_2,d_3,d_4}\mathrm{Li}_2(q_1^{d_1}q_2^{d_2}q_3^{d_3}q_4^{d_4})\\
\mathcal{W}_{\mathrm{brane}}^{II}&=\frac{1}{2} t_1t_3+\sum_{N^{II}}N^{II}_{d_1,d_2,d_3,d_4}\mathrm{Li}_2(q_1^{d_1}q_2^{d_2}q_3^{d_3}q_4^{d_4})
\end{aligned}
\end{equation*}.Invariants of the form $N_\mathrm{m,n,k,k}$ are summarized in Table\ref{table:20} \ref{table:21} \ref{table:22} \ref{table:23}, where the rows and columns are labelled by m and n, respectively.

\section{One Closed and Two Open Moduli Case}
\subsection{Open-Closed GKZ-system: Branes on Complete Intersections $\mathbb{P}^{(111|111)}_{[3,3]}$}
\subsubsection{Five Branes Wrapping Lines and Blowing Up Geometry}

The underlying manifold $X^*$ we are considering in the A-model is
the intersection of two cubics in $\mathbb{P}^5$ , whose mirror manifold $X$ can be represented a one-parameter family of bicubics, with group $G=\mathbb{Z}_3^2\times\mathbb{Z}_9$ acting on them,
\begin{equation}\label{eq:4.1}
\begin{cases}
P_1=x_1^3+x_2^3+x_3^3+\psi x_4x_5x_6\\
P_2=x_4^3+x_5^3+x_6^3+\psi x_1x_2x_3
\end{cases}
\end{equation}
where $\psi$ is  the complex structure modulus.

Turning to the specification of D-brane configurations, we consider the curve $S$ on $X$ 
\begin{equation*}
\begin{aligned}
P_1&=P_2=0, \\
h_1 & \equiv \beta^3 (x_1x_2x_3)-\alpha \beta \gamma x_2^3=0,\\ h_2& \equiv \gamma^3 (x_1x_2x_3)-\alpha\beta \gamma x_3^3=0
\end{aligned}
\end{equation*}

 An equivalent and convenient form is easy to obtained,
\begin{equation}
P_1=P_2=0,\quad\alpha^3x_2^3-\beta^3x_1^3=0,\quad \alpha^3x_3^3-\gamma^3x_1^3=0
\end{equation}

For generic values of the moduli , the $S$ is an irreducible higher genus Riemann surface. But we can always make a linearization by inserting $h_1$ and $h_2$ into $P_1,P_2$,
\begin{equation}
x_4^3+x_5^3+x_6^3-\Psi x_4x_5x_6=0,\quad \Psi=\frac{\alpha \beta \gamma}{\alpha^3+\beta^3+\gamma^3}\psi^2
\end{equation}
This is an one dimensional family of cubic plane elliptic curves in $\mathbb{P}^2$, called the  Hesse pencil. For special value of $\Psi$, it degenerate into 12 lines \cite{Artebani2006},
\begin{equation}\label{eq:6.5}
x_4+\eta_1 x_5+\eta_2 x_6=0,\quad \eta_1^3=\eta_2^3=1
\end{equation} 
Upon the action of group G, they are identified as a single line.

Thus the  deformation space of \ref{eq:6.5} is embedded in the deformation space of $S$. And away from that special locus, the obstructed deformation is identified with the unobstructed deformation of $S$, which means that we can use the obstructed deformation of that line to describe the unobstructed deformation of $S$.

 The polyhedron corresponding to the A-model manifold $X^*$ is denoted as $\Delta^*$ , the A-brane charge vectors  $l$, and A-brane charge vectors  $\hat{l}^1$, $\hat{l}^2$. The toric data in A-model side is as following table.

 \begin{table}[H]
\centering
 \begin{tabular}{c|ccccc|c|c|cc}
 $\quad$&\multicolumn{5}{|c|}{$\Delta^*$}&$l$&$\quad$&$\hat{l}^1$&$\hat{l}^2$\\
 \hline
 $v^*_0$&$0$&$0$&$0$&$0$&$0$&$-3$&$x_1x_2x_3$&$-1$&$-1$\\
 $v^*_0$&$0$&$0$&$0$&$0$&$0$&$-3$&$x_4x_5x_6$&$0$&$0$\\
 $v^*_1$&$-1$&$-1$&$-1$&$-1$&$-1$&$1$&$x_1^3$&$0$&$0$\\
 $v^*_2$&$1$&$0$&$0$&$0$&$0$&$1$&$x_2^3$&$1$&$0$\\
 $v^*_3$&$0$&$1$&$0$&$0$&$0$&$1$&$x_3^3$&$0$&$1$\\
 $v^*_4$&$0$&$0$&$1$&$0$&$0$&$1$&$x_4^3$&$0$&$0$\\
 $v^*_5$&$0$&$0$&$0$&$1$&$0$&$1$&$x_5^3$&$0$&$0$\\
 $v^*_6$&$0$&$0$&$0$&$0$&$1$&$1$&$x_6^3$&$0$&$0$\\
  \end{tabular}
 \caption{Toric Data of A-model side}
\end{table}

In toric coordinates, the Calabi-Yau threefold $X$ and the curve $S$ are described as
\begin{equation*}
\begin{aligned}
X: \quad P_1&=a_{0,1}+a_1(X_2X_3X_4X_5X_6)^{-1}+a_2X_2+a_3X_3=0,\\
 P_2&=a_{0,2}+a_4X_4+a_5X_5+a_6X_6=0 \\
S:\quad h_1&=a_7+a_8X_2=0,\quad h_2=a_9+a_{10}X_3=0
\end{aligned}
\end{equation*}
with $a_i$'s are free complex-valued coefficients. The GKZ-system as follow by \ref{eq:2.21},
\begin{equation}
\begin{gathered}
\mathcal{Z}_0=\sum_{i=0}^6\vartheta_i+1,\quad\mathcal{Z}_i=-\vartheta_1+\vartheta_{i+1},i=1,...,5\\
\mathcal{L}_1=\prod^6_{i=1}\frac{\partial}{\partial a_i}-(\frac{\partial}{\partial a_{0,1}})^3(\frac{\partial}{\partial a_{0,2}})^3
\end{gathered}
\end{equation}
where  $\vartheta_i=a_i \frac{\partial}{\partial a_i}$'s are the logarithmic derivative. $\mathcal{Z}_0$ represents the invariance of $P$ under overall rescaling and $\mathcal{Z}_{i}$'s relate to the torus symmetry,
\begin{equation*}
\mathcal{Z}_i:\quad X_{i+1} \mapsto \lambda X_{i+1},\quad (a_1,a_{i+1})\mapsto(\lambda a_1,\lambda^{-1} a_{i+1}),\quad i=1,2,3,4
\end{equation*} 
For $\mathcal{L}_i$'s, the represents the relations among Laurent monomials in $P_1$ and $P_2$ in \ref{eq:4.1} . 
\begin{equation*}
\mathcal{L}_1 :\quad (X_2^{-1})(X_3^{-1})(X_4^{-1})(X_5^{-1})(X_6^{-1})X_2X_3X_4X_5X_6=1
\end{equation*}
And all  GKZ operators annihilate the period matrix and determine the mirror maps and superpotentials.

After blowing up $X$ along $S$, the blow-up manifold is 
\begin{equation}\label{eq:4.66}
X^\prime:P_1=P_2=0,\quad Q=y_1(a_9+a_{10}X_3)-y_2(a_7+a_8X_2)
\end{equation}

By careful observations on the defining equations \ref{eq:4.66}, the GKZ system of $X^\prime$ is obtained as follow
\begin{equation}\label{eq:6.11}
\begin{gathered}
\mathcal{Z}_0^\prime=\sum_{i=0}^6 \vartheta_i+1,\quad\mathcal{Z}_1^\prime=\sum_{i=7}^{10} \vartheta_i,\quad\mathcal{Z}_2^\prime=-\vartheta_1+\vartheta_2+\vartheta_8,\quad\mathcal{Z}_3^\prime=-\vartheta_1+\vartheta_3+\vartheta_{10},\\
\mathcal{Z}_i^\prime=-\vartheta_1+\vartheta_i,\quad i=4,5,6,\quad\mathcal{Z}_7^\prime=-\vartheta_7-\vartheta_8+\vartheta_9+\vartheta_{10}\\
\mathcal{L}_1^\prime=\prod^3_{i=1}\frac{\partial}{\partial a_i}-(\frac{\partial}{\partial a_0})^3, \\
\mathcal{L}_2^\prime=\frac{\partial}{\partial a_2} \frac{\partial}{\partial a_7}-\frac{\partial}{\partial a_{0,1}} \frac{\partial}{\partial a_8},\quad \mathcal{L}_3^\prime=\frac{\partial}{\partial a_3} \frac{\partial}{\partial a_9}-\frac{\partial}{\partial a_{0,1}} \frac{\partial}{\partial a_{10}}
\end{gathered}
\end{equation}where $\mathcal{Z}_0^\prime,\mathcal{Z}_1^\prime$ are associated with the overall rescaling with respect to $P=0,Q=0$ respectively. $\mathcal{Z}_i^\prime,i=2,...,5$ are related to the torus symmetry. 
\begin{equation*}
\begin{aligned}
\mathcal{Z}_2^\prime:&\quad X_2\mapsto \lambda X_2,\quad(a_1,a_2,a_8)\mapsto(\lambda a_1,\lambda^{-1}a_2,\lambda^{-1} a_{8})\\
 \mathcal{Z}_3^\prime:&\quad X_3\mapsto \lambda X_3,\quad(a_1,a_3,a_{10})\mapsto (\lambda a_1,\lambda^{-1}a_3,\lambda^{-1} a_{10}) \\
\mathcal{Z}_i^\prime:&\quad X_i\mapsto \lambda X_i,\quad(a_1,a_i)\mapsto (\lambda a_1,\lambda^{-1}a_i),\quad i=4,5,6\\  
 \end{aligned}
\end{equation*} 
 In addition,$\mathcal{Z}_6^\prime$ is related to the torus symmetry $(y_1,y_2)\mapsto(\lambda y_1,\lambda^{-1} y_2) $. 
The  $\mathcal{L}_2^\prime,\mathcal{L}_3^\prime$ incorporate the parameter $a_7,...,a_{10}$ that are associated with the moduli of the curve $S$.

 Now, we formulate  GKZ system \ref{eq:6.11} on an enhanced polyhedron $\Delta^\prime$,
\begin{table}[H]
\centering
 \begin{tabular}{c|cccccccc|ccc|l}
 $\quad$& \multicolumn{8}{|c|}{$\Delta^\prime$}&$l^\prime_1$&$l^\prime_2$&$l^\prime_3$\\
 \hline
$v^\prime_0$&$1$&$0$&$0$&$0$&$0$&$0$&$0$&$0$&$0$&$-1$&$-1$&$w^\prime_{0,1}=x_1x_2x_3$\\
$v^\prime_0$&$1$&$0$&$0$&$0$&$0$&$0$&$0$&$0$&$-3$&$0$&$0$&$w^\prime_{0,2}=x_4x_5x_6$\\
$v^\prime_1$&$1$&$0$&$-1$&$-1$&$-1$&$-1$&$-1$&$0$&$1$&$0$&$0$&$w^\prime_1=x_1^3$\\
$v^\prime_2$&$1$&$0$&$1$&$0$&$0$&$0$&$0$&$0$&$-1$&$1$&$0$&$w^\prime_2=x_2^3$\\
$v^\prime_3$&$1$&$0$&$0$&$1$&$0$&$0$&$0$&$0$&$0$&$0$&$1$&$w^\prime_3=x_3^3$\\
$v^\prime_4$&$1$&$0$&$0$&$0$&$1$&$0$&$0$&$0$&$1$&$0$&$0$&$w^\prime_4=x_4^3$\\
$v^\prime_5$&$1$&$0$&$0$&$0$&$0$&$1$&$0$&$0$&$1$&$0$&$0$&$w^\prime_5=x_5^3$\\
$v^\prime_6$&$1$&$0$&$0$&$0$&$0$&$0$&$1$&$0$&$1$&$0$&$0$&$w^\prime_6=x_6^3$\\
$v^\prime_7$&$0$&$1$&$0$&$0$&$0$&$0$&$0$&$-1$&$-2$&$1$&$0$&$w^\prime_7=y_1 w^\prime_0$\\
$v^\prime_8$&$0$&$1$&$1$&$0$&$0$&$0$&$0$&$-1$&$2$&$-1$&$0$&$w^\prime_8=y_1 w^\prime_2$\\
$v^\prime_9$&$0$&$1$&$0$&$0$&$0$&$0$&$0$&$1$&$-1$&$0$&$1$&$w^\prime_9=y_2 w^\prime_0$\\
$v^\prime_{10}$&$0$&$1$&$0$&$1$&$0$&$0$&$0$&$1$&$1$&$0$&$-1$&$w^\prime_{10}=y_2 w^\prime_3$\\
 \end{tabular}
\end{table}
where $v^\prime_i$'s are the integral vertices and $w^\prime_i$,their corresponding monomials. The A-model closed string charge vectors and A-branes charge vectors satisfy the relations $l^1=l^\prime_1+2 l^\prime_2+l^\prime_3$, $\hat{l}^1=l^\prime_3$, $\hat{l}^2=l^\prime_4$.

The coordinates $z_i$ by \ref{eq:2.22}on the complex structure moduli space of $X^\prime$,
\begin{equation}
z_1=\frac{a_1 a_4 a_5 a_6 a_8 a_{10}}{a_{0,2}^3 a_2 a_7^2 a_9},\quad z_2=\frac{a_2 a_7}{a_{0,1} a_8},\quad z_3=\frac{a_3 a_9}{a_{0,1} a_{10}}
\end{equation}

Next, we convert the $\mathcal{L}_i$ operators in \ref{eq:6.11} to Picard-Fuchs operators $\mathcal{D}_i$, from differential equations about $a_j$ to those about  $z_j$,
\begin{equation}\label{eq:6.14}
\begin{aligned}
\mathcal{D}_1&=\theta_1^4 (2\theta_1-\theta_2)^2(\theta_1 -\theta_3)-z_1
 (-\theta_1+\theta_2)(-2\theta_1+\theta_2)^2(-\theta_1+\theta_3) \prod^3_{i=1}(-3\theta_1-i),\\
 \mathcal{D}_2&= (-\theta_1+\theta_2) (-2\theta_1+\theta_2)-z_2
 (-\theta_2-\theta_3-1)  (2\theta_1-\theta_2)\\
 \mathcal{D}_3&=\theta_3(-\theta_1+\theta_3)-z_3 (-\theta_2-\theta_3-1) (\theta_1 -\theta_3),\\
 \cdots
 \end{aligned}
 \end{equation}
where $\theta_i=z_i \frac{\partial}{\partial z_i}$'s are the logarithmic derivatives and each operator $\mathcal{D}_a$ corresponds to a linear combination among the charge vectors $l^e_1,l^e_2,l^e_3$.

\subsubsection{Brane Superpotential and Disk Instantons}
Now, we  solve the Picard-Fuchs  equations \ref{eq:6.14}derived in the last section and identified the mirror maps and superpotentials. 
 By the methods introduced in \ref{sec:2.2}, at $z_i\rightarrow 0$, the fundamental period of $X$ as series expansion is
\begin{equation*}
\omega_0=1 + 36 z_1 z_2^2 z_3 + 8100 z_1^2 z_2^4 z_3^2 + 2822400 z_1^3 z_2^6 z_3^3 + 1200622500 z_1^4 z_2^8 z_3^4+\mathcal{O}(z^{16})
\end{equation*}
There are four the single logarithmic solutions 
\begin{equation*}
\begin{aligned}
\omega_{1,1}=&\omega_0\log(z_1)+2 z_2 - z_2^2 - 12 z_1 z_2^2 + \frac{2}{3} z_2^3 -\frac{1}{2} z_2^4 + \frac{2}{5} z_2^5 + z_3 - 
 24 z_1 z_2 z_3 + 90 z_1 z_2^2 z_3 \\
 &+ 144 z_1 z_2^3 z3 - \frac{1}{2}z_3^2 + 
 72 z_1 z_2^2 z_3^2 +\frac{1}{3} z_3^3 -\frac{1}{4} z_3^4 + \frac{1}{5}z_3^5+\mathcal{O}(z^5)\\
\omega_{1,2}=&\omega_0\log(z_2)-z2 + \frac{1}{2}z2^2 - \frac{1}{3}z2^3 + \frac{1}{4}z2^4 - \frac{1}{5}z2^5 + 12 z_1 z_2 z_3 + 30 z_1 z_2^2 z_3 - 72 z_1 z_2^3 z_3+\mathcal{O}(z^6)\\
\omega_{1,3}=&\omega_0\log(z_3)+12 z_1 z_2^2 - 270 z_1^2 z_2^4 - z_3 + 30 z_1 z_2^2 z_3 + \frac{1}{2}z3^2 - 
 72 z_1 z_2^2 z_3^2 -\frac{1}{3} z_3^3 +\frac{1}{4} z_3^4\\
 & - \frac{1}{5}z_3^5+\mathcal{O}(z^5)
\end{aligned}
\end{equation*}
such that they are consistent with the single logarithmic periods of $X$, 
$\Pi^1=\omega_{1,1}+2\omega_{1,2}+\omega_{1,3}$.

By the definition of the flat coordinates and mirror maps 
\begin{equation}
t_j=\frac{\omega_{1,j}}{\omega_0}
 \end{equation}
 \begin{equation}
q _j = e^{2\pi it_j}
 \end{equation}
we obtain the $z_j$  as a series of $q_j$ upon inversion of the mirror maps
\begin{equation*}
\begin{aligned}
z_1=&q_1 - 2 q_1 q_2 + q_1 q_2^2 + 12 q_1^2 q_2^2 - 24 q_1^2 q_2^3 - q_1 q_3 + 
 2 q_1 q_2 q_3 + 24 q_1^2 q_2 q_3 - q_1 q_2^2 q_3\\
 & - 150 q_1^2 q_2^2 q_3 - 
 24 q_1^2 q_2 q_3^2+\mathcal{O}(q^5)\\
z_2=&q_2 + q_2^2 + q_2^3 + q_2^4 + q_2^5 - 12 q_1 q_2^2 q_3 - 42 q_1 q_2^3 q_3+\mathcal{O}(q^5)\\
z_3=&q_3 - 12 q_1 q_2^2 q_3 + q_3^2 - 42 q_1 q_2^2 q_3^2 + q_3^3 + q_3^4 + q_3^5+\mathcal{O}(q^5)
\end{aligned}
\end{equation*}
In addition, there are also double logarithmic solutions with leading terms
\begin{equation*}
\begin{gathered}
\frac{1}{2} \ell_1^2-\ell_2^2,\quad\frac{3}{2}\ell_2^2+\ell_1\ell_2,\quad\frac{1}{2}\ell_3^2+\ell_1\ell_3,\quad\ell_2\ell_3
\end{gathered}
\end{equation*}
where $\log(z_i)$'s are abbreviated as $\ell_i$'s. According to above, a specific linear combination of double logarithmic solutions is constructed and its disk instantons expansions are extracted. 
\begin{equation}
\mathcal{W}_{\mathrm{brane}}=4t_1t_2+6t_2^2+4t_2t_3+\sum_{N}N_{d_1,d_2,d_3}\mathrm{Li}_2(q_1^{d_1}q_2^{d_2}q_3^{d_3})
\end{equation}
We present first a few invariants of the form $N_{m,m+n,m}$ in Table\ref{table:24}. 
\subsection{Branes on Complete Intersections $\mathbb{P}^{(112|112)}_{[4,4]}$ and  $\mathbb{P}^{(123|123)}_{[6,6]}$}
Similar to the last section, we also calculate the superpotential and extract the Ooguri-Vafa invariants at large volume phase for complete intersections Calabi-Yau manifolds of  $\mathbb{P}^{(112|112)}_{[4,4]}$ and  $\mathbb{P}^{(123|123)}_{[6,6]}$. We summarize the main formulas and tables in Appendix \ref{App:A} and Ooguri-Vafa invariants for first several orders in Appendix \ref{App:C}.
\section{One Closed and Three Open Moduli Case}
\subsection{Open-Closed GKZ-System: Branes on Sextic Hypersurface}
\subsubsection{Branes Wrapping Rational Curves and Blowing up Geometry}
In this section, we consider the A-model manifold $X^*$ with charge vector $l=(-6,1,1,1,1,2)$. The corresponding toric polyhedron for $X^*$ consists of following integral vertices
\begin{equation}
(0,0,0,0),\quad (-1,-1,-1,-2),\quad (1,0,0,0),\quad (0,1,0,0),\quad (0,0,1,0),\quad (0,0,0,1)
\end{equation}
  The mirror sextic hypersurface $X$ arises as the Calabi-Yau hypersurfacesis  a mirror pair of  Calabi-Yau hypersurface in $\mathbb{P}^4_{(1,1,1,1,2)}$ . Its defining equation in homogeneous coordinates is
 \begin{equation*}
 P=a_1x_1^6+a_2x_2^6+a_3x_3^6+a_4x_4^6+a_5x_5^3+a_0 x_1x_2x_3x_4x_5
\end{equation*}.
On $X$ ,we consider parallel branes $S$ which are described by 
\begin{equation*}
S : \quad P=0,\quad h_1 \equiv  a_6 (x_1x_2x_3x_4x_5)^2+a_7x_1x_2x_3x_4x_5^4+a_8 x_5^6=0,\quad h_2\equiv a_9 x_1^{6}+ a_{10} x_5^{3}=0
\end{equation*}

The GKZ operators are derived by \ref{eq:2.21}
\begin{equation}
\begin{gathered}
\mathcal{Z}_0=\sum_{i=0}^5 \vartheta_i+1,\quad\mathcal{Z}_i=\vartheta_{i+1}-\vartheta_1,i=1,2,3,\quad\mathcal{Z}_4=\vartheta_5-2\vartheta_1\\
\mathcal{L}_1=\frac{\partial}{\partial a_1} \frac{\partial}{\partial a_2}\frac{\partial}{\partial a_3}\frac{\partial}{\partial a_4}\frac{\partial}{\partial a_5}^2-(\frac{\partial}{\partial a_0})^6
\end{gathered}
\end{equation}
where $\mathcal{Z}_0$ represents the invariance of $P$ under overall rescaling, $\mathcal{Z}_{i}$'s relate to the torus symmetry, and $\mathcal{L}_i$'s relate to the symmetries among monomials consisting of $P$.  And all  GKZ operators above annihilate the period matrix and determine the mirror maps and superpotential.

After blowing up $X$ along $S$, the blow-up manifold is obtained as the complete intersection in the projective bundle
\begin{equation}
\begin{aligned}
X^\prime:P &= a_1x_1^{6}+x_2^{6}+x_3^6+x_4^6+x_5^3+a_0(x_1x_2x_3x_4x_5)=0\\
Q & =y_1(a_9 x_1^{6}+ a_{10} x_5^{3})-y_2(a_6(x_1x_2x_3x_4x_5)^2+a_7x_1x_2x_3x_4x_5^4+a_8 x_5^6)
\end{aligned}
\end{equation}
where $a_i$'s  are free complex-valued coefficients. By  observation on the symmetry of obove defining equations, we can obtain the GKZ system of $X^\prime$ as complement to GKZ system of $X$.
\begin{equation}\label{eq:7.5}
\begin{gathered}
\mathcal{Z}_0^\prime=\sum_{i=0}^5 \vartheta_i+1,\quad\mathcal{Z}_1^\prime=\sum_{i=6}^{10} \vartheta_i,\quad\mathcal{Z}_2^\prime=-\vartheta_1+\vartheta_2+\vartheta_7+2\vartheta_8-\vartheta_9,\quad\mathcal{Z}_3^\prime=-1\vartheta_1+\vartheta_3-\vartheta_{9},\\
\mathcal{Z}_4^\prime=-\vartheta_1+\vartheta_4-\vartheta_9,\quad\mathcal{Z}_5^\prime=-2\vartheta_{1}+\vartheta_5-2\vartheta_6+\vartheta_8-6\vartheta_9,\quad\mathcal{Z}_6^\prime=-\vartheta_6-\vartheta_7-\vartheta_8+\vartheta_9+\vartheta_{10}\\
\mathcal{L}_1^\prime=\prod^5_{i=2}\frac{\partial}{\partial a_i}(\frac{\partial}{\partial a_7})^2 \frac{\partial}{\partial a_9}-(\frac{\partial}{\partial a_0})^4\frac{\partial}{\partial a_6}\frac{\partial}{\partial a_{10}},\quad \mathcal{L}_2^\prime=
\frac{\partial}{\partial a_6}\frac{\partial}{\partial a_8}-(\frac{\partial}{\partial a_7})^2 \\
\mathcal{L}_3^\prime=\frac{\partial}{\partial a_5} \frac{\partial}{\partial a_7}-\frac{\partial}{\partial a_0} \frac{\partial}{\partial a_8},\quad \mathcal{L}_4^\prime=\frac{\partial}{\partial a_1} \frac{\partial}{\partial a_{10}}-\frac{\partial}{\partial a_5} \frac{\partial}{\partial a_{9}}
\end{gathered}
\end{equation}
where $\mathcal{Z}_0^\prime,\mathcal{Z}_1^\prime$ are associated with the overall rescaling with respect to $P=0,Q=0$, $\mathcal{Z}_i^\prime,i=2,...,6$ are related to the torus symmetry, and  $\mathcal{L}_3^\prime,\mathcal{L}_4^\prime$ incorporate the parameter $a_6,...,a_{10}$ that are associated with the moduli of the curve $S$.  The new manifold $X^\prime$ is describe by the following charge vectors
 \begin{table}[H]
\centering
 \begin{tabular}{c|cccccccccccc}
 $\quad$&$0$& $1$&$2$&$3$&$4$&$5$&$6$&$7$&$8$&$9$&$10$\\
 \hline
$l^\prime_1$&$-4$&$0$&$1$&$1$&$1$&$1$&$-2$&$2$&$0$&$1$&$-1$\\
$l^\prime_2$&$0$&$0$&$0$&$0$&$0$&$0$&$1$&$-2$&$1$&$0$&$0$\\
$l^\prime_3$&$-1$&$0$&$0$&$0$&$0$&$1$&$0$&$1$&$-1$&$0$&$0$\\
$l_4^\prime$&$0$&$-1$&$1$&$0$&$0$&$0$&$0$&$0$&$0$&$1$&$-1$\\
 \end{tabular}
\end{table}

  The  the local coordinates $z^i$'s  on the complex structure moduli space of $X^\prime$ by \ref{eq:2.22} is,
\begin{equation}
z_1=\frac{ a_2 a_3 a_4 a_5 a_7^2 a_9}{a_0^4 a_6^2 a_{10}},\quad z_2=\frac{a_6a_8}{ a_7^2},\quad z_3=\frac{a_5 a_{7}}{a_0 a_8},\quad z_4=\frac{a_1 a_{10}}{a_5 a_{9}}
\end{equation}
Next, we convert the $\mathcal{L}_i^\prime$ in \ref{eq:7.5}operators to Picard-Fuchs operators $\mathcal{D}_i$, from differential equations about $a_j(j=0,\dots,10)$ to those about  $z_j(j=1,\dots,4)$ ,
\begin{equation}\label{eq:7.7}
\begin{aligned}
\mathcal{D}_1&=\theta_1^3(\theta_1+\theta_3-\theta_4) (2\theta_1-2\theta_2+\theta_3)^2(\theta_1-\theta_4)-z_1
 (-2\theta_1+\theta_2)^2(-\theta_1+\theta_4)\prod^4_{i=1}(-4\theta_1-\theta_3-i),\\
 \mathcal{D}_2&=(-2\theta_1+\theta_2)(\theta_2-\theta_3)-z_2(2\theta_2-2\theta_2+\theta_3)^2\\
 \mathcal{D}_3&=(\theta_1+\theta_3-\theta_4)(2\theta_1-2\theta_2+\theta_3)-z_3 (\theta_2-\theta_3) (-4\theta_1-\theta_3-1),\\
 \mathcal{D}_4&=\theta_4 (-\theta_1+\theta_4)-z_4(\theta_1+\theta_3-\theta_4)(\theta_1-\theta_4)\\
 \cdots
 \end{aligned}
 \end{equation}
where $\theta_i=z_i \frac{\partial}{\partial z_i}$'s are the logarithmic derivatives.

\subsubsection{Brane Superpotential and Disk Instantons}
Along the line of \ref{sec:2.2}, we solve the differential equations \ref{eq:7.7}  at $z_i\rightarrow 0$ and identify the mirror maps and superpotential. The fundamental period of $S$  is
\begin{equation*}
\omega_0=1 + 360 z_1 z_2^2 z_3^2 z_4 + 1247400 z_1^2 z_2^4 z_3^4 z_4^2 + 
 6861254400 z_1^3 z_2^6 z_3^6 z_4^3 +\mathcal{O}(z^{10})
\end{equation*}. 
 The single logarithmic solutions are
\begin{equation*}
\begin{aligned}
\omega_{1,1}=& \omega_0 \log(z_1)-2 z_2 - 3 z_2^2 - \frac{20}{3} z_2^3 - \frac{35}{2} z_2^4 - \frac{252}{5} z_2^5 + 2 z_2 z_3 + 
 2 z_2^2 z_3 + 4 z_2^3 z_3 + 10 z_2^4 z_3 - z_2^2 z_3^2\\
 & - 120 z_1 z_2^2 z_3^2 - 
 2 z_2^3 z_3^2 + 24 z_1 z_4 - 48 z_1 z_2 z_4 - 24 z_1 z_2^2 z_4 - 
 48 z_1 z_2^3 z_4 - 240 z_1 z_2 z_3 z_4 \\
 &+ 240 z_1 z_2^2 z_3 z_4 + 
 1260 z_1^2 z_4^2 - 5040 z_1^2 z_2 z_4^2++\mathcal{O}(z^5)\\
\omega_{1,2}=&\omega_0\log(z_2)+2 z_2 + 3 z_2^2 + \frac{20}{3} z_2^3 + \frac{35}{2} z_2^4 + \frac{252}{5} z_2^5 + z_3 - 
 2 z_2 z_3 - 2 z_2^2 z_3 - 4 z_2^3 z_3 - 10 z_2^4 z_3\\
 & - \frac{1}{2}z_3^2 + z_2 z_3^2 + 
 z_2^2 z_3^2 + 2 z_2^3 z_3^2 + \frac{1}{3}z_3^3 - z_2 z_3^3 - \frac{1}{4}z_3^4 + z_2 z_3^4 - 
 12 z_1 z_4 + 24 z_1 z_2 z_4 + 24 z_1 z_2^2 z_4\\
 & + 48 z_1 z_2^3 z_4 + 
 120 z_1 z_2 z_3 z_4 - 240 z_1 z_2^2 z_3 z_4 - 630 z_1^2 z_4^2 + 
 2520 z_1^2 z_2 z_4^2 +\mathcal{O}(z^5)\\
\omega_{1,3}=&\omega_0\log(z_3)-z_2 - \frac{3}{2} z_2^2 - \frac{10}{3} z_2^3 - \frac{35}{4} z_2^4 - \frac{126}{5} z_2^5 - z_3 + 
 z_2 z_3 + z_2^2 z_3 + 2 z_2^3 z_3 + 5 z_2^4 z_3 + \\
 &\frac{1}{2}z_3^2 - z_2 z_3^2 - \frac{
 1}{2}z_2^2 z_3^2 - z_2^3 z_3^2 - \frac{1}{3}z_3^3 + z_2 z_3^3 + \frac{1}{4}z_3^4 - 
 z_2 z_3^4 - \frac{1}{5}z_3^5 -12 z_1 z_2^2 z_4 - 24 z_1 z_2^3 z_4+\mathcal{O}(z^{5})\\
\omega_{1,4}=&\omega_0\log(z_4)+120 z_1 z_2^2 z_3^2 - 41580 z_1^2 z_2^4 z_3^4 + 180 z_1 z_2^2 z_3^2 z_4 + 
 498960 z_1^2 z_2^4 z_3^4 z_4 \\
 &- 360 z_1 z_2^2 z_3^2 z_4^2 + 
 727650 z_1^2 z_2^4 z_3^4 z_4^2 + 60 z_1 z_2^2 z_3^2 z_4^3+\mathcal{O}(z^{10})
\end{aligned}
\end{equation*}
by which the open-closed mirror maps are inverse series of flat coordinates     
\begin{equation*}
\begin{aligned}
z_1=&q_1 + 2 q_1 q_2 + q_1 q_2^2 - 4 q_1 q_2 q_3 - 4 q_1 q_2^2 q_3 + 4 q_1 q_2^2 q_3^2 +\mathcal{O}(q^5)\\
z_2=&q_2 - 2 q_2^2 + 3 q_2^3 - 4 q_2^4 + 5 q_2^5 - q_2 q_3 + 5 q_2^2 q_3 - 
 13 q_2^3 q_3 + 25 q_2^4 q_3 - 3 q_2^2 q_3^2 + 18 q_2^3 q_3^2+\mathcal{O}(q^5)\\
z_3=&q_3 + q_2 q_3 + q_3^2 + q_2^2 q_3^2 + q_3^3 + q_3^4 + q_3^5+\mathcal{O}(q^{5})\\
z_4=&q_4 - 120 q_1 q_2^2 q_3^2 q_4 - 180 q_1 q_2^2 q_3^2 q_4^2+\mathcal{O}(q^{7})
\end{aligned}
\end{equation*}
Then we construct a linear combination of double logarithmic  solutions and  insert the inverse mirror maps .
\begin{equation}
\begin{aligned}
\mathcal{W}_{\mathrm{brane}}&=3 t_1^2+ 6 t_1t_4+3 t_4^2+\sum_{N}N_{d_1,d_2,d_3,d_4}\mathrm{Li}_2(q_1^{d_1}q_2^{d_2}q_3^{d_3}q_4^{d_4})
\end{aligned}
\end{equation}Invariants of the form $N_\mathrm{m,n,n,m}$ are summarized in Table\ref{table:26}, where the rows and columns are labelled by m and n, respectively. 

When $ a_7^2=a_6 a_8$, the two individual branes coincide. We obtain a new set of charge vectors,
 \begin{table}[H]
\centering
 \begin{tabular}{c|ccccccccccc}
 $\quad$&$0$& $1$&$2$&$3$&$4$&$5$&$6$&$8$&$9$&$10$\\
 \hline
$l^c_1$&$-4$&$0$&$1$&$1$&$1$&$1$&$-1$&$1$&$1$&$-1$\\
$l^c_2$&$-2$&$0$&$0$&$0$&$0$&$2$&$1$&$-1$&$0$&$0$\\
$l^c_3$&$0$&$-1$&$1$&$0$&$0$&$0$&$0$&$0$&$1$&$-1$\\
 \end{tabular}
\end{table}
by which the new complex structure moduli space coordinates are
\begin{equation}
z_1^c=\frac{a_2a_3a_4a_5a_8a_9}{a_0^4 a_6 a_{10}},\quad z_2^c=\frac{a_5^2a_6}{a_0^2 a_8},\quad z_3^c=\frac{a_2a_9}{a_1 a_{10}}
\end{equation}
Similar to the separate case, the superpotential is constructed as  linear combination of double logarithmic solutions and Ooguri-Vafa invariants are exacted in \ref{table:26}.
\subsection{Open-Closed GKZ-System: Branes on Sextic Hypersurface}
\subsubsection{Branes Wrapping Rational Curves and Blowing up Geometry}
 The mirror octic hypersurface $X$ arises as the Calabi-Yau hypersurfaces in $\mathbb{P}^4_{(1,1,1,1,4)}$ 
 \begin{equation*}
 P=a_1x_1^8+a_2x_2^8+a_3x_3^8+a_4x_4^8+a_5x_5^2+a_0 x_1x_2x_3x_4x_5
\end{equation*} on which  we consider parallel branes which are described by intersections of divisors
\begin{equation*}
S : \quad P=0,\quad h_1 \equiv  a_6 (x_1x_2x_3x_4x_5)^2+a_7x_1x_2x_3x_4x_5^3+a_8 x_5^4=0,\quad h_2\equiv a_9 x_1^{8}+ a_{10} x_5^{2}=0
\end{equation*}

The  GKZ operators are derived by \ref{eq:2.21}
\begin{equation}
\begin{gathered}
\mathcal{Z}_0=\sum_{i=0}^5 \vartheta_i+1,\quad\mathcal{Z}_i=\vartheta_{i+1}-\vartheta_1,i=1,2,3,\quad\mathcal{Z}_4=\vartheta_5-4\vartheta_1\\
\mathcal{L}_1=\frac{\partial}{\partial a_1} \frac{\partial}{\partial a_2}\frac{\partial}{\partial a_3}\frac{\partial}{\partial a_4}\frac{\partial}{\partial a_5}^4-(\frac{\partial}{\partial a_0})^8
\end{gathered}
\end{equation}
where $\mathcal{Z}_0$ represents the invariance of $P$ under overall rescaling, $\mathcal{Z}_{i}$'s relate to the torus symmetry, and $\mathcal{L}_i$'s relate to the symmetries among monomials consisting of $P$.  And all  GKZ operators above annihilate the period matrix and determine the mirror maps and superpotential.

After blowing up $X$ along $S$, the blow-up manifold $X^\prime$ is obtained as the complete intersection
\begin{equation}
\begin{aligned}
X^\prime:P &= a_1x_1^{8}+x_2^{8}+x_3^8+x_4^8+x_5^2+a_0(x_1x_2x_3x_4x_5)=0\\
Q & =y_1(a_9 x_1^{8}+ a_{10} x_5^{2})-y_2(a_6(x_1x_2x_3x_4x_5)^2+a_7x_1x_2x_3x_4x_5^3+a_8 x_5^4)
\end{aligned}
\end{equation}
where $a_i$'s  are free complex-valued coefficients. By observation on above defining equations,  GKZ system of $X^\prime$ is obtained as complement to GKZ system of $X$.
\begin{equation}\label{eq:8.5}
\begin{gathered}
\mathcal{Z}_0^\prime=\sum_{i=0}^5 \vartheta_i+1,\quad\mathcal{Z}_1^\prime=\sum_{i=6}^{10} \vartheta_i,\quad\mathcal{Z}_2^\prime=-\vartheta_1+\vartheta_2+\vartheta_7+2\vartheta_8-\vartheta_10,\quad\mathcal{Z}_3^\prime=-\vartheta_1+\vartheta_3-\vartheta_{10},\\
\mathcal{Z}_4^\prime=-\vartheta_1+\vartheta_4-\vartheta_{10},\quad\mathcal{Z}_5^\prime=-4\vartheta_{1}+\vartheta_5-4\vartheta_{10},\quad\mathcal{Z}_6^\prime=-\vartheta_6-\vartheta_7-\vartheta_8+\vartheta_9+\vartheta_{10}\\
\mathcal{L}_1^\prime=\prod^4_{i=2}\frac{\partial}{\partial a_i}(\frac{\partial}{\partial a_7})^4 \frac{\partial}{\partial a_{10}}-(\frac{\partial}{\partial a_0})^3(\frac{\partial}{\partial a_6})^4 \frac{\partial}{\partial a_{9}},\quad \mathcal{L}_2^\prime=
\frac{\partial}{\partial a_6}\frac{\partial}{\partial a_8}-(\frac{\partial}{\partial a_7})^2 \\
\mathcal{L}_3^\prime=\frac{\partial}{\partial a_5} \frac{\partial}{\partial a_7}-\frac{\partial}{\partial a_0} \frac{\partial}{\partial a_8},\quad \mathcal{L}_4^\prime=\frac{\partial}{\partial a_1} \frac{\partial}{\partial a_{9}}-\frac{\partial}{\partial a_0} \frac{\partial}{\partial a_{10}}
\end{gathered}
\end{equation}
where $\mathcal{Z}_0^\prime,\mathcal{Z}_1^\prime$ are associated with the overall rescaling with respect to $P=0,Q=0$, $\mathcal{Z}_i^\prime,i=2,...,5$ are related to the torus symmetry, and  $\mathcal{L}_3^\prime,\mathcal{L}_4^\prime$ incorporate the parameter $a_6,...,a_{10}$ that are associated with the moduli of the curve $S$.The new manifold $X^\prime$ is describe by the following charge vectors
 \begin{table}[H]
\centering
 \begin{tabular}{c|cccccccccccc}
 $\quad$&$0$& $1$&$2$&$3$&$4$&$5$&$6$&$7$&$8$&$9$&$10$\\
 \hline
$l^\prime_1$&$-3$&$0$&$1$&$1$&$1$&$0$&$-4$&$4$&$0$&$-1$&$1$\\
$l^\prime_2$&$0$&$0$&$0$&$0$&$0$&$0$&$1$&$-2$&$1$&$0$&$0$\\
$l^\prime_3$&$-1$&$0$&$0$&$0$&$0$&$1$&$0$&$1$&$-1$&$0$&$0$\\
$l_4^\prime$&$-1$&$1$&$0$&$0$&$0$&$0$&$0$&$0$&$0$&$-1$&$1$\\
 \end{tabular}
\end{table}

  The  local coordinates $z^i$'s  on the complex structure moduli space of $X^\prime$ by \ref{eq:2.22},
\begin{equation}
z_1=\frac{  a_3 a_4 a_5 a_7^4 a_{10}}{a_0^4  a_{9}},\quad z_2=\frac{a_6a_8}{ a_7^2},\quad z_3=\frac{a_5 a_{7}}{a_0 a_8},\quad z_4=\frac{a_1 a_{9}}{a_0 a_{10}}
\end{equation}
Next, we convert the $\mathcal{L}_i^\prime$ in \ref{eq:5.10}operators to Picard-Fuchs operators $\mathcal{D}_i$, from differential equations about $a_j(j=0,\dots,10)$ to those about  $z_j(j=1,\dots,4)$ ,
\begin{equation}\label{eq:8.7}
\begin{aligned}
\mathcal{D}_1&=\theta_1^3(4\theta_1-2\theta_2+\theta_3)^4 (\theta_1-\theta_4)-z_1
 (-4\theta_1+\theta_2)^4(-\theta_1+\theta_4)\prod^4_{i=1}(-3\theta_1-\theta_3-\theta_4-i),\\
 \mathcal{D}_2&=(-2\theta_1+\theta_2)(\theta_2-\theta_3)-z_2(2\theta_2-2\theta_2+\theta_3)^2\\
 \mathcal{D}_3&=(\theta_1+\theta_3-\theta_4)(2\theta_1-2\theta_2+\theta_3)-z_3 (\theta_2-\theta_3) (-4\theta_1-\theta_3-1),\\
 \mathcal{D}_4&=\theta_4 (-\theta_1+\theta_4)-z_4(\theta_1+\theta_3-\theta_4)(\theta_1-\theta_4)\\
 \cdots
 \end{aligned}
 \end{equation}
where $\theta_i=z_i \frac{\partial}{\partial z_i}$'s are the logarithmic derivatives.

\subsubsection{Brane Superpotential and Disk Instantons}
Along the line of \ref{sec:2.2}, we solve the differential equations \ref{eq:8.7}  at $z_i\rightarrow 0$ and identify the mirror maps and superpotential. The fundamental period of $X$  is
\begin{equation*}
\omega_0=1 + 1680 z_1 z_2^4 z_3^4 z_4 + 32432400 z_1^2 z_2^8 z_3^8 z_4^2 +\mathcal{O}(z^{20})
\end{equation*}.
 The single logarithmic solutions are
\begin{equation*}
\begin{aligned}
\omega_{1,1}=& \omega_0 \log(z_1)-4 z_2 - 6 z_2^2 - \frac{40}{3} z2^3 + 4 z_2 z_3 + 4 z_2^2 z_3  + z_4 + 24 z_1 z_4 - 96 z_1 z_2 z_4 
+ 48 z_1 z_2^2 z_4  \\
&-\frac{1}{2}z_4^2 + \frac{1}{3}z_4^3  +\mathcal{O}(z^3)\\
\omega_{1,2}=&\omega_0\log(z_2)+2 z_2 + 3 z_2^2 + \frac{20}{3} z_2^3+ + z_3 - 2 z_2 z_3 - 
 2 z_2^2 z_3  -\frac{1}{2} z_3^2 + z_2 z_3^2  + \frac{1}{3}z_3^3   - 6 z_1 z_4 \\
 &+ 24 z_1 z_2 z_4 +\mathcal{O}(z^3)\\
\omega_{1,3}=&\omega_0\log(z_3)-z_2 - \frac{3}{2} z_2^2 - \frac{10}{3} z_2^3 - \frac{35}{4} z_2^4 - z_3 + z_2 z_3 + z_2^2 z_3 + 
 2 z_2^3 z_3 + \frac{1}{2}z_3^2 - z_2 z_3^2 \\
 &- \frac{1}{2}z_2^2 z_3^2 - \frac{1}{3}z_3^3 + 
 z_2 z_3^3 + \frac{1}{4}z3^4+\mathcal{O}(z^{4})\\
\omega_{1,4}=&\omega_0\log(z_4)-z_4 + \frac{1}{2}z_4^2 - \frac{1}{3}z_4^3 + \frac{1}{4}z_4^4+\mathcal{O}(z^{4})
\end{aligned}
\end{equation*}
by which the open-closed mirror maps are inverse series of flat coordinates     
\begin{equation*}
\begin{aligned}
z_1=&q_1 + 4 q_1 q_2 + 6 q_1 q_2^2 + 4 q_1 q_2^3 + q_1 q_2^4 - 8 q_1 q_2 q_3 - 
 24 q_1 q_2^2 q_3 - q_1 q_4 - 24 q_1^2 q_4 - 4 q_1 q_2 q_4 \\
 &- 72 q_1^2 q_2 q_4 - 
 6 q_1 q_2^2 q_4 + 8 q_1 q_2 q_3 q_4 + 24 q_1^2 q_4^2 +\mathcal{O}(q^4)\\
z_2=&q_2 - 2 q_2^2 + 3 q_2^3 - 4 q_2^4 - q_2 q_3 + 5 q2^2 q3 - 13 q_2^3 q_3 - 
 3 q_2^2 q_3^2 + 6 q_1 q_2 q_4 - 24 q_1 q_2^2 q_4 +\mathcal{O}(q^4)\\
z_3=&q_3 + q_2 q_3 + q_3^2 + q_2^2 q_3^2 + q_3^3 + q_3^4 + 6 q_1 q_2 q_3 q_4+\mathcal{O}(q^{4})\\
z_4=&q_4 + q_4^2 + q_4^3 + q_4^4+\mathcal{O}(q^{4})
\end{aligned}
\end{equation*}
Then we construct a linear combination of double logarithmic  solutions and  insert the inverse mirror maps,
\begin{equation}
\begin{aligned}
\mathcal{W}_{\mathrm{brane}}&=2 t_1^2+ 4 t_1t_4+2 t_4^2+\sum_{N}N_{d_1,d_2,d_3,d_4}\mathrm{Li}_2(q_1^{d_1}q_2^{d_2}q_3^{d_3}q_4^{d_4})
\end{aligned}
\end{equation}Invariants of the form $N_\mathrm{m,n,n,m}$ are summarized in Table\ref{table:26}, where the rows and columns are labelled by m and n, respectively. 

When $ a_7^2=a_6 a_8$, the two individual branes coincide. We obtain a new set of charge vectors,
 \begin{table}[H]
\centering
 \begin{tabular}{c|ccccccccccc}
 $\quad$&$0$& $1$&$2$&$3$&$4$&$5$&$6$&$8$&$9$&$10$\\
 \hline
$l^c_1$&$-3$&$0$&$1$&$1$&$1$&$0$&$-2$&$2$&$-1$&$1$\\
$l^c_2$&$-2$&$0$&$0$&$0$&$0$&$2$&$1$&$-1$&$0$&$0$\\
$l^c_3$&$-1$&$1$&$0$&$0$&$0$&$0$&$0$&$0$&$1$&$-1$\\
 \end{tabular}
\end{table}
by which the new complex structure moduli space coordinates are
\begin{equation}
z_1^c=\frac{a_2a_3a_4a_8^2 a_{10}}{a_0^3 a_6^2 a_{9}},\quad z_2^c=\frac{a_5^2a_6}{a_0^2 a_8},\quad z_3^c=\frac{a_1a_9}{a_0 a_{10}}
\end{equation}
Similar to the separate case, the superpotential is constructed as  linear combination of double logarithmic solutions and Ooguri-Vafa invariants are exacted \ref{table:27}.

\section{Summary and Conclusions}
In this work, we calculate the superpotentials in d=4 N=1 supersymmetric field theories arising from IIA D6-branes wrapping special Lagrangian three cycles of Calabi-Yau threefold. The special Lagrangian three-cycles with non-trivial topology are mirror to obstructed rational curves, which correspond to the brane excitation about the supersymmetric minimum.
We consider a five brane wrapping a rational curve that coincides with a toric curve $S$ at certain locus of the deformation space $\mathcal{M}(S)$. $S$ is described by the intersection of two divisors $D_1 \cap D_2$ and its  unobstructed deformation space  match with the obstructed deformation space of the rational curve wrapped by the five brane.  After blowing up, the toric curve $S$ is replaced by a exceptional divisor $E$ without introducing new degree of freedom. All the complex structure moduli and brane moduli are embedded into the complex moduli space of the blow-up new manifold, given as the complete intersection in the projective bundle $\mathcal{W}=\mathbb{P}(\mathcal{O}(D_1)\oplus \mathcal{O}(D_2))$.

From  observation on the defining equation of $X^\prime$, we obtain  the Picard-Fuchs equations that annihilate the period matrix defined by the natural pairing between the elements of relative homology $H_*(X,S)$ and cohomology $H^*(X,S)$. Via GKZ system of $X^\prime$, the system of Picard-Fuchs equations are solved at $z_i \rightarrow 0$. The single logarithmic solutions are interpreted as  mirror maps and  specific linear combinations of double logarithmic solutions are the B-brane superpotentials. Using multi-cover formula and inverse mirror maps, the Ooguri-Vafa invariants are extracted from A-model side and interpreted as counting disk instantons,i.e. holomorphic disks with boundary in a nontrivial homology class on a special Lagrangian submanifold. It would be interesting to directly extract these  invariants on the  A-model side directly by adequate localization techniques.

\newpage
\appendix
\section{ Blow-up Geometry of $X^{(112|112)}_{[4,4]}$ and $X^{(123|123)}_{[6,6]}$ }\label{App:A}
\subsection{$X^{(112|112)}_{[4,4]}$}
Polyhedron vertices and charge vectors associated to  A-model manifold of $X^*$ and A-branes on it:
 \begin{table}[H]
\centering
 \begin{tabular}{c|ccccc|c|c|cc}
 $\quad$&\multicolumn{5}{|c|}{$\Delta^*$}&$l$&$\quad$&$\hat{l}^1$&$\hat{l}^2$\\
 \hline
 $v^*_{0,1}$&$0$&$0$&$0$&$0$&$0$&$-4$&$x_1x_2x_3$&$-1$&$-1$\\
 $v^*_{0,2}$&$0$&$0$&$0$&$0$&$0$&$-4$&$x_4x_5x_6$&$0$&$0$\\
 $v^*_1$&$-1$&$-2$&$-1$&$-1$&$-2$&$1$&$x_1^4$&$0$&$0$\\
 $v^*_2$&$1$&$0$&$0$&$0$&$0$&$1$&$x_2^4$&$1$&$0$\\
 $v^*_3$&$0$&$1$&$0$&$0$&$0$&$2$&$x_3^4$&$0$&$1$\\
 $v^*_4$&$0$&$0$&$1$&$0$&$0$&$1$&$x_4^4$&$0$&$0$\\
 $v^*_5$&$0$&$0$&$0$&$1$&$0$&$1$&$x_5^4$&$0$&$0$\\
 $v^*_6$&$0$&$0$&$0$&$0$&$1$&$2$&$x_6^2$&$0$&$0$\\
  \end{tabular}
 \caption{toric data of A-model manifold}
\end{table} 
The GKZ system of $X$ as follows
\begin{equation}
\begin{gathered}
\mathcal{Z}_0=\sum_{i=0}^6 \vartheta_i+1,\quad\mathcal{Z}_1=-\vartheta_1+\vartheta_2,\quad\mathcal{Z}_2=-2\vartheta_1+\vartheta_3,\\
\mathcal{Z}_3=-\vartheta_1+\vartheta_4,\quad\mathcal{Z}_4=-2\vartheta_1+\vartheta_5,\quad\mathcal{Z}_5=-2\vartheta_1+\vartheta_6\\
\mathcal{L}_1=\frac{\partial}{\partial a_1}\frac{\partial}{\partial a_2}(\frac{\partial}{\partial a_3})^2 \frac{\partial}{\partial a_4}\frac{\partial}{\partial a_5}(\frac{\partial}{\partial a_6})^2-(\frac{\partial}{\partial a_{0,1}})^3(\frac{\partial}{\partial a_{0,2}})^3
\end{gathered}
\end{equation}

After blowing up $X$ along the  curve specified by $\hat{l}^1,\hat{l}^2$, the  GKZ system of blow-up manifold $X^\prime$:
\begin{equation}
\begin{gathered}
\mathcal{Z}_0^\prime=\sum_{i=0}^6 \vartheta_i+1,\quad\mathcal{Z}_1^\prime=\sum_{i=7}^{10} \vartheta_i,\quad\mathcal{Z}_2^\prime=-\vartheta_1+\vartheta_2-\vartheta_8+\vartheta_{10},\quad\mathcal{Z}_3^\prime=-2\vartheta_1+\vartheta_3-2\vartheta_8,\\
\mathcal{Z}_4^\prime=-\vartheta_1+\vartheta_4-\vartheta_{8},\quad\mathcal{Z}_5^\prime=-\vartheta_{1}+\vartheta_5-\vartheta_8,\quad\mathcal{Z}_6^\prime=-2\vartheta_1+\vartheta_6-2\vartheta_8,\quad\mathcal{Z}_7^\prime=-\vartheta_7-\vartheta_8+\vartheta_9+\vartheta_{10}\\
\mathcal{L}_1^\prime=\frac{\partial}{\partial a_1}\frac{\partial}{\partial a_2}(\frac{\partial}{\partial a_3})^2 \frac{\partial}{\partial a_4}\frac{\partial}{\partial a_5}(\frac{\partial}{\partial a_6})^2-(\frac{\partial}{\partial a_{0,1}})^3(\frac{\partial}{\partial a_{0,2}})^3, \\
\mathcal{L}_2^\prime=\frac{\partial}{\partial a_1} \frac{\partial}{\partial a_7}-\frac{\partial}{\partial a_{0,1}} \frac{\partial}{\partial a_8},\quad \mathcal{L}_3^\prime=\frac{\partial}{\partial a_2} \frac{\partial}{\partial a_9}-\frac{\partial}{\partial a_{0,1}} \frac{\partial}{\partial a_{10}}
\end{gathered}
\end{equation}

Toric data for the enhanced polyhedron associated to the $X^\prime$  :
\begin{table}[H]
\centering
 \begin{tabular}{c|cccccccc|ccc|l}
 $\quad$& \multicolumn{8}{|c|}{$\Delta^\prime$}&$l^\prime_1$&$l^\prime_2$&$l^\prime_3$\\
 \hline
$v^\prime_{0,1}$&$1$&$0$&$0$&$0$&$0$&$0$&$0$&$0$&$-1$&$-1$&$-1$&$w^\prime_{0,1}=x_1x_2x_3$\\
$v^\prime_{0,2}$&$1$&$0$&$0$&$0$&$0$&$0$&$0$&$0$&$-4$&$0$&$0$&$w^\prime_{0,2}=x_4x_5x_6$\\
$v^\prime_1$&$1$&$0$&$-1$&$-2$&$-1$&$-1$&$-2$&$0$&$-1$&$1$&$0$&$w^\prime_1=x_1^4$\\
$v^\prime_2$&$1$&$0$&$1$&$0$&$0$&$0$&$0$&$0$&$0$&$0$&$1$&$w^\prime_2=x_2^4$\\
$v^\prime_3$&$1$&$0$&$0$&$1$&$0$&$0$&$0$&$0$&$2$&$0$&$0$&$w^\prime_3=x_3^4$\\
$v^\prime_4$&$1$&$0$&$0$&$0$&$1$&$0$&$0$&$0$&$1$&$0$&$0$&$w^\prime_4=x_4^2$\\
$v^\prime_5$&$1$&$0$&$0$&$0$&$0$&$1$&$0$&$0$&$1$&$0$&$0$&$w^\prime_5=x_5^4$\\
$v^\prime_6$&$1$&$0$&$0$&$0$&$0$&$0$&$1$&$0$&$2$&$0$&$0$&$w^\prime_6=x_6^4$\\
$v^\prime_7$&$0$&$1$&$0$&$0$&$0$&$0$&$0$&$-1$&$-2$&$1$&$0$&$w^\prime_7=y_1 w^\prime_0$\\
$v^\prime_8$&$0$&$1$&$-1$&$-2$&$-1$&$-1$&$-2$&$-1$&$2$&$-1$&$0$&$w^\prime_8=y_1 w^\prime_2$\\
$v^\prime_9$&$0$&$1$&$0$&$0$&$0$&$0$&$0$&$1$&$-1$&$0$&$1$&$w^\prime_9=y_2 w^\prime_0$\\
$v^\prime_{10}$&$0$&$1$&$1$&$0$&$0$&$0$&$0$&$1$&$1$&$0$&$-1$&$w^\prime_{10}=y_2 w^\prime_3$\\
 \end{tabular}
\end{table}

Picard-Fuchs operators from above table:
\begin{equation}
\begin{aligned}
\mathcal{D}_1&=(2\theta_1)^2(\theta_1)^2(2\theta_1)^2(2\theta_1-\theta_2)^2(\theta_1-\theta_3)\\&-z_1(-\theta_1+\theta_2)(-2\theta_1+\theta_2)^2(-\theta_1+\theta_3)(-\theta_1-\theta_2-\theta_3-1)\prod_{i=1}^4(-4\theta_1-i)\\
\mathcal{D}_2&=(-\theta_1+\theta_2)(-2\theta_1+\theta_2)-z_2(-\theta_1-\theta_2-\theta_3-1)(2\theta_1-\theta_2)\\
\mathcal{D}_3&=\theta_3(-\theta_1+\theta_3)-z_3(-\theta_1-\theta_2-\theta_3-1)(\theta_1-\theta_3)\\
\cdots
\end{aligned}
\end{equation}

Solving above equations by GKZ-system, the unique power series solution,as well as the fundamental period of $X$, is
\begin{equation}
\begin{aligned}
\omega_0=&1 + 144 z_1 z_2^2 z_3 + 176400 z_1^2 z_2^4 z_3^2 + 
 341510400 z_1^3 z_2^6 z_3^3 + 811620810000 z_1^4 z_2^8 z_3^4\\
 &+\mathcal{O}(z^{16})
 \end{aligned}
\end{equation}
and four single logarithmic solutions
\begin{equation}
\begin{aligned}
\omega_{1,1}=&\omega_0\log(z_1)+2 z_2 - z_2^2 - 36 z_1 z_2^2 + \frac{2}{3} z_2^3 - \frac{1}{2}z_2^4 + 3150 z_1^2 z_2^4 + \frac{2}{5}
 z_2^5 + z_3 \\
 &- 72 z_1 z_2 z_3 + 492 z_1 z_2^2 z_3 + 
 720 z_1 z_2^3 z_3 - \frac{1}{2}z_3^2 + 360 z_1 z_2^2 z_3^2 +\frac{1}{3} z_3^3 - \frac{1}{4}z_3^4 \\
 &+\frac{1}{5} z_3^5+\mathcal{O}(z^5)\\
 \omega_{1,2}=&\omega_0\log(z_2)-z_2 + \frac{1}{2}z_2^2 -\frac{1}{3} z_2^3 +\frac{1}{4} z_2^4 -\frac{1}{5} z_2^5+ 36 z_1 z_2 z_3 + 156 z_1 z_2^2 z_3 \\
 &- 360 z_1 z_2^3 z_3+\mathcal{O}(z^5) \\
 \omega_{1,3}=&\omega_0\log(z_3)+36 z_1 z_2^2 - z_3 + 156 z_1 z_2^2 z_3 + \frac{1}{2}z_3^2 - 
 360 z_1 z_2^2 z_3^2 - \frac{1}{3}z3^3\\
 & + 360 z_1 z_2^2 z_3^3 + \frac{1}{4}z_3^4 - \frac{1}{5}z3^5+\mathcal{O}(z^5)
 \end{aligned}
\end{equation}

Upon the fundamental period and single logarithmic solutions, the inverse mirror maps are
\begin{equation}
\begin{aligned}
z_1 =&q_1 - 2 q_1 q_2 + q_1 q_2^2 + 36 q_1^2 q_2^2 - 72 q_1^2 q_2^3 - q_1 q_3 + 
 2 q_1 q_2 q_3 + 72 q_1^2 q_2 q_3 - q_1 q_2^2 q_3\\
 & - 672 q_1^2 q_2^2 q_3 - 
 72 q_1^2 q_2 q_3^2+\mathcal{O}(q^5)\\
 z_2=&q_2 + q_2^2 + q_2^3 + q_2^4 + q_2^5 - 36 q_1 q_2^2 q_3 - 192 q_1 q_2^3 q_3+\mathcal{O}(q^5)\\
 z_3=&q_3 - 36 q_1 q_2^2 q_3 + q_3^2 - 192 q_1 q_2^2 q_3^2 + q_3^3+\mathcal{O}(q^5)
 \end{aligned}
\end{equation}
The brane superpotential as linear combination of double logarithmic solutions is
\begin{equation}
\mathcal{W}_{\mathrm{brane}}=4t_1t_2+6t_2^2+4t_2t_3+\sum_{N}N_{d_1,d_2,d_3}\mathrm{Li}_2(q_1^{d_1}q_2^{d_2}q_3^{d_3})
\end{equation} and the disk instantons are presented in Table \ref{table:24}

\subsection{$X^{(123|123)}_{[6,6]}$}
Polyhedron vertices and charge vectors A-model manifold $X^*$ and A-branes on it:
 \begin{table}[H]
\centering
 \begin{tabular}{c|ccccc|c|c|cc}
 $\quad$&\multicolumn{5}{|c|}{$\Delta^*$}&$l$&$\quad$&$\hat{l}^1$&$\hat{l}^2$\\
 \hline
 $v^*_{0,1}$&$0$&$0$&$0$&$0$&$0$&$-6$&$x_1x_2x_3$&$0$&$0$\\
 $v^*_{0,2}$&$0$&$0$&$0$&$0$&$0$&$-6$&$x_4x_5x_6$&$0$&$0$\\
 $v^*_1$&$-2$&$-3$&$-1$&$-2$&$-3$&$1$&$x_1^6$&$-1$&$-1$\\
 $v^*_2$&$1$&$0$&$0$&$0$&$0$&$2$&$x_2^3$&$1$&$0$\\
 $v^*_3$&$0$&$1$&$0$&$0$&$0$&$3$&$x_3^2$&$0$&$1$\\
 $v^*_4$&$0$&$0$&$1$&$0$&$0$&$1$&$x_4^6$&$0$&$0$\\
 $v^*_5$&$0$&$0$&$0$&$1$&$0$&$2$&$x_5^3$&$0$&$0$\\
 $v^*_6$&$0$&$0$&$0$&$0$&$1$&$3$&$x_6^2$&$0$&$0$\\

  \end{tabular}
 \caption{Toric Data of A-model side}
\end{table}
GKZ system of $X$ as follows
\begin{equation}
\begin{gathered}
\mathcal{Z}_0=\sum_{i=0}^6 \vartheta_i+1,\quad\mathcal{Z}_1=-2\vartheta_1+\vartheta_2,\quad\mathcal{Z}_2=-3\vartheta_1+\vartheta_3,\\
\mathcal{Z}_3=-\vartheta_1+\vartheta_4,\quad\mathcal{Z}_4=-2\vartheta_1+\vartheta_5,\quad\mathcal{Z}_5=-3\vartheta_5+\vartheta_6\\
\mathcal{L}_1=(\frac{\partial}{\partial a_1})(\frac{\partial}{\partial a_2})^2(\frac{\partial}{\partial a_3})^3 (\frac{\partial}{\partial a_4})(\frac{\partial}{\partial a_5})^2(\frac{\partial}{\partial a_6})^3-(\frac{\partial}{\partial a_{0,1}})^6(\frac{\partial}{\partial a_{0,2}})^6
\end{gathered}
\end{equation}

After blowing up along the  curve specified by $\hat{l}^1,\hat{l}^2$, the  GKZ system of blow-up manifold $X^\prime$:
\begin{equation}
\begin{gathered}
\mathcal{Z}_0^\prime=\sum_{i=0}^6 \vartheta_i+1,\quad\mathcal{Z}_1^\prime=\sum_{i=7}^{10} \vartheta_i,\quad\mathcal{Z}_2^\prime=-2\vartheta_1+\vartheta_2-2\vartheta_7+\vartheta_8-2\vartheta_9,\\
\mathcal{Z}_3^\prime=-3\vartheta_1+\vartheta_3-3\vartheta_7-3\vartheta_9+\vartheta_{10},
\quad \mathcal{Z}_4^\prime=-\vartheta_1+\vartheta_4-\vartheta_{7}-\vartheta_9,\quad\mathcal{Z}_5^\prime=-2\vartheta_1+\vartheta_5-2\vartheta_7-2\vartheta_9,\\
\mathcal{Z}_6^\prime=-3\vartheta_1+\vartheta_6-3\vartheta_7-3\vartheta_9,\quad\mathcal{Z}_7^\prime=-\vartheta_7-\vartheta_8+\vartheta_9+\vartheta_{10}\\
\mathcal{L}_1^\prime=(\frac{\partial}{\partial a_1})(\frac{\partial}{\partial a_2})^2(\frac{\partial}{\partial a_3})^3( \frac{\partial}{\partial a_4})(\frac{\partial}{\partial a_5})^2(\frac{\partial}{\partial a_6})^3-(\frac{\partial}{\partial a_{0,1}})^6(\frac{\partial}{\partial a_{0,2}})^6, \\
\mathcal{L}_2^\prime=\frac{\partial}{\partial a_1} \frac{\partial}{\partial a_8}-\frac{\partial}{\partial a_{3}} \frac{\partial}{\partial a_7},\quad \mathcal{L}_3^\prime=\frac{\partial}{\partial a_1} \frac{\partial}{\partial a_{10}}-\frac{\partial}{\partial a_{2}} \frac{\partial}{\partial a_{9}}
\end{gathered}
\end{equation}

Toric data for the enhanced polyhedron associated to the $X^\prime$ 's mirror manifold :
\begin{table}[H]
\centering
 \begin{tabular}{c|cccccccc|ccc|l}
 $\quad$& \multicolumn{8}{|c|}{$\Delta^\prime$}&$l^\prime_1$&$l^\prime_2$&$l^\prime_3$\\
 \hline
$v^\prime_{0,1}$&$1$&$0$&$0$&$0$&$0$&$0$&$0$&$0$&$-6$&$0$&$0$&$w^\prime_{0,1}=x_1x_2x_3$\\
$v^\prime_{0,2}$&$1$&$0$&$0$&$0$&$0$&$0$&$0$&$0$&$-6$&$0$&$0$&$w^\prime_{0,2}=x_4x_5x_6$\\
$v^\prime_1$&$1$&$0$&$-2$&$-3$&$-1$&$-2$&$-3$&$0$&$-2$&$1$&$1$&$w^\prime_1=x_1^6$\\
$v^\prime_2$&$1$&$0$&$1$&$0$&$0$&$0$&$0$&$0$&$3$&$0$&$-1$&$w^\prime_2=x_2^3$\\
$v^\prime_3$&$1$&$0$&$0$&$1$&$0$&$0$&$0$&$0$&$5$&$-1$&$0$&$w^\prime_3=x_3^2$\\
$v^\prime_4$&$1$&$0$&$0$&$0$&$1$&$0$&$0$&$0$&$1$&$0$&$0$&$w^\prime_4=x_4^6$\\
$v^\prime_5$&$1$&$0$&$0$&$0$&$0$&$1$&$0$&$0$&$2$&$0$&$0$&$w^\prime_5=x_5^3$\\
$v^\prime_6$&$1$&$0$&$0$&$0$&$0$&$0$&$1$&$0$&$3$&$0$&$0$&$w^\prime_6=x_6^2$\\
$v^\prime_7$&$0$&$1$&$-2$&$-3$&$-1$&$-2$&$-3$&$-1$&$1$&$0$&$-1$&$w^\prime_7=y_1 w^\prime_0$\\
$v^\prime_8$&$0$&$1$&$1$&$0$&$0$&$0$&$0$&$-1$&$-1$&$0$&$1$&$w^\prime_8=y_1 w^\prime_2$\\
$v^\prime_9$&$0$&$1$&$-2$&$-3$&$-1$&$-2$&$-3$&$1$&$2$&$-1$&$0$&$w^\prime_9=y_2 w^\prime_0$\\
$v^\prime_{10}$&$0$&$1$&$0$&$1$&$0$&$0$&$0$&$1$&$-2$&$1$&$0$&$w^\prime_{10}=y_2 w^\prime_3$\\
 \end{tabular}
\end{table}

Picard-Fuchs equations from above table:
\begin{equation}
\begin{aligned}
\mathcal{D}_1&=(3\theta_1-\theta_3)^3(5\theta_1-\theta_2)^5\theta_1(2\theta_1)^2(3\theta_1)^3(2\theta_1-\theta_2)^2(\theta_1-\theta_3)\\&-z_1(-2\theta_1+\theta_2+\theta_3)^2(-2\theta_1+\theta_2)^2(-\theta_1+\theta_3)\prod_{i=1}^6\prod_{j=1}^6(-6\theta_1-i)(-6\theta_1-j)\\
\mathcal{D}_2&=(-2\theta_1+\theta_2+\theta_3)(-2\theta_1+\theta_2)-z_2(5\theta_1-\theta_2)(2\theta_1-\theta_2)\\
\mathcal{D}_3&=(-2\theta_1+\theta_2+\theta_3)(-\theta_1+\theta_3)-z_3(3\theta_1-\theta_3)(\theta_1-\theta_3)\\
\cdots
\end{aligned}
\end{equation}

Soling above equations by GKZ-system, the unique power series solution,as well as the fundamental period of $X$, is
\begin{equation}
\begin{aligned}
\omega_0=&1 + 3600 z_1 z_2^2 z_3 + 192099600 z_1^2 z_2^4 z_3^2 + 
 16679709446400 z_1^3 z_2^6 z_3^3 \\
 &+ 1791735431214128400 z_1^4 z_2^8 z_3^4+\mathcal{O}(z^{16})
\end{aligned}
\end{equation}
and the single logarithmic solutions are
\begin{equation}
\begin{aligned}
\omega_{1,1}=&\omega_0\log(z_1)-1200 z_1 z_2^2 - 1800 z_1 z_2 z_3 + 29640 z_1 z_2^2 z_3 + 10800 z_1 z_2^3 z_3\\
& + 
 3600 z_1 z_2^2 z_3^2+\mathcal{O}(z^5)\\
 \omega_{1,2}=&\omega_0\log(z_2)+900 z_1 z_2 z_3 + 3000 z_1 z_2^2 z_3 - 5400 z_1 z_2^3 z_3+\mathcal{O}(z^5)\\
 \omega_{1,3}=&\omega_0\log(z_3)+1200 z_1 z_2^2 + 1800 z_1 z_2^2 z_3 - 3600 z_1 z_2^2 z_3^2+\mathcal{O}(z^5)
\end{aligned}
\end{equation}

By the fundamental period and single logarithmic solutions,the inverse mirror maps are
\begin{equation}
\begin{aligned}
z_1=&q_1 + 1200 q_1^2 q_2^2 - 4243320 q_1^3 q_2^4 + 1800 q_1^2 q_2 q_3 - 
 29640 q_1^2 q_2^2 q_3+\mathcal{O}(q^5)\\
 z_2=&q_2 - 900 q_1 q_2^2 q_3 - 3000 q_1 q_2^3 q_3+\mathcal{O}(q^5)\\
 z_3=&q_3 - 1200 q_1 q_2^2 q_3 + 5683320 q_1^2 q_2^4 q_3 - 1800 q_1 q_2^2 q_3^2+\mathcal{O}(q^5)
\end{aligned}
\end{equation}
The brane superpotential is
\begin{equation}
\mathcal{W}_{\mathrm{brane}}=4t_1t_2+6t_2^2+4t_2t_3+\sum_{N}N_{d_1,d_2,d_3}\mathrm{Li}_2(q_1^{d_1}q_2^{d_2}q_3^{d_3})
\end{equation} and the disk instantons are presented in Table \ref{table:24}

\section{Ooguri-Vafa Invariants for Two Closed and Two Open Moduli}
\subsection{Ooguri-Vafa Invariants for $X^{(1,1,3,3,3)}_9$}
\begin{table}[H]\def\r{\color{red}}\def\b{\color{blue}}
\centering
 \begin{tabular}{|c|cccccc|}
 \hline
 \multicolumn{7}{|c|}{k=0: Ooguri-Vafa Invariants $N_{k,m,k,m+n}$ }\\
 \hline
 $N_{0,m,0,m+n}$&$n=1$&$2$&$3$&$4$&$5$&$6$\\
 \hline
 $m=0$&\b$3$&\b$0$&\b$0$&\b$0$&\b$0$&$0$\\
 $1$&\b$-3$&\b$-3$&\b$-3$&\b$-3$&$-3$&$-3$\\
 $2$&\b$15$&\b$21$&\b$27$&$36$&$45$&$57$\\
 $3$&\b$-120$&\b$-183$&$-279$&$-420$&$-618$&$*$\\
 $4$&\b$1179$&$1944$&$3210$&$5250$&$*$&$*$\\
 $5$&$-13572$&$-22983$&$-39771$&$*$&$*$&$*$\\
 \hline
 \multicolumn{7}{c}{$\quad$}\\
 \hline
 \multicolumn{7}{|c|}{k=1: Ooguri-Vafa Invariants $N_{k,m,k,m+n}$ }\\
 \hline
 $N_{1,m,1,m+n}$&$n=1$&$2$&$3$&$4$&$5$&$6$\\
 \hline
 $m=0$&\b$27$&\b$0$&\b$0$&$0$&$0$&$0$\\
 $1$&\b$90$&\b$90$&$90$&$90$&$90$&$90$\\
 $2$&\b$-684$&$-954$&$-1314$&$-1764$&$-2304$&$-2934$\\
 $3$&$7470$&$11736$&$18486$&$28620$&$43218$&$*$\\
 $4$&$-94644$&$-158022$&$-267768$&$-450270$&$*$&$*$\\
  $5$&$1302120$&$2254968$&$3998718$&$*$&$*$&$*$\\
 \hline
  \multicolumn{7}{c}{$\quad$}\\
 \hline
 \multicolumn{7}{|c|}{k=2: Ooguri-Vafa Invariants $N_{k,m,k,m+n}$ }\\
 \hline
 $N_{2,m,2,m+n}$&$n=1$&$2$&$3$&$4$&$5$&$6$\\
 \hline
 $m=0$&\b$81$&\b$108$&\b$0$&$0$&$0$&$0$\\
 $1$&\b$-1539$&\b$-1377$&$-1377$&$-1377$&$-1377$&$-1377$\\
 $2$&\b$13554$&$19728$&$29079$&$41112$&$55917$&$73404$\\
 $3$&$-204120$&$-333612$&$-553878$&$898587$&$-1412154$&$*$\\
 $4$&$3351969$&$5781384$&$10209969$&$17848764$&$*$&$*$\\
  $5$&$-56886543$&$-101222919$&$-185715828$&$*$&$*$&$*$\\
 \hline
  \multicolumn{7}{c}{$\quad$}\\
 \hline
 \multicolumn{7}{|c|}{k=3: Ooguri-Vafa Invariants $N_{k,m,k,m+n}$ }\\
 \hline
 $N_{3,m,3,m+n}$&$n=1$&$2$&$3$&$4$&$5$&$6$\\
 \hline
 $m=0$&$255$&$-984$&$729$&$0$&$0$&$0$\\
 $1$&$18150$&$16764$&$14796$&$14796$&$14796$&$14796$\\
 $2$&$-132492$&$-221262$&$-370224$&$-580212$&$-864306$&$-1168506$\\
 $3$&$3063876$&$5345640$&$9647424$&$16885548$&$28230822$&$*$\\
 $4$&$-68640885$&$-123720696$&$-231991668$&$-428397570$&$*$&$*$\\
  $5$&$1903950048$&$2812011138$&$5315645308$&$*$&$*$&$*$\\
 \hline
  \hline
 \end{tabular}
 \caption{Ooguri-Vafa Invariants $N_{k,m,k,m+n}$ for $X_9^{(1,1,1,3,3)}$ with brane I at large volume. $k$ and $m$ label the class $t_1,t_2$ of $X_9$ and $n$ labels the brane winding. Blue entries agree with Table 5 of \cite{Alim2009} and entries $*$ exceed in this table the order of our calculation}
 \label{table:13}
\end{table}

\begin{table}[H]\def\r{\color{red}}\def\b{\color{blue}}
\centering
 \begin{tabular}{|c|cccccc|}
 \hline
 \multicolumn{7}{|c|}{k=0: Ooguri-Vafa Invariants $N_{k,m,k,m+n}$ }\\
 \hline
 $N_{0,m,0,m+n}$&$n=1$&$2$&$3$&$4$&$5$&$6$\\
 \hline
 $m=0$&\b$54$&\b$0$&\b$0$&\b$0$&\b$0$&$0$\\
 $1$&\b$54$&\b$-18$&\b$0$&\b$0$&$0$&$0$\\
 $2$&\b$36$&\b$0$&\b$0$&$0$&$0$&$0$\\
 $3$&\b$54$&\b$0$&$0$&$0$&$0$&$*$\\
 $4$&\b$54$&$-18$&$0$&$0$&$*$&$*$\\
 $5$&$36$&$0$&$0$&$*$&$*$&$*$\\
 \hline
 \multicolumn{7}{c}{$\quad$}\\
 \hline
 \multicolumn{7}{|c|}{k=1: Ooguri-Vafa Invariants $N_{k,m,k,m+n}$ }\\
 \hline
 $N_{1,m,1,m+n}$&$n=1$&$2$&$3$&$4$&$5$&$6$\\
 \hline
 $m=0$&\b$0$&\b$0$&\b$0$&\b$0$&$0$&$0$\\
 $1$&\b$-108$&\b$36$&\b$0$&$0$&$0$&$0$\\
 $2$&\b$2772$&\b$-1026$&$0$&$18$&$0$&$0$\\
 $3$&\b$243756$&$-193050$&$100548$&$-33588$&$6696$&$*$\\
 $4$&$2947320$&$-2801070$&$2212272$&$-1340010$&$*$&$*$\\
 $5$&$23798376$&$-22631562$&$19965852$&$*$&$*$&$*$\\
 \hline
 \multicolumn{7}{c}{$\quad$}\\
 \hline
 \multicolumn{7}{|c|}{k=2: Ooguri-Vafa Invariants $N_{k,m,k,m+n}$ }\\
 \hline
 $N_{2,m,2,m+n}$&$n=1$&$2$&$3$&$4$&$5$&$6$\\
 \hline
 $m=0$&\b$0$&\b$0$&\b$0$&$0$&$0$&$0$\\
 $1$&\b$270$&\b$-90$&$0$&$0$&$0$&$0$\\
 $2$&\b$-11160$&$4104$&$0$&$-72$&$0$&$0$\\
 $3$&$174960$&$-74358$&$162$&$2916$&$0$&$*$\\
 $4$&$-7067304$&$6235344$&$-4430376$&$2634444$&$*$&$*$\\
 $5$&$92080530$&$-71321472$&$602803296$&$*$&$*$&$*$\\
 \hline
 \multicolumn{7}{c}{$\quad$}\\
 \hline
 \multicolumn{7}{|c|}{k=3: Ooguri-Vafa Invariants $N_{k,m,k,m+n}$ }\\
 \hline
 $N_{3,m,3,m+n}$&$n=1$&$2$&$3$&$4$&$5$&$6$\\
 \hline
 $m=0$&$0$&$0$&$0$&$0$&$0$&$0$\\
 $1$&$0$&$0$&$0$&$0$&$0$&$0$\\
 $2$&$0$&$0$&$0$&$0$&$0$&$0$\\
 $3$&$0$&$0$&$12$&$0$&$0$&$*$\\
 $4$&$0$&$0$&$0$&$0$&$*$&$*$\\
 $5$&$0$&$194724$&$-541296$&$*$&$*$&$*$\\
 \hline
 \end{tabular}
 \caption{Ooguri-Vafa Invariants $N_{k,m,k,m+n}$ for $X_9^{(1,1,1,3,3)}$ with brane II at large volume. $k$ and $m$ label the class $t_1,t_2$ of $X_9$ and $n$ labels the brane winding. Blue entries agree with Table 6 of \cite{Alim2009} and entries $*$ exceed in this table the order of our calculation}
 \label{table:15}
\end{table}

 \begin{table}[H]\def\r{\color{red}}\def\b{\color{blue}}
\centering
 \begin{tabular}{|c|cccccc|}
 \hline
 \multicolumn{7}{|c|}{k=1: Ooguri-Vafa Invariants $N_{k,m+n,k,n}$ }\\
 \hline
 $N_{1,m+n,1,n}$&$n=0$&$1$&$2$&$3$&$4$&$5$\\
 \hline
 $m=1$&\b$72$&\b$-1728$&\b$-80460$&\b$-1075140$&$-9482724$&$-65006280$\\
 $2$&\b$-36$&\b$17280$&\b$340092$&$3488940$&$25895214$&$156528234$\\
 $3$&\b$-1224$&\b$-64800$&$-977832$&$-8913456$&$-61353288$&$-350009424$\\
 $4$&\b$5508$&$176688$&$2301588$&$19441008$&$127166868$&$*$\\
 $5$&$-15336$&$-398304$&$-4742280$&$-38004912$&$*$&$*$\\
 $6$&$33948$&$787968$&$8853948$&$*$&$*$&$*$\\
 \hline
 \multicolumn{7}{c}{$\quad$}\\
 \hline
 \multicolumn{7}{|c|}{k=2: Ooguri-Vafa Invariants $N_{k,m+n,k,n}$ }\\
 \hline
 $N_{2,m+n,2,n}$&$n=0$&$1$&$2$&$3$&$4$&$5$\\
 \hline
 $m=1$&\b$-180$&\b$7020$&\b$-97686$&\b$2643372$&$-37415520$&$-6097896648$\\
 $2$&\b$108$&\b$-5832$&\b$-588276$&$3924954$&$1797212016$&$-467071781328$\\
 $3$&\b$-108$&\b$133488$&$368388$&$-441631440$&$-17699810580$&
 $-467071781328$\\
 $4$&\b$-10944$&$-411264$&$75854232$&$4282520544$&$107873699400$&$*$\\
 $5$&$61308$&$-6824952$&$-730782756$&$-24887701224$&$*$&$*$\\
 $6$&$153828$&$74008728$&$4105498284$&$*$&$*$&$*$\\
 \hline
 \multicolumn{7}{c}{$\quad$}\\
 \hline
 \multicolumn{7}{|c|}{k=3: Ooguri-Vafa Invariants $N_{k,m+n,k,n}$ }\\
 \hline
 $N_{3,m+n,3,n}$&$n=0$&$1$&$2$&$3$&$4$&$5$\\
 \hline
 $m=1$&$0$&$0$&$0$&$0$&$0$&$0$\\
 $2$&$0$&$0$&$0$&$0$&$0$&$0$\\
 $3$&$-8$&$0$&$0$&$192$&$0$&
 $0$\\
 $4$&$0$&$0$&$0$&$0$&$0$&$*$\\
 $5$&$0$&$0$&$0$&$0$&$*$&$*$\\
 $6$&$4$&$0$&$0$&$*$&$*$&$*$\\
 \hline
 \end{tabular}
 \caption{Ooguri-Vafa Invariants $N_{k,m+n,k,n}$ for $X_9^{(1,1,1,3,3)}$ with brane II at large volume. $k$ and $m$ label the class $t_1,t_2$ of $X_9$ and $n$ labels the brane winding. Blue entries agree with Table 6 of \cite{Alim2009} and entries $*$ exceed in this table the order of our calculation}
  \label{table:16}
\end{table}
\subsection{Ooguri-Vafa Invariants for $X^{(1,1,2,2,2)}_8$}
\newpage
\begin{table}[H]
\rotatebox{90}{
\centering
 \begin{tabular}{|c|ccccc|}
 \hline
 \multicolumn{6}{|c|}{k=0: Ooguri-Vafa Invariants $N_{m+n,k,n,k}$ }\\
 \hline
 $N_{0,m+n,0,n}$&$n=0$&$1$&$2$&$3$&$4$\\
 \hline
 $m=1$&$-18$&$-216$&$-4892$&$-151264$&$-5681870$\\
 $2$&$72$&$2624$&$100832$&$4214112$&$191601760$\\
 $3$&$-486$&$-33048$&$-1797228$&$-94486824$&$-5033832642$\\
 $4$&$4608$&$449280$&$31446656$&$1982842880$&$120829061120$\\
 $5$&$-50850$&$-6447800$&$-550590500$&$-40388001000$&$*$\\
 $6$&$614304$&$96158016$&$9679385952$&$*$&$*$\\
 \hline
  \multicolumn{6}{c}{$\quad$ }\\
 \hline
 \multicolumn{6}{|c|}{k=1: Ooguri-Vafa Invariants $N_{m+n,k,n,k}$ }\\
 \hline
 $N_{1,m+n,1,n}$&$n=0$&$1$&$2$&$3$&$4$\\
 \hline
 $m=1$&$-42$&$-6192$&$-733644$&$-79606336$&$-8211597214$\\
 $2$&$774$&$137232$&$17797280$&$2037691872$&$218057272396$\\
 $3$&$-14136$&$-2930112$&$-416393532$&$-50570319792$&$-5639510827530$\\
 $4$&$258054$&$61211088$&$9500484384$&$1226211853424$&$142918383992888$\\
 $5$&$-4712292$&$-1258590240$&$-212363598836$&$-29126855887056$&$*$\\
 $6$&$86082552$&$25561830528$&$4666034521152$&$*$&$*$\\
 \hline
  \multicolumn{6}{c}{$\quad$ }\\
 \hline
 \multicolumn{6}{|c|}{k=2: Ooguri-Vafa Invariants $N_{m+n,k,n,k}$ }\\
 \hline
 $N_{2,m+n,2,n}$&$n=0$&$1$&$2$&$3$&$4$\\
 \hline
 $m=1$&$0$&$-1368$&$-1077372$&$-365059512$&$-86759396282$\\
 $2$&$204$&$198576$&$75466800$&$19307887296$&$3955750079488$\\
 $3$&$-19302$&$-11637432$&$-3670150356$&$-847235304168$&$-162420251412438$\\
 $4$&$903132$&$485877600$&$144775961928$&$32293413751824$&$6042541334443616$\\
 $5$&$-32358618$&$-16917337440$&$-4997647584648$&$-1110817713352920$&$*$\\
 $6$&$995862804$&$525552662880$&$157318976664144$&$*$&$*$\\
 \hline
  \multicolumn{6}{c}{$\quad$ }\\
 \hline
 \multicolumn{6}{|c|}{k=3: Ooguri-Vafa Invariants $N_{m+n,k,n,k}$ }\\
 \hline
 $N_{3,m+n,3,n}$&$n=0$&$1$&$2$&$3$&$4$\\
 \hline
 $m=1$&$0$&$0$&$0$&$-55820$&$-150679552$\\
 $2$&$0$&$11744$&$32336576$&$24359333568$&$10806351662272$\\
 $3$&$-1566$&$-5076$&$-4570885188$&$-2261841187392$&$-932100832928880$\\
 $4$&$413766$&$596077920$&$374635723$&$152201297231776$&$57296094052931103$\\
 $5$&$-38831778$&$-42101500720$&$-22526450326120$&$-9573014306940528$&$*$\\
  $6$&$2350574802$&$2241417150432$&$1107534104102880$&$*$&$*$\\
 \hline
 \end{tabular}}
 \caption{Ooguri-Vafa Invariants $N_{k,m+n,k,n}$ for $X^{(1,1,2,2,2)}_8$. $k$ and $m$ label the class $t_1,t_2$ of $X_8$ and $n$ labels the brane winding. Entries $*$ exceed in this table the order of our calculation}
  \label{table:19}
\end{table}
\subsection{Ooguri-Vafa Invariants for $X_{12}^{(1,1,2,2,6)}$}\def\r{\color{red}}\def\b{\color{blue}}
\begin{table}[H]
\centering
\rotatebox{90}{
 \begin{tabular}{|c|ccccccc|}
 \hline
 \multicolumn{8}{|c|}{k=0: Ooguri-Vafa Invariants $N^I_{m,n,k,k}$}\\
 \hline
 $N_{m,n,0,0}$&$n=0$&$1$&$2$&$3$&$4$&$5$&$6$\\
 \hline
 $m=0$&\b$0$&\b$3$&\b$-12$&\b$81$&$-768$&$8475$&$-102384$\\
 $1$&\b$6$&\b$24$&\b$-150$&\b$1536$&$-18876$&$255192$&$-3658740$\\
 $2$&\b$-3$&\b$51$&\b$-828$&\b$12789$&$-210525$&$3567759$&$-61486848$\\
 $3$&\b$0$&\b$0$&\b$-2448$&\b$62424$&$-1424304$&$30845952$&$-647281800$\\
 $4$&\b$0$&\b$-51$&\b$-4284$&\b$200175$&$-6560640$&$185328198$&$-4798234056$\\
 $5$&$0$&$-24$&$-3822$&$446208$&$-21903294$&$824369400$&$-26693892732$\\
 $6$&$0$&$-3$&$0$&$707832$&$-55037712$&$2824014780$&$-115993486620$\\
 $7$&$0$&$0$&$3822$&$785040$&$-106603500$&$7651196928$&$-404555630886$\\
 $8$&$0$&$0$&$4284$&$527133$&$-161311872$&$16706608653$&$-1155008094300$\\
 \hline
  \multicolumn{8}{c}{$\quad$}\\
  \hline
 \multicolumn{8}{|c|}{k=1: Ooguri-Vafa Invariants $N^I_{m,n,k,k}$}\\
 \hline
 $N_{m,n,1,1}$&$n=0$&$1$&$2$&$3$&$4$&$5$&$6$\\
 \hline
 $m=0$&\b$0$&\b$7$&\b$-129$&\b$2356$&$-43009$&$785382$&$-14347092$\\
 $1$&\b$0$&\b$60$&\b$-1812$&\b$46308$&$-1087560$&$24288552$&$-524645988$\\
 $2$&\b$0$&\b$231$&\b$-11688$&\b$422769$&$-12912432$&$355724019$&$-9146942640$\\
 $3$&\b$0$&\b$0$&\b$-44796$&\b$2369184$&$-95605692$&$3282800160$&$-101243194536$\\
 $4$&\b$0$&\b$-231$&\b$-118617$&\b$9172275$&$-495951000$&$21450278046$&$-799644722631$\\
 $5$&$0$&$-60$&$-233400$&$26397816$&$-1925808876$&$105855800388$&$-4805576016564$
\\
 $6$&$0$&$-7$&$0$&$59555823$&$-5856840424$&$411529509612$&$-22896432050424$\\
 $7$&$0$&$0$&$233400$&$111805968$&$-14462333760$&$1300043526960$&$-89053059573312$
\\
 $8$&$0$&$0$&$118617$&$179694201$&$-29956661529$&$3422531345265$&$-289181994799893$
\\
 \hline
 \end{tabular}}
 \caption{Ooguri-Vafa Invariants $N^I(m,n,k,k)$ for $X^{(1,1,2,2,6)}_{12}$ at large volume. Blue result agree with Table 1 of \cite{Jockers2009} and entries $*$ exceed in this table the order of our calculation }
 \label{table:20}
\end{table}

\begin{table}[H]\def\r{\color{red}}\def\b{\color{blue}}
\centering
\rotatebox{90}{
 \begin{tabular}{|c|ccccccc|}
 \hline
 \multicolumn{8}{|c|}{k=2: Ooguri-Vafa Invariants $N^I_{m,n,k,k}$}\\
 \hline
 $N_{m,n,2,2}$&$n=0$&$1$&$2$&$3$&$4$&$5$&$6$\\
 \hline
 $m=0$&\b$0$&\b$0$&\b$-34$&\b$3172$&$-150522$&$5393103$&$-165977134$\\
 $1$&\b$0$&\b$0$&\b$-438$&\b$62580$&$-3849906$&$168975888$&$-6150177030$\\
 $2$&\b$0$&\b$0$&\b$-2556$&\b$576165$&$-46702074$&$2531136435$&$-109516928532$\\
 $3$&\b$0$&\b$0$&\b$-8592$&\b$3273720$&$-356493366$&$24094632336$&$-1247189399850$\\
 $4$&\b$0$&\b$0$&\b$-20250$&\b$12926751$&$-1922497884$&$163696266396$&$-10205528854440$\\
 $5$&$0$&$0$&$-43854$&$38066712$&$-7813282584$&$846085289856$&$-63961350090048$\\
 $6$&$0$&$0$&$0$&$87483723$&$-24949597038$&$3465967380876$&$-319764160565976$\\
 $7$&$0$&$0$&$43854$&$166124856$&$-64549577412$&$11584119074256$&
 $-1712597040576252$
\\
  $8$&$0$&$0$&$20250$&$288193197$&$-139132564554$&$43650289626$&$*$\\
 \hline
  \multicolumn{8}{c}{$\quad$}\\
  \hline
 \multicolumn{8}{|c|}{k=3: Ooguri-Vafa Invariants $N^I_{m,n,k,k}$}\\
 \hline
 $N_{m,n,1,1}$&$n=0$&$1$&$2$&$3$&$4$&$5$&$6$\\
 \hline
 $m=0$&\b$0$&\b$0$&\b$0$&\b$261$&$68961$&$6471963$&$-391762467$\\
 $1$&\b$0$&\b$0$&\b$0$&\b$4992$&$-1750680$&$203030160$&$-14586604596$\\
 $2$&\b$0$&\b$0$&\b$0$&\b$42489$&$-20943672$&$3047732583$&$-261743302824$\\
 $3$&\b$0$&\b$0$&\b$0$&\b$215676$&$-156884940$&$29101608144$&$-3011532345432$\\
 $4$&\b$0$&\b$0$&\b$0$&\b$738051$&$-827067552$&$198520743132$&$-24958437292476$\\
 $5$&$0$&$0$&$0$&$1834944$&$-3278132892$&$1031301949$&$-191124519232428$\\
 $6$&$0$&$0$&$0$&$3543912$&$-10209175968$&$5273411151753$&$*$\\
 $7$&$0$&$0$&$0$&$5887176$&$-32716895652$&$*$&$*$\\
 $8$&$0$&$0$&$0$&$10856700$&$*$&$*$&$*$\\
 \hline
 \end{tabular}}
 \caption{Ooguri-Vafa Invariants $N^I(m,n,k,k)$ for $X_{(1,1,2,2,6)}$. Blue entries agree with Table 1 of \cite{Jockers2009} and entries $*$ in this table exceed the order of our calculation}
 \label{table:21}
\end{table}

\begin{table}[H]\def\r{\color{red}}\def\b{\color{blue}}
\centering
\rotatebox{90}{
 \begin{tabular}{|c|ccccccc|}
 \hline
 \multicolumn{8}{|c|}{k=0: Ooguri-Vafa Invariants $N^{II}_{m,n,k,k}$}\\
 \hline
 $N_{m,n,0,0}$&$n=0$&$1$&$2$&$3$&$4$&$5$&$6$\\
 \hline
$m=0$& \b$0$ &	\b$2$ & \b$-11$ & \b$90$ & $-956$	&$11470$	&$-148104$\\
$1$&\b$0$	&\b$18$	&\b$-144$&\b$	1728$&	$-23688$	&$347634$	&$-5319648$\\
$2$& \b$0$ &\b$90$&\b$-864$&\b$	14850$&$	-268794$&$	4914882$&$	-90121788$\\
$3$&\b$0$&\b$	0$&\b$	-3072$&\b$	76626$&$	-1869000$&$	43209696$&$	-959687352$\\
$4$&\b$0$&\b$-90$&\b$	-7983$&\b$	268938$&$	-8964558$&$	265751964	$&$-7224918462$\\
$5$&$0$&$	-18$&$	-20016	$&$694368	$&$-31713192$&$	1219991850$&$	-41011490232$\\
$6$&$0	$&$-2$&$	0$&$	1418040$&$	-86560046$&$	4358141460$&$	-182848393872$\\
$7$&$0$&$	0$&$	20016$&$	2551068$&$	-189151560$&$	12482067168$&$	-658806706584$\\
$8$&$0$&$	0	$&$7983$&$	5220558	$&$-343903860$&$	29357148258$&$	-1959719830479$	\\
 \hline
  \multicolumn{8}{c}{$\quad$}\\
  \hline
 \multicolumn{8}{|c|}{k=1: Ooguri-Vafa Invariants $N^{II}_{m,n,k,k}$}\\
 \hline
 $N_{m,n,1,1}$&$n=0$&$1$&$2$&$3$&$4$&$5$&$6$\\
 \hline
$m=0$&\b$0$&\b$	-2$&\b$	0$&\b$	408$&$	-12976$&$	318240$&$	-7064886$	\\
$1$&\b$0$&\b$	-18$&\b$	0$&\b$	8136	$&$-331560$&$	9921366$&$	-260033616$\\
$2$&\b$0$&\b$	-90$&\b$	0$&\b$	76698$&$	-4015620	$&$147332466$&$	-4580847000$\\
$3$&\b$0$&\b$	0$&\b$	0$&\b$	452268$&$	-30639624$&$	1387245600$&$	-51445509768$\\
$4$&$0$&$	90$&$	0$&$	1877238$&$	-165558276$&$	9310082616$&$	-414111436578	$\\
$5$&$0$&$	18$&$	0$&$	5834448	$&$-676162008$&$	47508506460$&$	-2548090964232$\\
$6$&$0$&$	2$&$	0$&$	13963950$&$	-2176167772$&$	192171282588$&$	-12488328860112$\\
$7$&$0$&$	0$&$	0$&$	27159444	$&$-5689981368$&$	634779163584$&$	-50184535778280$\\
$8$&$0$&$	0$&$	0$&$	54249498$&$	-12403534176	$&$1752461446878$&$	-169030073870352$\\
 \hline
 \end{tabular}}
 \caption{Ooguri-Vafa Invariants $N^{II}(m,n,k,k)$ for $X^{(1,1,2,2,6)}_{12}$. Blue entries agree with Table 2 of \cite{Jockers2009} and entries $*$ in this table exceed the order of our calculation }
 \label{table:22}
\end{table}

\begin{table}[H]\def\r{\color{red}}\def\b{\color{blue}}
\centering
\rotatebox{90}{
 \begin{tabular}{|c|ccccccc|}
 \hline
 \multicolumn{8}{|c|}{k=2: Ooguri-Vafa Invariants $N^{II}_{m,n,k,k}$}\\
 \hline
 $N_{m,n,2,2}$&$n=0$&$1$&$2$&$3$&$4$&$5$&$6$\\
 \hline
$m=0$&\b$0$&\b$	0	$&\b$11$&\b$	-408$&$	0$&$	539430$&$	-30383709$\\
$1$&\b$0$&\b$	0$&\b$	144$&\b$	-8136$&$	0$&$	17027136	$&$-1132913808$\\
$2$&\b$0$&\b$	0$&\b$	864$&\b$	-76698	$&$0$&$	258298074$&$	-20374948188$\\
$3$&\b$0$&\b$	0$&\b$	3072$&\b$	-452268$&$	0$&$	2503487904$&$	-235230388200$\\
$4$&\b$0$&\b$	0$&\b$	7983	$&\b$-1877238$&$	0$&$	17412238368$&$	-1958962807170$\\
$5$&$0$&$	0$&$	20016$&$	-5834448$&$	0$&$	92608329600	$&$-12543118517016$\\
$6$&$0$&$	0$&$	0$&$	-13963950$&$	0$&$	391990737708$&$	-64292612674320$\\
$7$&$0$&$	0$&$	-20016$&$	-27159444$&$	0$&$	1357019627616$&$	-289834848988800$\\
$8$&$0$&$	0$&$	-7983$&$	-54249498$&$	0$&$	4136493648546$&$	*$\\
 \hline
  \multicolumn{8}{c}{$\quad$}\\
  \hline
 \multicolumn{8}{|c|}{k=3: Ooguri-Vafa Invariants $N^{II}_{m,n,k,k}$}\\
 \hline
 $N_{m,n,3,3}$&$n=0$&$1$&$2$&$3$&$4$&$5$&$6$\\
 \hline
$m=0$&\b$0$&\b$	0$&\b$	0$&\b$	-90$&$	12976$&$	-539430$&$	0$\\
$1$&\b$0$&\b$	0$&\b$	0$&\b$	-1728$&$	331560$&$	-17027136$&$	0$\\
$2$&\b$0$&\b$	0$&\b$	0$&\b$	-14850$&$	4015620	$&$-258298074$&$	0$\\
$3$&\b$0$&\b$	0$&\b$	0$&\b$	-76626	$&$30639624$&$	-2503487904$&$	0$\\
$4$&\b$0$&\b$	0$&\b$	0$&\b$	-268938$&$	165558276$&$	-17412238368	$&$0$\\
$5$&$0$&$	0$&$	0$&$	-694368$&$	676162008$&$	-92608329600$&$	-2309467987008$\\
$6$&$0$&$	0$&$	0$&$	-1418040$&$	2176167772$&$	-351774546534$&$	-22545027404271$\\
$7$&$0$&$	0$&$	0$&$	-2551068	$&$5200228704$&$	-1014274703904$&$	*$\\
$8$&$0$&$	0$&$	0$&$	-6391956$&$	8448740559$&$	*$&$	*$\\
 \hline
 \end{tabular}}
 \caption{Ooguri-Vafa Invariants $N^{II}(m,n,k,k)$ for $X^{(1,1,2,2,6)}_{12}$. Blue entries agree with Table 2 of \cite{Jockers2009} and entries $*$ in this table exceed the order of our calculation }
 \label{table:23}
\end{table} 
\section{Ooguri-Vafa Invariants for One Closed and Two Open Moduli}\label{App:C}
\begin{table}[H]
\centering
\rotatebox{90}{
 \begin{tabular}{|c|cccc|}
 \hline
 \multicolumn{5}{|c|}{ Ooguri-Vafa Invariants $N_{m,m+n,m}$ for $\mathbb{P}_{11111}[3,3]$}\\
 \hline
  $N_{m,m+n,m}$&$n=0$&$1$&$2$&$3$\\
 \hline
 $m=0$&$0$&$	-12$&$	0$&$	0$\\
$1$&$120$&$	108$&$	-288$&$	60$\\
$2$&$-2100$&$	11664$&$	9612$&$	-33876$\\
$3$&$70440$&$	-525264$&$	2066280$&$	1751112$\\
$4$&$-3191280$&$	29727324$&$	-139852800$&$	478494708	$\\
 \hline
 \multicolumn{5}{c}{ $\quad$}\\
 \hline
 \multicolumn{5}{|c|}{ Ooguri-Vafa Invariants $N_{m,m+n,m}$ for $\mathbb{P}_{112112}[4,4]$}\\
 \hline
 $N_{m,m+n,m}$&$n=0$&$1$&$2$&$3$\\
 \hline
$m=0$&$ 0$&$	-16$&$	0	$&$0$\\
$1$&$552$&$	1120$&$	-2544$&$	1056$\\	
$2$&$-39264	$&$281280$&$	539392$&$	-1693632$\\
$3$&$5532984	$&$-52350528$&$	280580256$&$	557960544	$\\
$4$&$-1043002176$&$	12346010064$&$	-77352996864$&$	372579061488$	\\
\hline
\multicolumn{5}{c}{$\quad$}\\
\hline
\multicolumn{5}{|c|}{ Ooguri-Vafa Invariants $N_{m,m+n,m}$ for $\mathbb{P}_{123123}[6,6]$}\\
 \hline
 $N_{m,m+n,m}$&$n=0$&$1$&$2$&$3$\\
 \hline
$m=1$&$0$&$	-16$&$	0$&$	0$\\
$1$&$27240$&$	72800$&$	-144240$&$	53280$\\
$2$&$-93024240$&$	697855680$&$	1746803968$&$	-4807419840$\\
$3$&$633116600568$&$	-6085353582144$&$	34589091686304$&$	86687436861792$\\
$4$&$-5741552410002720$&$	68585896764953040$&$	-442591883444977920$&$	2274867202698155760$\\
\hline
 \end{tabular}}
 \caption{Ooguri-Vafa Invariants for $\mathbb{P}^{(111|111)}_{[3,3]}$,$\mathbb{P}^{[112|112]}_{[4,4]}$ and $\mathbb{P}^{[123|123]}_{[6,6]}$ at large volume. The entries $*$ in this table exceed the order of our calculation.}
 \label{table:24}
\end{table}

\section{Ooguri-Vafa Invariants for One Closed and Three Open Moduli}
\begin{table}[H]
\centering
\begin{tabular}{|c|cccccc|}
 \hline
 \multicolumn{7}{|c|}{$U(1)$ Ooguri-Vafa Invariants $N_{m,m,m,n}$ }\\
 \hline
  $N_{m,m,m,n}$&$n=0$&$1$&$2$&$3$&$4$&$5$\\
 \hline
$m=0$&$0$&$0$&$	0$&$	0$&$	0$&$	0$\\
$1$&$0$&$	-4896$&$	0$&$	0$&$	0$&$	0$\\
$2$&$0$&$	0$&$	1729728$&$	0$&$	0$&$	0$\\
$3$&$0$&$	0$&$	0$&$	-1530550656	$&$0$&$	0$\\
$4$&$0$&$	0$&$	0$&$	0	$&$1882007669502$&$	0$\\

\hline
\multicolumn{7}{c}{ $\quad$}\\
\multicolumn{7}{c}{ $\quad$}\\
\hline
\multicolumn{7}{|c|}{$U(2)$ Ooguri-Vafa Invariants $N_{m,m,n}$}\\
\hline
 $N_{m,m,n}$&$n=0$&$1$&$2$&$3$&$4$&$5$\\
\hline
$m=0$&$0$ & $0$ & $0$ & $0$ & $0$& $0$ \\
$1$&$0$ & $4832$ & $0$ & $0$ & $0$ & $0$ \\
$2$&$0$ & $0$ & $6266104$ & $0$ & $0$ & $0$\\
$3$&$0$ & $0$ & $0$ & $\frac{55336185568}{3}$ & $0$& $0$\\
$4$&$0$ & $0$ & $0$ & $0$ & $75779426010282$ & $-3443691801600$\\
\hline
\end{tabular}
\caption{Ooguri-Vafa invariant of Sextic Hypersurface}
 \label{table:26}
\end{table}

\begin{table}[H]
\centering
\begin{tabular}{|c|cccccc|}
\hline
 \multicolumn{7}{|c|}{$U(1)$ Ooguri-Vafa Invariants $N_{m,m,m,n}$ }\\
 \hline
$N_{m,m,m,n}$ & $n=0$ & $n=1$& $n=2$& $n=3$ & $n=4$&$n=5$\\
\hline
$m=0$&$0$ & $0$ & $0$& $0$& $0$ & $0$\\
$1$&$0$ & $-3328$ & $0$ & $0$ & $0$ & $0$\\
$2$&$0$ & $0$ & $2030336$ &$0$ & $0$ & $0$\\
$3$&$0$ & $0$ & $0$ & $	-3115785728$ & $0$ & $0$\\
$4$&$0$ & $0$ & $0$ & $237081600$	&$*$ &$	*$\\
\hline
\multicolumn{7}{c}{$\quad$}\\
\multicolumn{7}{c}{$\quad$}\\
\hline
\multicolumn{7}{|c|}{$U(2)$ Ooguri-Vafa Invariants $N_{m,m,n}$ }\\
 \hline
$N_{m,m,n}$ & $n=0$ & $n=1$& $n=2$& $n=3$ & $n=4$&$n=5$\\
\hline
$m=0$&$0$&$	0$&$	0$&$	0$&$	0$&$	0$\\
$1$&$0$&$	-5504$&$	0$&$	0$&$	0$&$	0$\\
$2$&$0$&$	0$&$	9520016$&$	0$&$	0$&$	0$\\
$3$&$0$&$	0$&$	0$&$	\frac{-430429613824}{15}$&$	0$&$	0$\\
$4$&$0$&$	0$&$	0$&$	0$&$	\frac{1000605023457608}{7}$&$	-9744720529920$\\
\hline
\end{tabular}
\caption{Ooguri-Vafa invariant of Octic Hypersurface}
\label{table:27}
\end{table}

\printbibliography
\end{document}